\documentclass[aps,prd,groupedaddress,float,amsmath,amssymb,nofootinbib]{revtex4-2}

\usepackage[dvipdfmx]{graphicx}% Include figure files
\usepackage{dcolumn}% Align table columns on decimal point
\usepackage{bm}% bold math
\usepackage{braket}
\usepackage{breqn}
\usepackage{color}
\usepackage{hyperref}

\renewcommand{\Re}[1]{\mathrm{Re}\left[ #1 \right]}
\renewcommand{\Im}[1]{\mathrm{Im}\left[ #1 \right]}

\newcommand{\ti}{t_{\rm ini}}

\begin{document}

% Use the \preprint command to place your local institutional report
% number in the upper righthand corner of the title page in preprint mode.
% Multiple \preprint commands are allowed.
% Use the 'preprintnumbers' class option to override journal defaults
% to display numbers if necessary
%\preprint{}

%Title of paper
\title{Impact of multiple modes on the evolution of\\ self-interacting axion condensate around rotating black holes}

% repeat the \author .. \affiliation etc. as needed
% \email, \thanks, \homepage, \altaffiliation all apply to the current
% author. Explanatory text should go in the []'s, actual e-mail
% address or url should go in the {}'s for \email and \homepage.
% Please use the appropriate macro foreach each type of information

% \affiliation command applies to all authors since the last
% \affiliation command. The \affiliation command should follow the
% other information
% \affiliation can be followed by \email, \homepage, \thanks as well.
\author{Hidetoshi Omiya}
\email{omiya@tap.scphys.kyoto-u.ac.jp}
\affiliation{Department of Physics$,$ Kyoto University$,$ Kyoto 606-8502$,$ Japan}
\author{Takuya Takahashi}
\email{t.takahashi@tap.scphys.kyoto-u.ac.jp}
\affiliation{Department of Physics$,$ Kyoto University$,$ Kyoto 606-8502$,$ Japan}
\author{Takahiro Tanaka}
\email{t.tanaka@tap.scphys.kyoto-u.ac.jp}
\affiliation{Department of Physics$,$ Kyoto University$,$ Kyoto 606-8502$,$ Japan}
\affiliation{Center for Gravitational Physics and Qunatum Information$,$ Yukawa Institute for Theoretical Physics$,$ Kyoto University$,$ Kyoto 606-8502$,$ Japan}
\author{Hirotaka Yoshino}
\affiliation{Department of Physics$,$ Osaka Metropolitan University$,$ Osaka 558-8585$,$ Japan}
\email{hyoshino@omu.ac.jp}

%Collaboration name if desired (requires use of superscriptaddress
%option in \documentclass). \noaffiliation is required (may also be
%used with the \author command).
%\collaboration can be followed by \email, \homepage, \thanks as well.
%\collaboration{}
%\noaffiliation

\date{\today}

\begin{abstract}
Ultra-light particles, such as axions, form a macroscopic condensate around a highly spinning black hole by the superradiant instability. Due to its macroscopic nature, the condensate opens the possibility of detecting the axion through gravitational wave observations. However, the precise evolution of the condensate must be known for the actual detection. For future observation, we numerically study the influence of the self-interaction, especially interaction between different modes, on the evolution of the condensate in detail. First, we focus on the case when condensate starts with the smallest possible angular quantum number. For this case, we perform the non-linear calculation and show that the dissipation induced by the mode interaction is strong enough to saturate the superradiant instability, even if the secondary cloud starts with quantum fluctuations. Our result indicates that explosive phenomena such as bosenova do not occur in this case. 
 We also show that the condensate settles to a quasi-stationary state mainly composed of two modes, one with the smallest angular quantum number for which the superradiant instability occurs and the other with the adjacent higher angular quantum number. We also study the case when the condensate starts with the dominance of the higher angular quantum number. We show that the dissipation process induced by the mode coupling does not occur for small gravitational coupling. Therefore, bosenova might occur in this case.
\end{abstract}

% insert suggested keywords - APS authors don't need to do this
%\key words{}

%\maketitle must follow title, authors, abstract, and keywords

\maketitle

\section{Introduction}

Axions are well-motivated field contents in a variety of contexts. For example, they are invoked to solve the strong CP problem~\cite{Peccei:1977hh,Weinberg:1977ma, Wilczek:1977pj,Kim:1979if,Shifman:1979if,Zhitnitsky:1980tq,Dine:1981rt}, can be a good candidate for dark matter~\cite{Dine:1982ah,Preskill:1982cy,Abbott:1982af,Hui:2016ltb,Adams:2022pbo}, and are naturally and copiously derived from string theory~\cite{Svrcek:2006yi,Arvanitaki:2009fg,Cicoli:2012sz}. The axions derived from string theory can have Compton wavelengths comparable to astrophysically relevant scales. In such cases, axions are expected to cause various interesting astrophysical phenomena~\cite{Arvanitaki:2009fg,Arvanitaki:2010sy}. 

When the Compton wavelength is approximately the same as the black hole (BH) radius, axions form a macroscopic condensate around a BH because of the superradiant instability~\cite{Zouros:1979iw,Detweiler:1980uk,Dolan:2007mj, Brito:2015oca}. Here, we refer to the macroscopic condensate occupying a single mode as an axion cloud. 
The superradiant instability is fast enough that, even if the axion condensate starts with a small amplitude such as the level unavoidable from quantum fluctuations, it grows to be sufficiently large within the age of the universe. 
As it grows, the rotational energy is extracted from the BH, which reduces the BH spin. Clouds also have the quadrupole moment and radiate continuous gravitational waves. 
Therefore, the existence of an axion field is potentially verified by observing the distribution of the spins of BHs and continuous gravitational waves~\cite{Brito:2014wla,Arvanitaki:2014wva, Arvanitaki:2016qwi,LIGOScientific:2021jlr,Saha:2022hcd}.

Detailed cloud dynamics need to be understood for future verification of axion field by observations. Possible influences on the cloud dynamics include the axion field self-interaction~\cite{Arvanitaki:2010sy,Yoshino:2012kn,Mocanu:2012fd, Yoshino:2015nsa,Gruzinov:2016hcq,Fukuda:2019ewf,Baryakhtar:2020gao, Omiya:2020vji, Omiya:2022mwv}, the tidal interaction in binary systems~\cite{Baumann:2018vus,Baumann:2019ztm,Takahashi:2021eso,Takahashi:2021yhy,Baumann:2021fkf,Tong:2022bbl,East:2022ppo}, and interactions with other fields~\cite{Rosa:2017ury,Ikeda:2018nhb}. Among these effects, the self-interaction would have a particularly large impact. For example, if the self-interaction is attractive, the cloud growth is accelerated~\cite{Omiya:2020vji,Omiya:2022mwv}, which eventually leads to the collapse of the condensate~\cite{Yoshino:2012kn,Yoshino:2015nsa}. This collapse, called bosenova, is thought to cause a burst of gravitational waves. It has also been suggested that the interaction between clouds may dissipate the energy of the condensate efficiently and forces it to settle into a quasi-stationary state~\cite{Baryakhtar:2020gao}. In such cases, explosive phenomena will not occur.

However, previous studies on the effects of self-interaction are not satisfactory. For example, most of the works treat the self-interaction only perturbatively~\cite{Gruzinov:2016hcq,Fukuda:2019ewf,Baryakhtar:2020gao, Omiya:2020vji}, but perturbative approach breaks down once the self-interaction starts to accelerate the instability~\cite{Omiya:2020vji}. Perturbative treatment usually misses the effects of the deformation of the cloud. Furthermore, the non-relativistic approximation, commonly used in many literatures, cannot be valid in analyzing relativistic cases. On the other hand, the non-perturbative and relativistic treatment in Ref.~\cite{Omiya:2022mwv} assumes that the condensate is composed of a single cloud, {\it i.e.}, all axion particles are in the same state, and misses the dissipation caused by the interaction between multiple clouds. Also, it is difficult to numerically track the whole evolution of the condensate starting from a very small amplitude until it reaches such a large amplitude that the self-interaction is important, by means of dynamical simulations because of the large discrepancy between the dynamical and the instability timescales.

In this paper, we extend the non-perturbative calculation scheme proposed in~\cite{Omiya:2022mwv} to include the presence of the secondary cloud. Our calculations show that the two clouds attract each other, which enhances the dissipation caused by the cloud-cloud interaction. 
As a result, the condensate settles to a quasi-stationary state
in most cases. Furthermore, thus realized quasi-stationary state turns out to be stable against excitations of other modes. 
We also find a parameter region where the dissipation does not work efficiently when the fastest growing mode is a higher multipole mode with a large specific angular momentum.
In this case, the condensate may become large enough to cause a bosenova.  

This paper is organized as follows. In Sec.~\ref{sec:2A}, we review the linear analysis of the evolution of axion clouds. In Sec.~\ref{sec:2B}, we review the perturbative analysis of self-interaction. In Sec.~\ref{sec:3}, we formulate a non-perturbative calculation method and discuss the non-linear evolution of the condensate, starting with a superposition of two different modes. In Sec.~\ref{sec:4}, we examine whether or not the other modes can be excited. In Sec.~\ref{sec:5}, we analyze the case in which the condensate starts with the dominance of a higher multipole mode. In Sec.~\ref{sec:6}, we summarize our results and comment on the spin-down of the central BH and gravitational wave emission from the clouds. In the rest of this paper, we take units $c = G = \hbar = 1$, unless otherwise stated.

\section{Perturbative evolution of a self-interacting axion condensate}

In this paper we adopt the following action for the axion
\begin{align}\label{eq:action}
	S = F_a^2 \int d^4\! x \sqrt{-g} \left\{ - \frac{1}{2}g^{\mu\nu}\partial_\mu \phi \,\partial_\nu \phi - \mu^2 \left(1 - \cos\phi\right)\right\}~,
\end{align}
where $\mu$ and $F_a$ give the axion mass and decay constant, respectively. 
%Since we set $G=1$, $F_a$ should be understood as normalized by $M_{pl}=G^{-1/2}$. 
Here, the background metric is given by the Kerr metric,
\begin{align}
\label{kerrmetric}
	ds^2 &=  - \left(1 - \frac{2 M r}{\rho^2}\right)dt^2 - \frac{4 a M r\sin^2\!\theta}{\rho^2} dt\, d\varphi \cr
	 &\ \ \ \ \ + \left[(r^2 + a^2)\sin^2\!\theta+ \frac{2 M r}{\rho^2} a^2 \sin^4\! \theta \right]d\varphi^2 + \frac{\rho^2}{\Delta} dr^2 + \rho^2 d\theta^2~,
\end{align}
with
\begin{align}
	\Delta &= r^2 - 2 Mr + a^2~,\qquad \rho^2 = r^2 + a^2\cos^2\!\theta~.
\end{align}
The two roots of $\Delta=0$ specify the positions of the event horizon $r_+$ and the Cauchy horizon $r_-$, respectively. 
Varying the action \eqref{eq:action} with respect to $\phi$, we obtain the equation of motion of the axion as
\begin{align}
	\square_g \phi - \mu^2 \sin\phi = 0~.
	\label{eq:eomfull}
\end{align}

\subsection{Axion cloud without self-interaction}\label{sec:2A}

First, we consider the case in which the amplitude of the axion is small ($|\phi| \ll 1$). In this case, the self-interaction of the axion is negligible and we can linearize Eq.~\eqref{eq:eomfull}, to obtain
\begin{align}
	(\square_g  - \mu^2) \phi = 0~.
	\label{eq:eomzeroth}
\end{align}
Following Ref.~\cite{Brill:1972xj}, we assume
\begin{align}
	\label{eq:sep}
	\phi  = \Re{e^{-i(\omega t - m \varphi)}S_{lm\omega}(\theta)R_{lm\omega}(r)}~.
\end{align}
The two integers $l$ and $m$ and one complex number $\omega$ specify a solution. 
The angular function $S_{lm\omega}$ and the radial function $R_{lm\omega}$ obey
\begin{align}
		\frac{1}{\sin\theta}\frac{d}{d\theta}\left(\sin\theta \frac{d{S}_{lm\omega}}{d\theta}\right) + \left[c^2(\omega) \cos^2\!\theta - \frac{m^2}{\sin^2\theta}\right]S_{lm\omega} = - \Lambda_{lm}(\omega) S_{lm\omega}~,\label{eq:EOMang}
\end{align}
\begin{align}
		\frac{d}{dr}\left(\Delta\frac{dR_{lm\omega}}{dr}\right)  + \left[\frac{K^2(\omega)}{\Delta} - \mu^2 r^2 -\lambda_{lm}(\omega) \right]R_{lm\omega} = 0~,\label{eq:EOMrad}
\end{align}
where
\begin{align}
		 c^2 (\omega) &= a^2 (\omega^2 - \mu^2)~, \qquad K(\omega) = (r^2+a^2)\omega - am~,\cr
	 \lambda_{lm}(\omega) & = -2am \omega +a^2\omega^2 +\Lambda_{lm}(\omega)~,
\end{align}
and $\Lambda_{lm}(\omega)$ is the separation constant. The physically appropriate boundary conditions for the radial function are the ingoing boundary condition at the event horizon and the exponentially decaying boundary condition at infinity, which are
\begin{align}\label{eq:bclinear}
	R_{lm\omega}(r) \to \begin{cases}
	(\frac{r-r_+}{M})^{-i \frac{2 M r_+}{r_+ - r_-}(\omega - m \Omega_H)}~, & (r\to r_+)~,\\
	\frac{1}{r/M}\left(\frac{r}{M}\right)^{-\frac{\mu^2 - 2 \omega^2}{\sqrt{\mu^2 - \omega^2}}} e^{-\sqrt{\mu^2 - \omega^2}r}~, & (r\to \infty)~,
	\end{cases}
\end{align}
with the horizon rotation velocity $\Omega_H = a/2Mr_+$. 

\begin{figure}[t]
        \centering
	\includegraphics[keepaspectratio,scale=0.4]{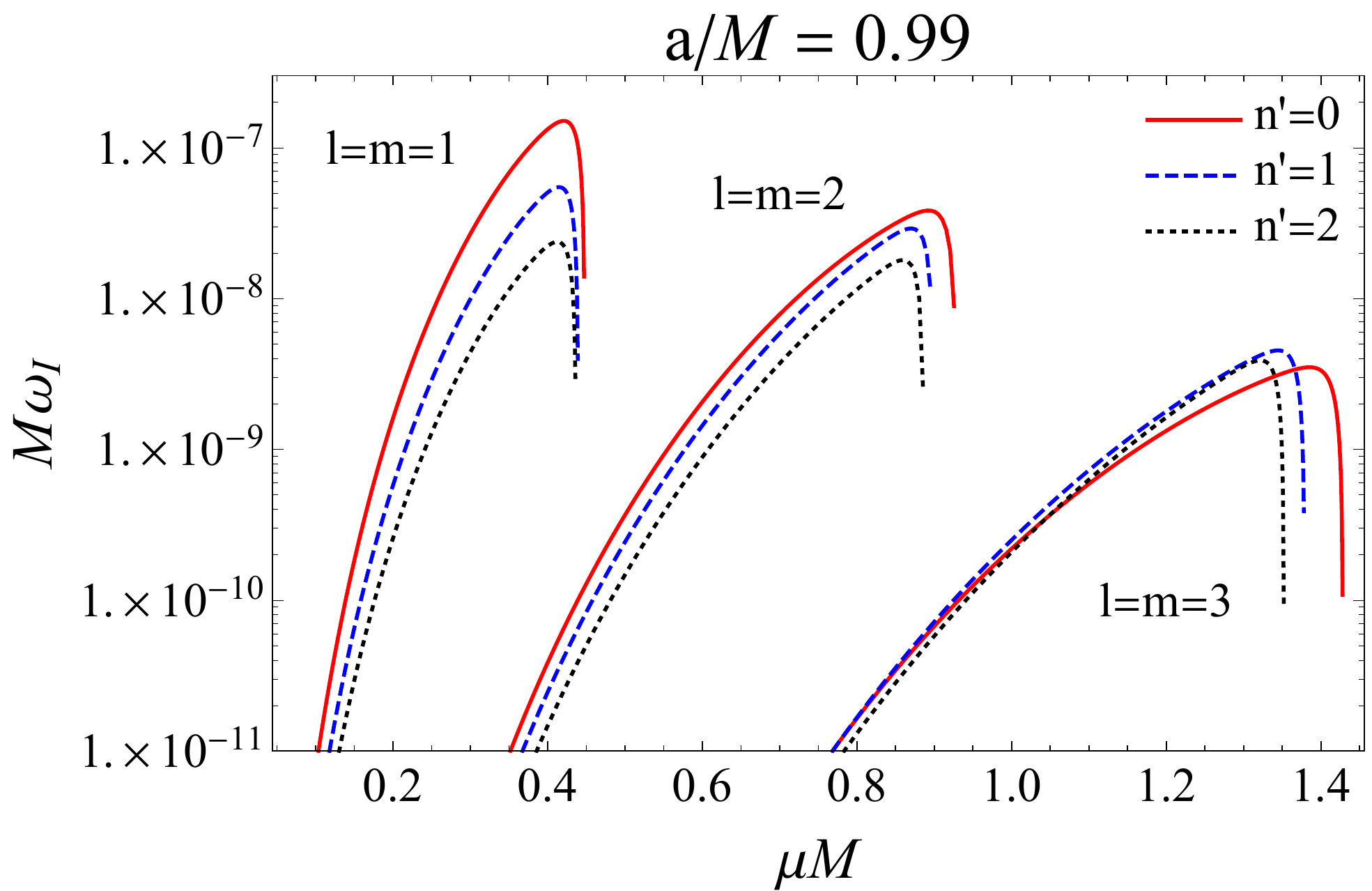}
	\caption{The growth rate of the superradiant instability as function of $\mu M$. The red solid, blue dashed, and black dotted lines correspond to the imaginary part of the frequencies $\omega^{(n)}$ for $n = l + 1 + n',$ with  $n' = 0,1,$ and $2$, respectively. The spin of the central BH is set to $a/M = 0.99$.}
	\label{fig:omegaI}
\end{figure}

\begin{figure}[t]
        \centering
	\includegraphics[keepaspectratio,scale=0.4]{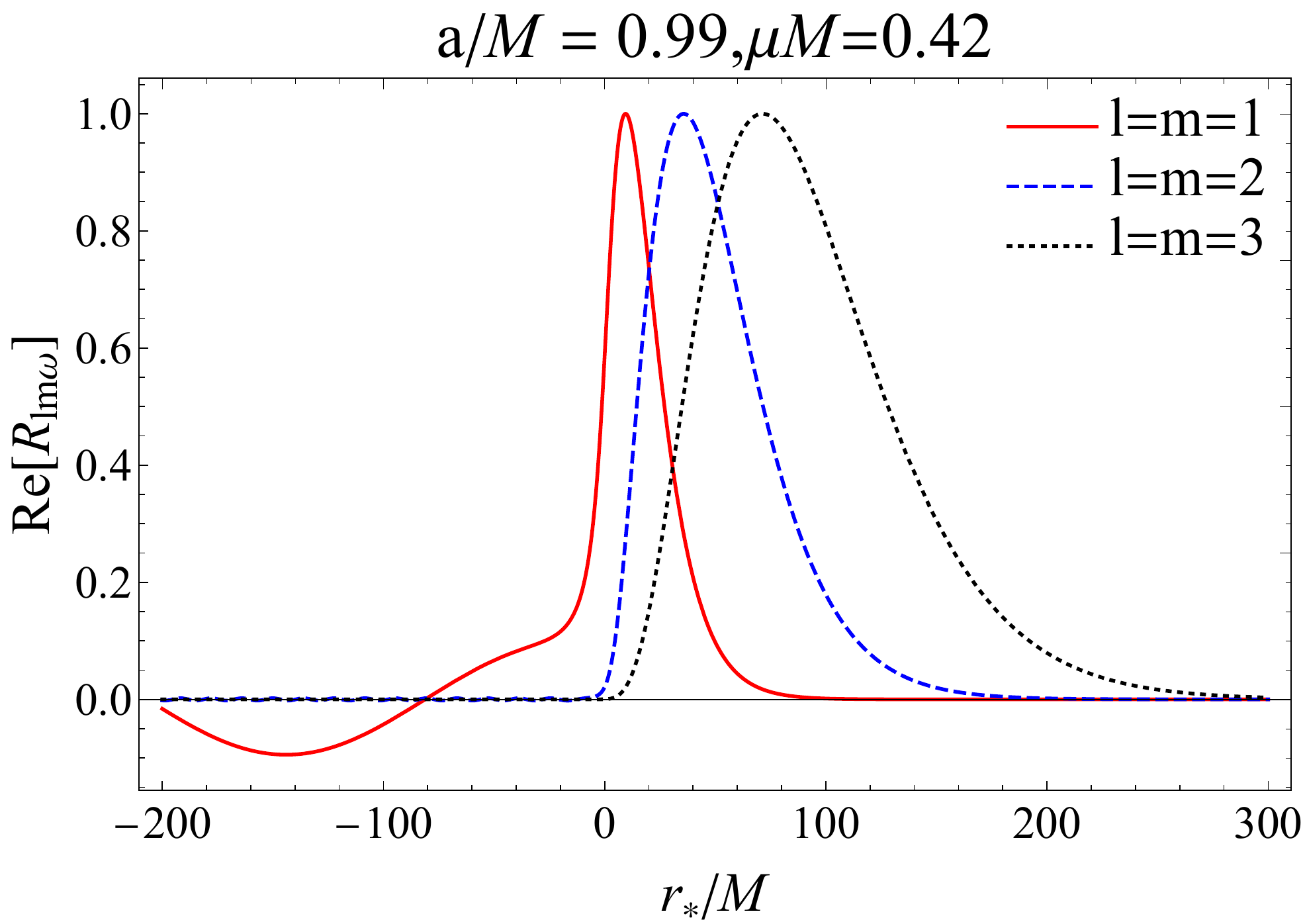}
	\caption{The real part of the radial profile of the bound states of the axion cloud when the non-linear self-interaction is absent. The red solid, blue dashed, and black dotted lines correspond to the mode functions of the fundamental modes with $l=m=1$, $l=m=2$, and $l=m=3$, respectively. The spin of the central BH and the mass of axion are set to $a/M = 0.99$ and $\mu M = 0.42$, respectively. We normalize the peak amplitude of $R_{lm\omega}$ to 1. The horizontal axis corresponds to the tortoise coordinate $r_*$, defined by $d r_* = (r^2+a^2)/\Delta dr$.}
	\label{fig:linconfig}
\end{figure}

By imposing the boundary conditions \eqref{eq:bclinear}, the frequency $\omega$ takes discrete values. The behavior of the radial function is similar to that of a hydrogen atom. 
Following the case of hydrogen atom, we label discrete frequencies by integers $n (\geq l + 1)$ and 
denote the corresponding solution as 
\begin{align}
	\phi_{nlm} &= \Re{\Phi_{nlm}}~,\\
	\Phi_{nlm} &= e^{-i(\omega^{(n)} t - m \varphi)}S_{lm\omega^{(n)}}(\theta)R_{lm\omega^{(n)}}(r)~.
\end{align}
We refer to the solution corresponding to the smallest $n(=l+1)$ as the fundamental mode and the ones corresponding to larger values of $n$ as overtone modes. 
We solve the radial and angular differential equations, \eqref{eq:EOMang} and \eqref{eq:EOMrad}, by the continued fraction method presented in~\cite{Dolan:2007mj}. 
In Fig.~\ref{fig:omegaI} we show the imaginary part of the frequency $\omega^{(n)}$ and in Fig.~\ref{fig:linconfig} we show the radial mode function of the fundamental mode obtained numerically. 

As can be observed in Fig.~\ref{fig:omegaI}, the imaginary part of $\omega^{(n)}$ is positive. This clearly indicates that these modes are exponentially growing in time. 
From various investigations~\cite{Zouros:1979iw, Detweiler:1980uk, Dolan:2007mj}, it is known that these unstable modes are superradiant
\begin{align}\label{eq:SRcondition}
	\omega^{(n)}_{R} < m \Omega_H~,
\end{align}
and bounded,
\begin{align}\label{eq:boundcondition}
	\omega^{(n)}_R <\mu~,
\end{align}
and evolve adiabatically,
\begin{align}\label{eq:adiabacticcondition}
	\omega^{(n)}_{I} \ll \omega^{(n)}_R~.
\end{align}
Here, the subscripts $R$ and $I$, respectively, represent the real and imaginary parts, {\it i.e.}, 
$\omega_R = \Re{\omega}$ and $\omega_I = \Im{\omega}$.
These bounded modes continue to extract the energy by the superradiance. 
Although the growth due to the superradiant instability is much slower compared with the dynamical timescale, 
it is fast enough for the axion to accumulate around BHs within the age of the universe~\cite{Arvanitaki:2010sy}. 
Recall that we denote the component of the condensate 
composed of each growing bounded mode as an axion cloud in this paper. 

For a fixed axion mass $\mu$, the instability rate gets smaller as the eigenvalue of the mode corresponding to the specific orbital angular momentum $l$ increases.  
This is because the axion need to tunnel through the angular momentum barrier to extract the energy from the BH. 
For a larger $l$, the angular momentum barrier gets higher and thus the tunneling rate is suppressed more. 
The presence of the angular momentum barrier also explains the suppression of mode functions near the event horizon in Fig.~\ref{fig:linconfig}. 
Another feature of clouds with a large $l$ is that they are spatially more extended to a large $r$ region. 
These observations will become important to understand the results in the following sections.

\subsection{Saturation of superradiant instability by self-interaction}\label{sec:2B}

In this subsection we take into account the effect of the self-interaction perturbatively. 
Keeping only the leading non-linear term in Eq.~\eqref{eq:eomfull}, we obtain
\begin{align}
	\square_g \phi - \mu^2 \phi = -\frac{\mu^2}{3!}\phi^3~.
	\label{eq:eomfull2}
\end{align}
We perturbatively expand the axion field as 
\begin{align}
	\phi = \phi^{(0)} + \phi^{(1)} + \cdots~,
\end{align}
and then Eq.~\eqref{eq:eomfull2} reads
\begin{align}
	(\square_g - \mu^2) \phi^{(0)} &= 0~,\label{eq:zeroth}\\
	(\square_g - \mu^2)\phi^{(1)} &= -\frac{\mu^2}{3!}\left(\phi^{(0)}\right)^3\label{eq:first}~, 
\end{align}
at the leading order and the next. 
We consider the case in which $\phi^{(0)}$ is given by a superposition of two different clouds,
\begin{align}
	\phi^{(0)} = \sqrt{E_1} \phi_{1} + \sqrt{E_2} \phi_{2}~, 
\end{align}
where $\phi_{1} = \phi_{n_1l_1m_1}$ and $\phi_2 = \phi_{n_2l_2m_2}$.
In the following, we normalize $R_i = R_{l_im_i\omega_i}$ in such a way that the energy of the $i$-th cloud calculated by
\begin{align}
	\int dr\, d\cos\theta\, d\varphi \ (r^2 + a^2 \cos^2\!\theta)\sqrt{g^{tt}}T_{\mu\nu}(\phi_i)\xi^{\mu}_{(t)} n^\nu_{(t)}~,
\end{align}
is equal to 1. Here, $T_{\mu\nu}(\phi_i)$ is the energy momentum tensor composed of $\phi_i$, $\xi^{\mu}_{(t)}$ is 
the time-like killing vector given by $\xi^\mu_{(t)} = (1,0,0,0)$, and $n^\mu_{(t)}$ is the future-directed unit normal to the $t = {\rm constant}$ surface. Substituting $\phi^{(0)}$ into Eq.~\eqref{eq:first}, we obtain
\begin{align}\label{eq:firstex}
	(\square_g - \mu^2)\phi^{(1)} = -\frac{\mu^2}{3!}\left(E_1^{3/2} \phi_{1}^3 +  E_2^{3/2}\phi_{2}^3 +3 E_1 E_2^{1/2}\phi_1^2\phi_2 +3 E_1^{1/2} E_2 \phi_1\phi_2^2 \right)~.
\end{align}

Before starting to solve the first order equation \eqref{eq:first}, 
we formally examine what kind of terms are contained in the right-hand side of this equation. 
Since each term is composed of a product of $\phi_{i}$, its $(t,\varphi)$-dependence is factorized as $\propto e^{- i (\omega t - m \varphi)}$, with some $(\omega, m)$. 
Hence, we classify the terms by their value of $(\omega,m)$, based on the criteria for the superradiance~\eqref{eq:SRcondition} and for the boundedness~\eqref{eq:boundcondition}.

For concreteness, we work with the combination of the fundamental mode of the lowest multipole with   $l_1=m_1=1, n_1= 2$, and that of the second lowest multipole with $l_2 = m_2 = 2, n_2 = 3$. We take the second mode not to be a $l=m=1$ overtone which has a larger growth rate than the $l=m=2$ modes, since $l=m=1$ overtones turn out to decay in the two mode approximation adopted in this section. This point will be discussed in Sec.~\ref{sec:4A1}.
The classification can be done in the following manner\footnote{Since $\phi_i$ is real, if the term with $(\omega,m)$ exists, so does the term with $(-\omega^*,-m)$. The latter is just the complex conjugate of the former, and hence it is abbreviated in the classification list of $(\omega,m)$.}:

\begin{figure}[t]
 \centering
 \includegraphics[keepaspectratio, scale=0.4]{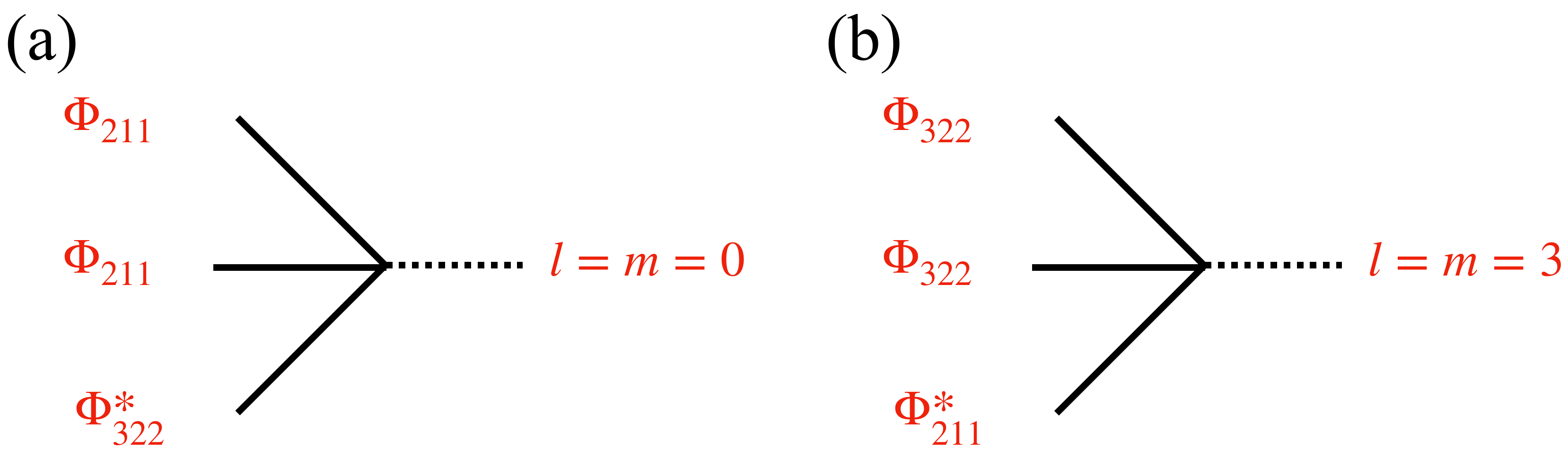}
 \caption{Diagrams of the dominant emission processes. Solid lines correspond to the modes in the source term, which form clouds, and dotted lines correspond to the mode which falls to the BH or escapes to infinity. The left diagram (a) corresponds to the excitation of $(\omega,m) = (2\omega_1 - \omega_2^*,0)$ mode. This process sends the energy of the condensate back into the horizon. While the right one (b) corresponds to the excitation of a low frequency radiation  $(\omega,m) = (2\omega_2 - \omega_1^*,3)$  and dissipates the energy to infinity. 
 The right one (b) gives the dominant contribution to the energy flux between these two processes. In principle, excitations of higher $l$ modes can also contribute, but their contributions are smaller, since the overlap of the wave function of higher $l$ modes with the modes that appear in the source is smaller compared to the mode with the next smallest value of $l$.}
 \label{fig:emissionleading}
\end{figure}

\begin{description}

\item[$(\omega,m) = (3\omega_1,3), (2\omega_1 + \omega_2,4), (\omega_1 + 2\omega_2,5), (3\omega_2,6)$]\mbox{}

These excited modes have a frequency around $\omega \sim 3 \mu$, which is much larger than the axion mass $\mu$. Therefore, these modes are not bounded and escape to infinity. 
However, as long as the self-interaction can be treated perturbatively, the energy flux through these modes is too small to terminate the growth of clouds~\cite{Baryakhtar:2020gao, Omiya:2020vji, Omiya:2022mwv}. For this reason, we neglect these modes in this section. 
The energy flux is suppressed because the wave length of these modes ($\sim 1/3\mu$) is much shorter than the spatial extension of the cloud ($\sim 1/\mu(\mu M)$, as long as $\mu M\lesssim 1$. 
). 
In this sense, we call these modes high frequency radiation.

\item[$(\omega,m) = (2\omega_2 - \omega_1^* , 3)$]\mbox{}

 This excited mode satisfies $\Re{\omega} > \mu$. Therefore, it also escapes to infinity as in the high frequency radiation. However, its wave length is longer ($\lambda \sim 1/\mu(\mu M)$) than the high frequency radiation. Thus, the energy flux can become large enough to be relevant for the saturation of the growth of clouds~\cite{Baryakhtar:2020gao}. 
 We call this mode low frequency radiation.

\item[$(\omega,m) = (2\omega_1 - \omega_2^* , 0)$]\mbox{}

The frequency of this excited mode is smaller than the axion mass, $\omega < \mu$. 
Thus, there is no emission to infinity. On the other hand, since modes with $m=0$ cannot satisfy the superradiance condition \eqref{eq:SRcondition}, such a non-superradiant excitation contributes to positive energy flux to the horizon and returns the energy of clouds back to the BH. In addition, $m=0$ means that the suppression of the flux due to the potential barrier is relatively suppressed compared to $m>0$ modes. We refer to modes of this kind as non-superradiant dissipative modes.

\item[$(\omega,m) = (2\omega_1- \omega_1^*,1),(\omega_1 + \omega_2 - \omega_2^*,1),(2\omega_2- \omega_2^*,2),(\omega_2 + \omega_1 - \omega_1^*,2)$]\mbox{}

These excited modes have almost the same $\omega$ and exactly the same $m$ as those of the superradiant modes that we start with (the difference is at most $\mathcal{O}(\omega_{1,I})$, which is small). Using the perturbative renormalization group analysis~\cite{Omiya:2020vji}, we can show that these modes contribute to accelerate the superradiant instability. 

\end{description}

\begin{figure}[t]
 \centering
 \includegraphics[keepaspectratio, scale=0.4]{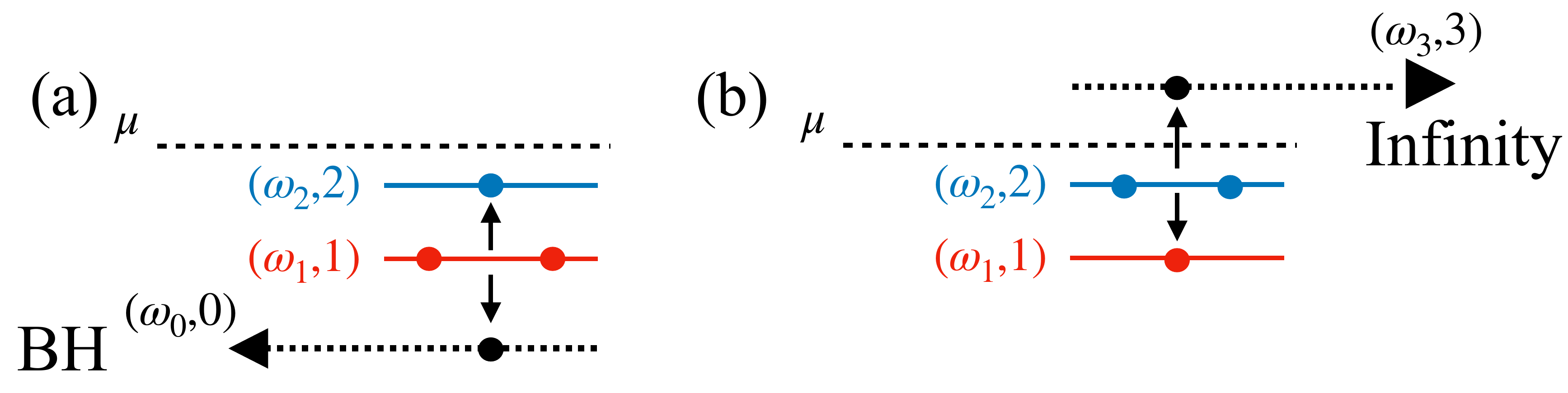}
 \caption{The left and right figures are the visualization of the emission processes shown in the diagrams (a)  and (b) in Fig.~\ref{fig:emissionleading}, respectively. The solid horizontal lines correspond to the energy levels of the bound states involved in the process. In the case of (a), one of the two axions in the $l=m=1$ cloud makes a transition to the $l=m=2$ cloud. Owing to the energy and angular momentum conservation laws, the other simultaneously generated axion occupies the $(\omega,m) = (\omega_0,0)$ mode with $\omega_0 = 2\omega_1 - \omega_2 < \mu$. Since the $m=0$ mode is non-superradiant, this axion particle transfers the energy to the BH. 
 In the case of (b), one of the two axions in the $l=m=2$ cloud makes a transition to the $l=m=1$ cloud. In this case, the axion after transition occupies the $(\omega,m) = (\omega_3,3)$ mode with $\omega_3 = 2\omega_2 - \omega_1 > \mu$, which carries the energy to infinity.}
 \label{fig:transition}
\end{figure}

We show diagrammatic expressions for relevant excitation processes in Fig.~\ref{fig:emissionleading}.  
In these diagrams the states on the left side simply represent components of the non-linear source. The right-hand side corresponds to the state excited by the non-linear source. In the following, we only consider the processes in which the excited state takes $l=m$, {\it i.e.}, the lowest orbital angular momentum $l$ for a given $m$, since they give the dominant contribution to the evolution of the clouds. 
Here, the transition proceeds in such a way that the state on the right side is generated, 
but the number of particles in a state on the left side does not always decrease: two of three decrease while one increases.  
One can read from the energy conservation if 
the number of particles in each state on the left side increases or decreases through the process.  We visualize the processes in Fig.~\ref{fig:transition}.

Keeping only the terms classified as the second and third classes, 
$\phi^{(1)}(x)$ is given by
\begin{align}
	\phi^{(1)}(x) =& - \frac{\mu^2}{2!}\left( E_1 \sqrt{E_2}\int dx' \sqrt{-g(x')} G_{\rm ret}(x,x') e^{-i(2\omega_1 - \omega_2^*) t'} R_1(r')^2 R_2^*(r') S_1(\theta')^2 S^*_2(\theta')\right.\cr
	& \left. + \sqrt{E_1} E_2\int dx' \sqrt{-g(x')} G_{\rm ret}(x,x') e^{-i(2\omega_2 - \omega_1^*) t + 3 i \varphi'} R_1(r')^* R_2(r')^2 S_1(\theta')^* S_2(\theta')^2\right) + {\rm c.c.}~.
\end{align}
Here, c.c. denotes the complex conjugate and $G_{\rm ret}(x,x')$ is the retarded Green's function that satisfies
\begin{align}
	 (\square_g - \mu^2)G_{\rm ret}(x,x') = \frac{\delta^{(4)}(x-x')}{\sqrt{-g(x)}}~.
\end{align}
Owing to the symmetry of the Kerr spacetime, one can decompose the Green's function as 
\begin{align}\label{specdecomp}
		G_{\rm ret}(x,x') = \frac{1}{2\pi}\sum_{l,m}\int_C\frac{d\omega}{2\pi}e^{-i\omega(t-t')}e^{im(\varphi-\varphi')}S_{lm\omega}(\theta)S_{lm\omega}(\theta')G_{lm\omega}(r,r')~,
\end{align}
where
\begin{align}
	\label{eq:modegreen}
	G_{lm\omega}(r,r') = \frac{1}{W_{lm}(\omega)}\left(R^{\rm in}_{lm\omega}(r)R^{\rm up}_{lm\omega}(r')\theta(r'-r) +R^{\rm in}_{lm\omega}(r')R^{\rm up}_{lm\omega}(r)\theta(r-r')  \right)\,,
\end{align}
and the function $W_{lm}(\omega)$ is the Wronskian between $R^{\rm in}_{lm\omega}$ and $R^{\rm up}_{lm\omega}$ defined by
\begin{align}
	W_{lm}(\omega) = \Delta\left(R^{\rm in}_{lm\omega}\partial_r R^{\rm up}_{lm\omega} - R^{\rm up}_{lm\omega}\partial_r R^{\rm in}_{lm\omega}\right)~.
\end{align}
We choose the integration contour $C$ in Eq.~\eqref{specdecomp} to pass above all zero points of the Wronskian $W_{lm}(\omega)$ in the complex $\omega$ plane. With this choice, $G_{\rm ret}$ satisfies the retarded boundary condition. 
We introduced $R^{\rm in}_{lm\omega}$ and $R^{\rm up}_{lm\omega}$ as the solutions of Eq.~\eqref{eq:EOMrad} satisfying the boundary conditions
\begin{subequations}
\label{eq:BC}
\begin{align}
\label{inhorizon}
	R_{lm\omega}^{\rm in } &\longrightarrow \begin{cases}
	e^{-i (\omega-m\Omega_H)r_*}~, & (r\to r_+)\\
	A_{\mathrm{in}}(\omega) \frac{e^{-i \sqrt{\omega^2-\mu^2}r_*}}{r/M} + A_{\mathrm{out}}(\omega)\frac{e^{+i\sqrt{\omega^2-\mu^2}r_*}}{r/M}~, & (r\to+\infty)
	\end{cases}~,\\
\label{outinf}
	R_{lm\omega}^{\rm up} &\longrightarrow \begin{cases}
	B_{\mathrm{in}}(\omega)e^{-i (\omega-m\Omega_H)r_*} + B_{\mathrm{out}}(\omega)e^{+i (\omega-m\Omega_H)r_*}~, & (r\to r_+)\\
	\frac{e^{+i\sqrt{\omega^2-\mu^2}r_*}}{r/M}~, & (r\to+\infty)
	\end{cases}~.
\end{align}
\end{subequations}
Here, we take the branch in which we have $\Im{\sqrt{\omega^2 - \mu^2}} > 0$ for $\omega^2<\mu^2$, which implies that the boundary conditions at infinity for $\omega^2<\mu^2$ are
\begin{align}
	R_{lm\omega}^{\rm in } &\longrightarrow 
	A_{\mathrm{in}}(\omega) \frac{e^{+ \sqrt{\mu^2 - \omega^2}r_*}}{r/M} + A_{\mathrm{out}}(\omega)\frac{e^{-\sqrt{\mu^2 - \mu^2}r_*}}{r/M}~, & (r\to+\infty)~,\\
	R_{lm\omega}^{\rm up} &\longrightarrow 
	\frac{e^{-\sqrt{\mu^2 - \omega^2}r_*}}{r/M}~, & (r\to+\infty)~, 
\end{align}
where we introduce $r_*:=\int dr(r^2+a^2)/\Delta$. 

To take into account the dissipation caused by $\phi^{(1)}$ in the cloud evolution, we promote the energies of the respective clouds, $E_1$ and $E_2$, 
to time-dependent variables. 
For consistent time evolution, 
the energy and angular momentum conservation laws 
\begin{align}\label{eq:enecons}
	\frac{d E_1}{dt} + \frac{d E_2}{dt} &= - F^{E}_{\rm tot}~,\\
	\frac{m_1}{\omega_{1,R}}\frac{d E_1}{dt} + \frac{m_2}{\omega_{2,R}}\frac{d E_2}{dt} &= - F^{J}_{\rm tot}~,\label{eq:angcons}
\end{align}
need to be satisfied, where $F^{E}_{\rm tot}$ and $F^{J}_{\rm tot}$ are the net outgoing energy and angular momentum 
fluxes evaluated at the boundaries substituting $\phi^{(1)}$. 
Here, we used the fact that, for a cloud which satisfies the linearized equation of motion, the energy and the angular momentum are related by
\begin{align}
	J = \frac{m}{\omega}E~.
\end{align}
The net energy flux can be calculated by the summation of the energy flux at the horizon $F_{\mathcal{H}^+}$ and that at infinity $F_{\mathcal{I}^+}$, which are, respectively, given by
\begin{align}
	F_{\mathcal{H}^+} &= \int d\cos\theta d\varphi\, 2 M r_+ T_{\mu\nu}(\phi)\xi^{\mu}_{(t)} l^{\nu}|_{r=r_+}~,\label{eq:enefluxH}\\
	F_{\mathcal{I}^+} &=\int d\cos\theta d\varphi\, (r^2 + a^2 \cos^2\!\theta)\sqrt{g^{rr}}T_{\mu\nu}(\phi)\xi^{\mu}_{(t)} n^{\nu}_{(r)}|_{r \to \infty}~.\label{eq:enefluxI}
\end{align}
Here, $l^{\mu}$ is the outgoing null vector specified by 
$l^{\mu}= \frac{1}{2}(1,\Delta/(r^2 + a^2),0,a/(r^2 + a^2))$%in the BL coordinates 
\cite{1974ApJ...193..443T}, and $n^\mu_{(r)}$ is the outward-pointing unit vector normal to the $r = {\rm constant}$ surface. The net angular momentum flux can be calculated by replacing $\xi^\mu_{(t)}$ with $-\xi^\mu_{(\varphi)} = (0,0,0,-1)$. 

After some algebra, we can show that the evolution equations for energies reduce to
\begin{align}
	\frac{d E_1}{dt} + \frac{d E_2}{dt} =& 2\omega_{1,I} E_1 + 2\omega_{2,I} E_2 - F_{0} E_1^2 E_2 - F_{3} E_1 E_2^2~,\label{eq:energycons}\\
	\frac{1}{\omega_{1,R}}\frac{d E_1}{dt} + \frac{2}{\omega_{2,R}}\frac{d E_2}{dt} =& 2\omega_{1,I}\frac{1}{\omega_{1,R}} E_1 + 2\omega_{2,I} \frac{2}{\omega_{2,R}} E_2 - \frac{3}{2\omega_{2,R} - \omega_{1,R}}F_{3} E_1 E_2^2~,\label{eq:angcons2}
\end{align}
with
\begin{align}
	F_0 &=\sum_{l} 2\pi M r_+ \omega_{0,R}^2\left|\frac{1}{W_{l0}(\omega_0)} \int_{r_+}^{\infty} dr\int_0^\pi d\theta\, (r^2 + a^2 \cos^2\!\theta) R^{\rm up}_{l0\omega_0}(r) R_1(r)^2 R_2(r)^* S_1(\theta)^2 S_2(\theta)^* \right|^2 ~,\label{eq:flux0}\\
	F_3 &= \sum_l \pi \omega_{3,R} \sqrt{\omega_{3,R}^2 - \mu^2}\left|\frac{1}{W_{l3}(\omega_3)} \int_{r_+}^{\infty} dr\int_0^\pi d\theta\, (r^2 + a^2 \cos^2\!\theta) R^{\rm in}_{l3\omega_3}(r) R_2(r)^2 R_1(r)^* S_2(\theta)^2 S_1(\theta)^*\right|^2 ~.\label{eq:flux3}
\end{align}
We present the derivation of the above formulae for the fluxes in Appendix \ref{sec:AppA}. In the following, we keep only the most dominant $l$ mode in the summation. 
In the present case, they are the $l=0$ mode in $F_0$ and the $l=3$ mode in $F_3$.
Simple algebra recasts Eqs.~\eqref{eq:energycons} and \eqref{eq:angcons2} into 
\begin{align}
    \frac{d E_1}{dt} &= 2 \omega_{1,I} E_1 - \frac{2 \omega_{1,R}}{\omega_{0,R}} F_0 E_1^2 E_2 + \frac{\omega_{1,R}}{\omega_{3,R}} F_3 E_1 E_2^2~,\label{eq:E1pert}\\
    \frac{d E_2}{dt} &= 2 \omega_{2,I} E_2 + \frac{\omega_{2,R}}{\omega_{0,R}} F_0 E_1^2 E_2 - \frac{2\omega_{2,R}}{\omega_{3,R}} F_3 E_1 E_2^2~.\label{eq:E2pert}
\end{align}
These equations describe the physical meaning of the diagrams in 
Fig.~\ref{fig:emissionleading} as well as in Fig.~\ref{fig:transition}. As visualized in Fig.~\ref{fig:transition}, the diagram on the left in Fig.~\ref{fig:emissionleading} represents the dissipation of the energy of the $l=m=1$ cloud feeding some energy to the $l=m=2$ cloud. The diagram on the right in Fig.~\ref{fig:emissionleading} represents the opposite, reducing the energy of the $l=m=2$ cloud and fattening the $l=m=1$ cloud.

The effect of the fourth class, the acceleration of the growth by the self-interaction, can be taken into account by the method introduced in Ref.~\cite{Omiya:2020vji}. 
Here, we consider the effect from the $(\omega,m) = (2\omega_1 - \omega_1^*,1)$ channel only. Effects from the other channels should be subdominant, since they involve the secondary $l=m=2$ cloud as the source. The large spatial extension of the $l=m=2$ cloud also reduces the strength of the interaction. Then, adding the contribution of the fourth class modifies the evolution equation \eqref{eq:E1pert} to
\begin{align}
    \frac{d E_1}{dt} &= 2 \omega_{1,I}\left(1 - \mu^2 \Re{C^{(1)}}E_1 + \frac{1}{3}\mu^4 \Re{\hat{C}^{(2)}} E_1^2\right) E_1 - \frac{2 \omega_{1,R}}{\omega_{0,R}} F_0 E_1^2 E_2 + \frac{\omega_{1,R}}{\omega_{3,R}} F_3 E_1 E_2^2~.\label{eq:E1pertacc}
\end{align}
Here, $C^{(1)}$ is given by
\begin{align}
    C^{(1)} = -\frac{1}{4 \omega_{1,I}\sqrt{\omega_1^2 - \mu^2} \alpha_{\omega_1}A_{\rm out}(\omega_1)} \int_{r_+}^{\infty} dr\int_0^\pi d\theta\, (r^2 + a^2 \cos^2\!\theta) R_{1}(r) |R_1(r)|^2 S_1(\theta)|S_1(\theta)|^2~,
\end{align}
with $\alpha_{\omega_1}$ given by the leading term in the Taylor expansion of 
the Wronskian %$W_{lm}(\omega)$ 
around $\omega =\omega_1$,
\begin{align}
    W_{11}(\omega_1) \sim 2i\sqrt{\omega_1^2 - \mu^2} \alpha_{\omega_1} (\omega -\omega_1) + \cdots~.
\end{align}
The coefficient $\hat{C}^{(2)}$ is given by a similar but more complicated expression (See Ref.~\cite{Omiya:2020vji} for the detail and the actual value of these coefficients).

\begin{figure}[t]
 \centering
 \includegraphics[keepaspectratio, scale=0.4]{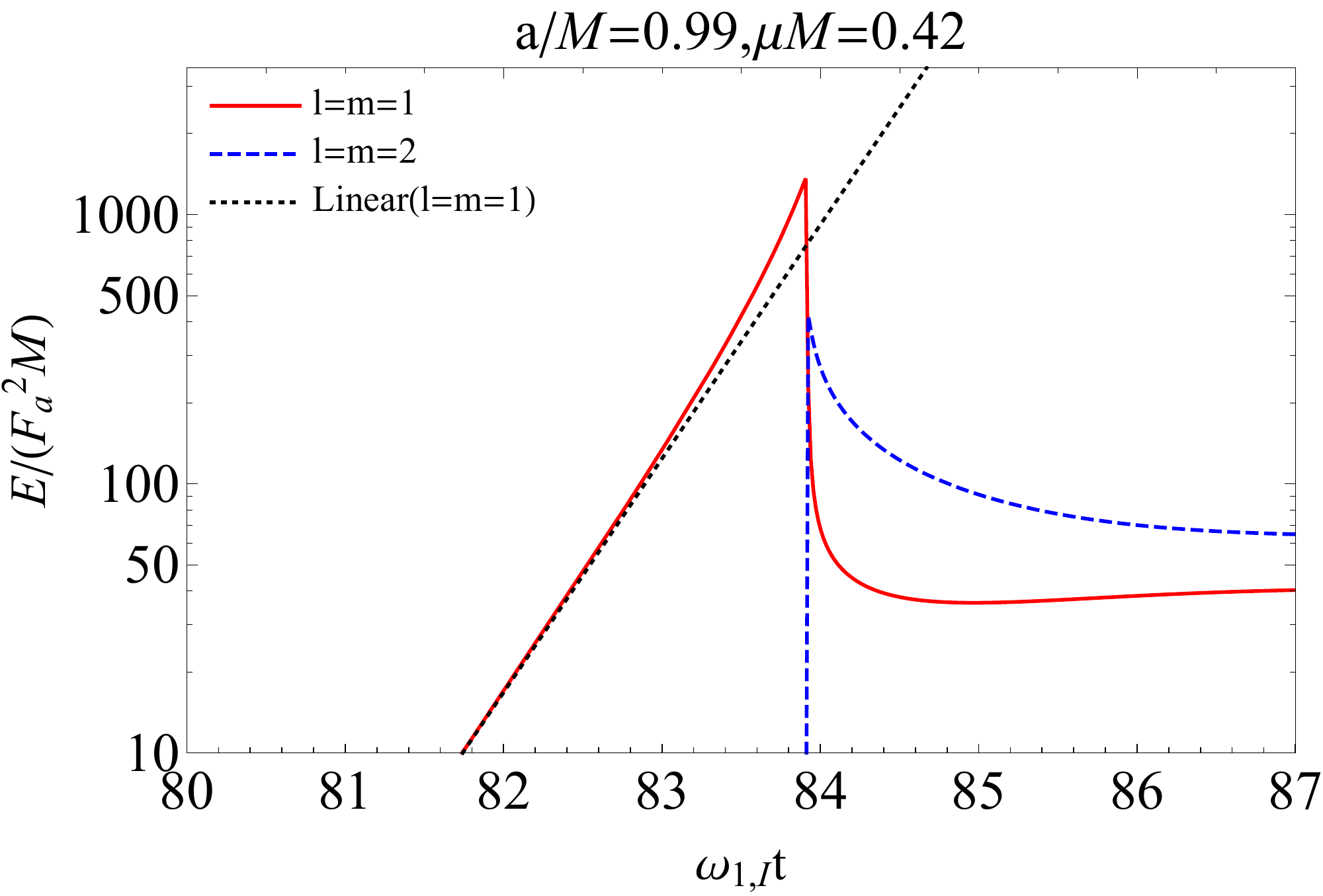}
 \includegraphics[keepaspectratio, scale=0.4]{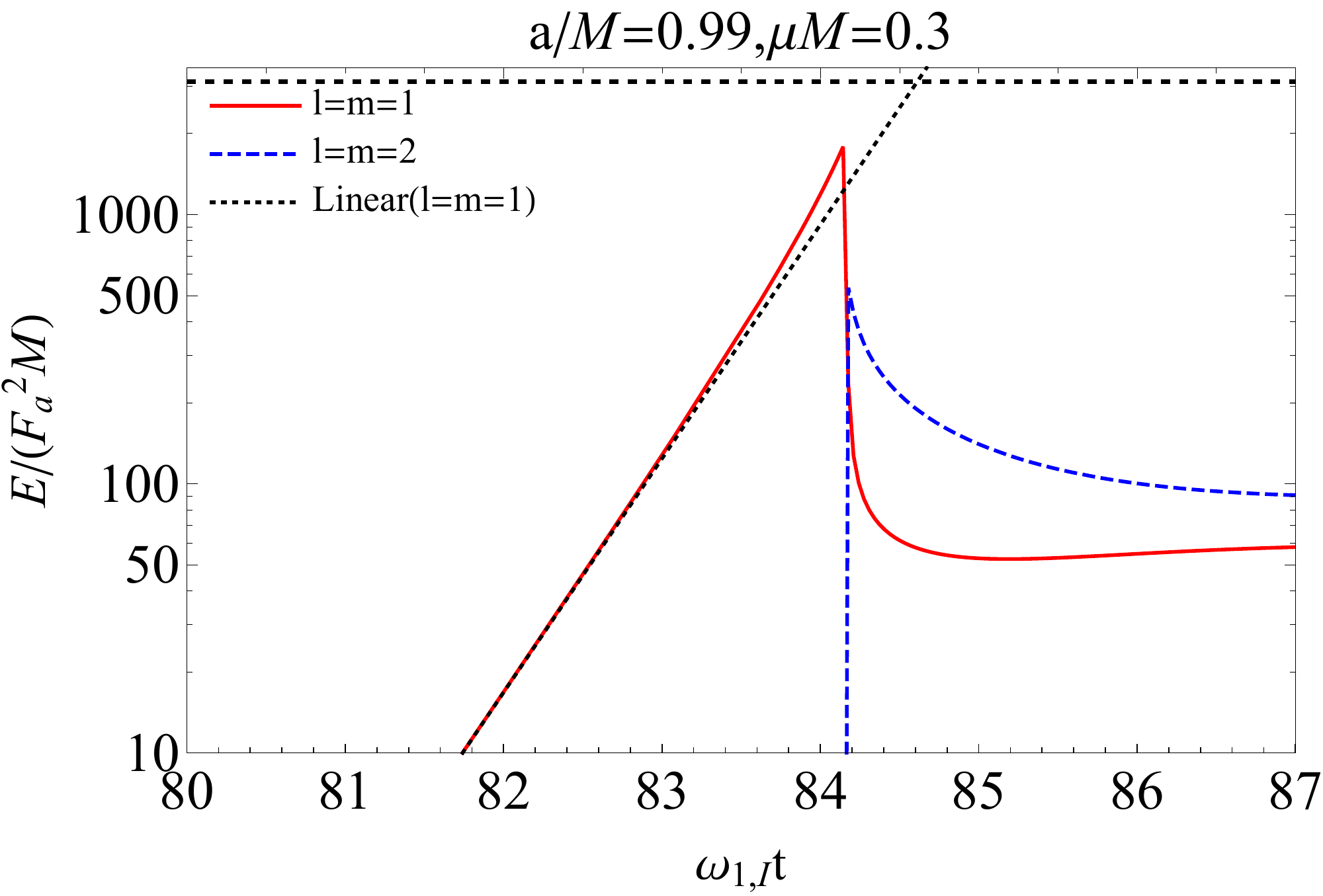}
 \caption{Time evolution of the energy of each mode when the axion cloud is composed of two modes $l=m=1$ and $l=m=2$ in the perturbtive approach. The left (right) panel shows the time evolution of the energy of the $l=m=1$ cloud, $E_1 $(red solid), and that of the $l=m=2$ cloud, $E_2$ (blue dotted), with the axion mass $\mu M = 0.42 (0.3)$. 
 In both panels, black dotted lines are the time evolution of the $l=m=1$ cloud energy in the linear approximation. We set the spin of the BH to $a/M = 0.99$. We take the initial energies of the two clouds to be $E_{1} = E_{2} = 10^{-70} F_a^2 M$. We normalized the time in the horizontal axis by the superradiant growth rate of the $l=m=1$ cloud, $\omega_{1,I}^{-1}$. The horizontal dashed line in the right panel describes the energy where the dynamical instability sets in.}
 \label{fig:pertevol}
\end{figure}

In Fig.~\ref{fig:pertevol}, we show examples of the time evolution for $\mu M = 0.42$ and $\mu M = 0.3$. Since the cloud should start with an extremely small energy, we take the initial condition as 
\begin{align}\label{eq:initialcond}
    E_1(\ti ) = E_2(\ti ) =\mu \sim \frac{10^{-70}}{G\mu M} \left(\frac{\mu}{10^{-10}{\rm eV}}\right)^2 \left(\frac{F_a/M_{pl}}{10^{-3}}\right)^2  M~,
\end{align}
which corresponds to the case in which each axion cloud starts with a single axion particle. Here in Eq.~\eqref{eq:initialcond}, 
we recovered $G=1/M_{pl}^2$ to avoid unnecessary confusion. The choice of parameters, $\mu M = 0.42$, gives an instability rate close to the maximum value~\cite{Dolan:2007mj}, and the $\mu M = 0.3$ is in the range where bosenova occurs if we consider the evolution of the condensate composed of a single cloud~\cite{Omiya:2022mwv}. In both cases, qualitative behavior of the obtained time evolution is the same. In early times, both $l=m=1$ and $l=m=2$ clouds grow owing to the superradiant instability, independently of the effect of self-interaction. Since the instability timescale of the $l=m=2$ cloud is much longer than that of the $l=m=1$ cloud, only the $l=m=1$ cloud grows effectively. 

As the cloud grows, the instability is accelerated by the self-interaction. After the energy of the $l=m=1$ cloud gets close to the maximum value, overshooting the value at which the first and second terms in the right-hand side of Eq.~\eqref{eq:E1pertacc} balance in the end, the rapid dissipation of $l=m=1$ cloud and simultaneously the excitation of the $l=m=2$ cloud happen. Overshooting occurs because the energy dissipation through the non-superradiant dissipative mode (the left panel of Fig.~\ref{fig:emissionleading}) is proportional to the energy of the $l=m=2$ cloud, and 
hence suppressed before the $l=m=2$ cloud becomes large. After this rapid energy dissipation, eventually, the whole condensate settles to a superposition of the two clouds, sharing comparable energies between them. Interestingly, the excitation of the $l=m=2$ cloud occurs more rapidly than the initial growth rate of the $l=m=1$ cloud, which is much faster than the initial rate of the superradiant instability of the $l=m=2$ cloud.

The overall picture so far is based on the perturbation theory. 
However, because of the presence of the overshooting, 
the cloud inevitably enters the non-linear regime, where the deformation 
of the cloud configuration from the linear one cannot be appropriately taken into account perturbatively. 
In the fully non-linear description, the numerical coefficients in Eqs.~\eqref{eq:E2pert} and \eqref{eq:E1pertacc} are altered, leading to a modification of the time evolution. 
In the next section, we perform non-perturbative calculations to examine whether more accurate treatment of the non-linear effects can qualitatively change the evolution or not.

\section{Non-Perturbative evolution of self-interacting axion clouds}\label{sec:3}

In this section we treat the problem non-perturbatively, without truncating the non-linear terms in \eqref{eq:eomfull2}. 
Our basic strategy is to take advantage of the fact that the timescale for the evolution of clouds is much longer than the dynamical timescale (see Eq.~\eqref{eq:adiabacticcondition}). 
In Ref.~\cite{Omiya:2022mwv}. This method was used to track the evolution 
of axion condensate in the non-linear regime, in a simpler setup where it starts with  
a single superradiant cloud. 
Here, we extend the method to include another initially subdominant cloud composed of a higher multipole superradiant mode.

\subsection{Formulation}

Since the evolution is adiabatic, the cloud configuration at each moment 
can be well approximated by a stationary one with a given amplitude. 
The axion field composed of the $l=m=1$ cloud with the amplitude $A_1$ and the $l=m=2$ cloud with the amplitude $A_2$ can be written as 
\begin{align}
	\phi = \phi_{1}(A_1,A_2) +\phi_{2}(A_1,A_2) + \phi_{r}(A_1,A_2)~.
 \label{eq:solsummed}
\end{align}
Here, $\phi_{1}$, $\phi_{2}$, and $\phi_{r}$ corresponds to the $l=m=1$ cloud, the $l=m=2$ cloud, and the radiative modes ({\it e.g.} $(\omega,m) = (2\omega_2 - \omega_1^*,3)$ and $(2\omega_1 - \omega_2^*,0)$ modes), respectively.

As the basic picture obtained by the perturbative analysis should be correct, 
in the situation of our interest the amplitude of the $l=m=1$ cloud is larger than that of the $l=m=2$ cloud. In such a situation, the self-interaction of the $l=m=2$ cloud is weak, and we would be able to safely 
neglect the deformation of the $l=m=1$ cloud caused by the influence of the $l=m=2$ cloud. 
Of course, we cannot neglect the non-linear effect of the $l=m=1$ cloud itself. 
Under this understanding, we approximate the configuration as
\begin{align}\label{eq:assump1}
	\phi_1(A_1,A_2) &= \phi_1(A_1)~,\\
	\phi_2(A_1,A_2) &= A_2 \phi_2(A_1)~,\label{eq:assump2}
\end{align}
neglecting all the non-linear effects sourced by the $\phi_2$ mode, 
except for excitations of the radiative modes.
Then, the configuration of $\phi_1$ is independently determined by solving
\begin{align}\label{eq:phi1}
	\square_g \phi_{1} - \mu^2 \sin\phi_{1} = 0~. 
\end{align}
With $\phi_1$ given as a background, $\phi_2,$ and $\phi_r$ are determined by solving linearized equations of motion,
\begin{align}
	\square_g\phi_{2} - \mu^2 \left(\cos\phi_{1}\right) \phi_{2}&=0~,\label{eq:phi2}\\
	\square_g\phi_{r} - \mu^2 \left(\cos\phi_{1}\right) \phi_{r}&= S_{r}\label{eq:phirad}~.
\end{align}
Here, $S_{r}$ represents the source terms for the radiative modes given by
\begin{align}
	S_{r} = \mu^2( \cos \phi_{1} ) \phi_{2} - \frac{\mu^2}{2} (\sin\phi_{1}) \phi_{2}^2  + \cdots~.
\end{align}

We first solve Eq.~\eqref{eq:phi1} for $\phi_{1}(A_1)$, assuming
\begin{align}\label{eq:adiabaticansatz}
	\phi_{1}(A_1) =& \sum_{k = 1}^{\infty}\sum_{l\ge k m_1}^{\infty} e^{-i k (\omega_1(A_1) t - m_1 \varphi)} \tilde{R}^{1}_{kl}(r;A_1) Y_{l\,k\times m_1}(\cos\theta) + {\rm c.c.}~.
\end{align}
Here, $Y_{lm}(x)$ is the elevation angle part of the spherical harmonics defined as 
\begin{align}
	Y_{lm}(x) \equiv N^m_{l} P^m_l(x)~,
\end{align}
where $P^m_{l}(x)$ is the associated Legendre polynomial and
\begin{align}
 N^m_{l}=\sqrt{\frac{(l - m)! (2l+1)}{2(l+m)!}}~.
\end{align}
We choose $N^m_{l}$ such that $Y_{lm}(x)$ is normalized as
\begin{align}
	\int_{-1}^{1} dx\, Y_{lm}(x) Y_{l'm}(x) = \delta_{ll'}~.
\end{align}
For definiteness, we define the amplitude $A_1$ as a parameter that specifies the amplitude of the fundamental 
mode at a large radius, {\it i.e.},
\begin{align}
	\tilde{R}^1_{1m_1}(r;A_1) &\to A_1 \frac{e^{- \sqrt{\mu^2 - \omega_1^2} r}}{r/M} \left(\frac{r}{M}\right)^{-M \frac{\mu^2 - 2 \omega_1^2}{\sqrt{ \mu^2 - \omega_1^2}}}(1 + \mathcal{O}(r^{-1}))~, & (r &\to \infty)~.
 \label{eq:expatInfinity}
\end{align}
After projecting Eq.~\eqref{eq:phi1} to each $l$ and $m$ harmonics, we obtain a set of coupled ordinary differential equations for $\tilde{R}^1_{l}$, which can be solved for a given $A_1$ with the outgoing/exponentially decaying boundary condition at infinity 
\begin{align}\label{eq:bcinf}
	\tilde{R}^1_{kl} \to & A^{\rm (out)}_{kl}\frac{e^{+i \sqrt{n^2\omega_1^2 - \mu^2}r}}{r/M} \left(\frac{r}{M}\right)^{- i M\frac{\mu^2 - 2 n^2 \omega_1^2}{\sqrt{n^2 \omega_1^2 - \mu^2}}} \left(1 + \frac{a_1}{r/M} +\frac{a_2}{(r/M)^2} +\cdots \right) ~, & (r&\to \infty)~,
\end{align}
and the ingoing boundary condition at the event horizon~\cite{Omiya:2022mwv}
\begin{align}\label{eq:bchor}
	\tilde{R}^1_{kl} \to &  A^{\rm (in)}_{kl}\left(\frac{r-r_-}{M}\right)^{i n\frac{2 M r_-}{r_+-r_-}(\omega_1 - m\frac{a}{2M r_-}) - (1-2kM\omega_1 i)- i M \frac{\mu^2 - 2 k^2 \omega_1^2}{\sqrt{k^2 \omega_1^2 - \mu^2}}}\cr
	&\qquad \times e^{i \sqrt{k^2\omega_1^2 - \mu^2} (r-r_-)} \left(\frac{r-r_+}{M}\right)^{- i n\frac{2 M r_+}{r_+ - r_-}(\omega_1 - m_1 \Omega_H)}~, & (r &\to r_+)~.
\end{align}
For an expansion near infinity given in Eq.~\eqref{eq:bcinf}, 
we expand $\tilde{R}_{kl}^1$ up to $a_7$, and numerically determine the coefficients order by order. Then, we solve the coupled non-linear equations numerically, imposing boundary conditions Eqs. \eqref{eq:bcinf} and \eqref{eq:bchor} at finite $r_*$ with large absolute values, and obtain the complex frequency $\omega_1$ and the complex amplitudes $A_{kl}^{\rm (in),(out)}$. The actual values of the $r_*$ to place the boundary conditions \eqref{eq:bcinf} and \eqref{eq:bchor} are appropriately chosen for each $\mu M$. For example we take $r_* = 100$ for the boundary condition \eqref{eq:bcinf} and $r_* = -100$ for the boundary condition \eqref{eq:bcinf}, for $\mu M = 0.42$. So far, we repeat the computations performed in Ref.~\cite{Omiya:2022mwv}, and 
hence we do not go into the details of the numerical method, here. 

After $\phi_{1}$ is determined, we solve Eq.~\eqref{eq:phi2}. We adopt an ansatz similar to $\phi_1$ for $\phi_2$ as
\begin{align}
	\phi_{2} = \sum_{l} e^{- i (\omega_2 t - m_2 \varphi)} \tilde{R}^{2}_{l}(r) Y_{lm_2}(x) + {\rm c.c.}~.
\end{align}
Substituting the ansatz to Eq.~\eqref{eq:phi2} and projecting it to the $(l, m_2)$ harmonics, we obtain
\begin{align}
	\frac{d}{dr}\left(\Delta\frac{d \tilde{R}^{2}_{l}}{dr}\right) &+ \left[\frac{(\omega_2 (r^2 + a^2) - a m_2)^2}{\Delta} - \mu^2 r^2 + 2 a \omega_2 m_2 - a^2\omega_2^2  -l(l+1) \right. \cr
	&\qquad\left. + a^2 (\omega_2^2 - \mu^2)\frac{1-2l(l+1) + 2 (2m_2)^2}{3-4l(l+1)} \right]R^{2}_{l}\cr
	& + a^2 (\omega_2^2 - \mu^2)\left(\frac{(l-1-2m_2)(l-2m_2)}{(2l-3)(2l-1)}\frac{N^{m_2}_{l-2}}{N^{m_2}_{l}}R^{2}_{l-2} \right.
 %\cr
%	&\qquad\qquad\qquad\qquad 
\left.+ \frac{(l+2 + 2m_2)(l+1+2m_2)}{(2l+3)(2l+5)}\frac{N^{m_2}_{l+2}}{N^{m_2}_{l}}R^{2}_{l+2} \right)\cr
	& 
  +\int_{0}^{2\pi}\frac{d\varphi}{2\pi} \int _{-1}^1 dx \ Y_{lm_2}(x)e^{i m_2 \varphi} (r^2 + a^2 x^2)(1- \cos\phi_{1}(A_1))\phi_{2}= 0 \label{eq:Rlmeq}~.
\end{align}
This set of coupled ordinary differential equations defines a two point boundary value problem that determines the eigenvalue $\omega_2$, under the assumption that the $l=2$ component is dominant. 
(The solution describes only a single mode. It is composed of a superposition of various $l$ harmonics, simply because the elevation angle dependence of the mode is not given by a single component of the ordinary spherical harmonics.)
The boundary condition is similar to Eqs.~\eqref{eq:bcinf} and \eqref{eq:bchor} but with $(\omega_1,m_1)$ replaced by $(\omega_2,m_2)$. Normalization of the solution is arbitrary, but for later convenience, we choose the normalization such that the energy calculated for $\phi_2$ becomes unity in the unit of $F_a^2 M$.

Now that we have $\phi_1(A_1)$ and $\phi_2(A_1)$, we can solve the equations for the radiative modes \eqref{eq:phirad} in a method similar to the one used in determining $\phi_2(A_1)$. Since we are interested in the excitation of the $m=0$ and $m=3$ modes, we project the equation to the respective components. This automatically sets $\omega$ to  $2\omega_1-\omega_2$ and $2\omega_2 - \omega_1$, respectively\footnote{Here, we neglect the imaginary part of the frequency, since it is much smaller than its real part due to the adiabatic nature of the evolution.}.

The obtained solution $\phi$ by solving Eq.~\eqref{eq:solsummed} is valid only for a short period,   
and cannot remain to be valid over the instability timescale. 
To obtain a solution which can describe the instability, 
we promote the amplitudes $A_1$ and $A_2$ to time-dependent variables, $A_1(t)$ and $A_2(t)$, which are to be determined by the equations derived from the energy and angular momentum conservation laws, as in the previous section. Namely, the time evolution of the amplitudes is determined by solving 
\begin{align}
	\frac{d E(A_1,A_2)}{dt} = \frac{\partial E(A_1,A_2)}{\partial A_1} \frac{dA_1}{dt} + \frac{\partial E(A_1,A_2)}{\partial A_2} \frac{dA_2}{dt} = -F_E(A_1,A_2)~,\label{eq:energyevonl}\\
	\frac{d J(A_1,A_2)}{dt} = \frac{\partial J(A_1,A_2)}{\partial A_1} \frac{dA_1}{dt} + \frac{\partial J(A_1,A_2)}{\partial A_2} \frac{dA_2}{dt} = -F_J(A_1,A_2)~.\label{eq:angmomevonl}
\end{align}
Here, the energy $E(A_1, A_2)$ and the energy flux $F_E(A_1,A_2)$ are evaluated by
\begin{align}
 & E(A_1,A_2) = \int dr\, d\cos\theta\, d\varphi \ (r^2 + a^2 \cos^2\!\theta)\sqrt{g^{tt}}\,T_{\mu\nu}(A_1,A_2)\xi^{\mu}_{(t)} n^\nu_{(t)}~,\label{eq:totenergy}\\
  & F_{E}(A_1,A_2)  = F_{E}^{H}(A_1,A_2) + F_{E}^{{\infty}}(A_1,A_2)~,\\
  & F_E^H(A_1,A_2) =  \int d\cos\theta d\varphi\, 2 M r_+ T_{\mu\nu}(A_1,A_2)\xi^{\mu}_{(t)} l^{\nu}|_{r=r_+}~,\\
  & F^{\infty}_E(A_1,A_2) = \int d\cos\theta d\varphi\, (r^2 + a^2 \cos^2\!\theta)\sqrt{g^{rr}}\,T_{\mu\nu}(A_1,A_2)\xi^{\mu}_{(t)} n^{\nu}_{(r)}|_{r \to \infty}~,
\end{align}
where $T_{\mu\nu}(A_1,A_2)$ is the energy momentum tensor evaluated for $\phi(A_1,A_2)$. The expressions for the angular momentum $J(A_1,A_2)$ and its flux $F_J(A_1,A_2)$ are obtained by replacing the time-translation killing vector $\xi_{(t)}$ with the rotational killing vector $-\xi_{(\varphi)}$ in the above formulae. 
Note that under the assumptions \eqref{eq:assump1} and \eqref{eq:assump2}, the total energy \eqref{eq:totenergy} is the summation of the energy of the $l=m=1$ cloud and that of $l=m=2$ cloud
\begin{align}
    E(A_1,A_2) &= E_1(A_1) + E_2(A_1,A_2)~.
\end{align}
In addition, $E_2(A_1,A_2)$ is quadratic in $A_2$, {\it i.e.},
\begin{align}
    E_2(A_1,A_2) = \hat{E}_2(A_1)|A_2|^2~.
\end{align}

\begin{figure}[t]
 \centering
 \includegraphics[keepaspectratio, scale=0.4]{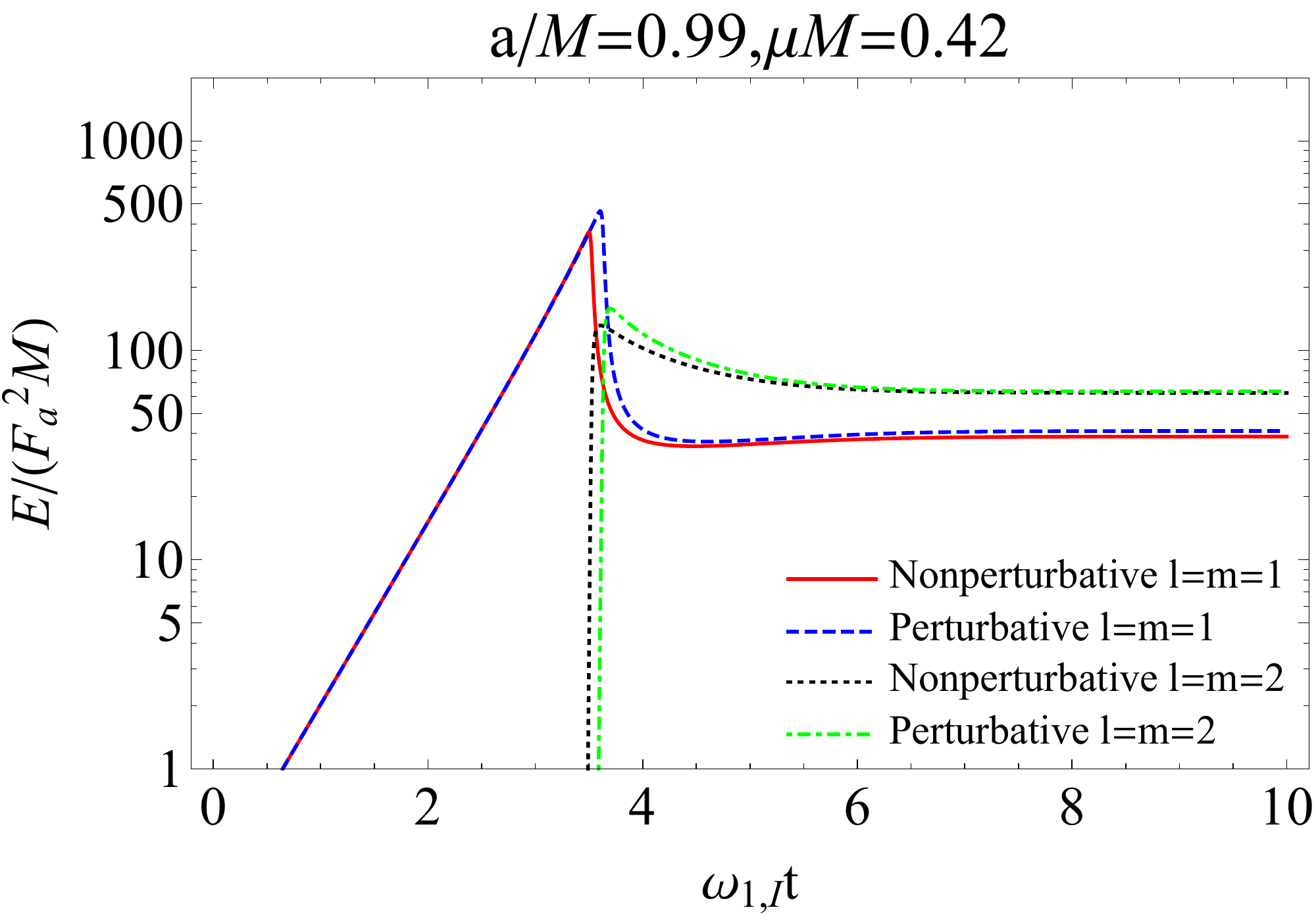}
  \includegraphics[keepaspectratio, scale=0.4]{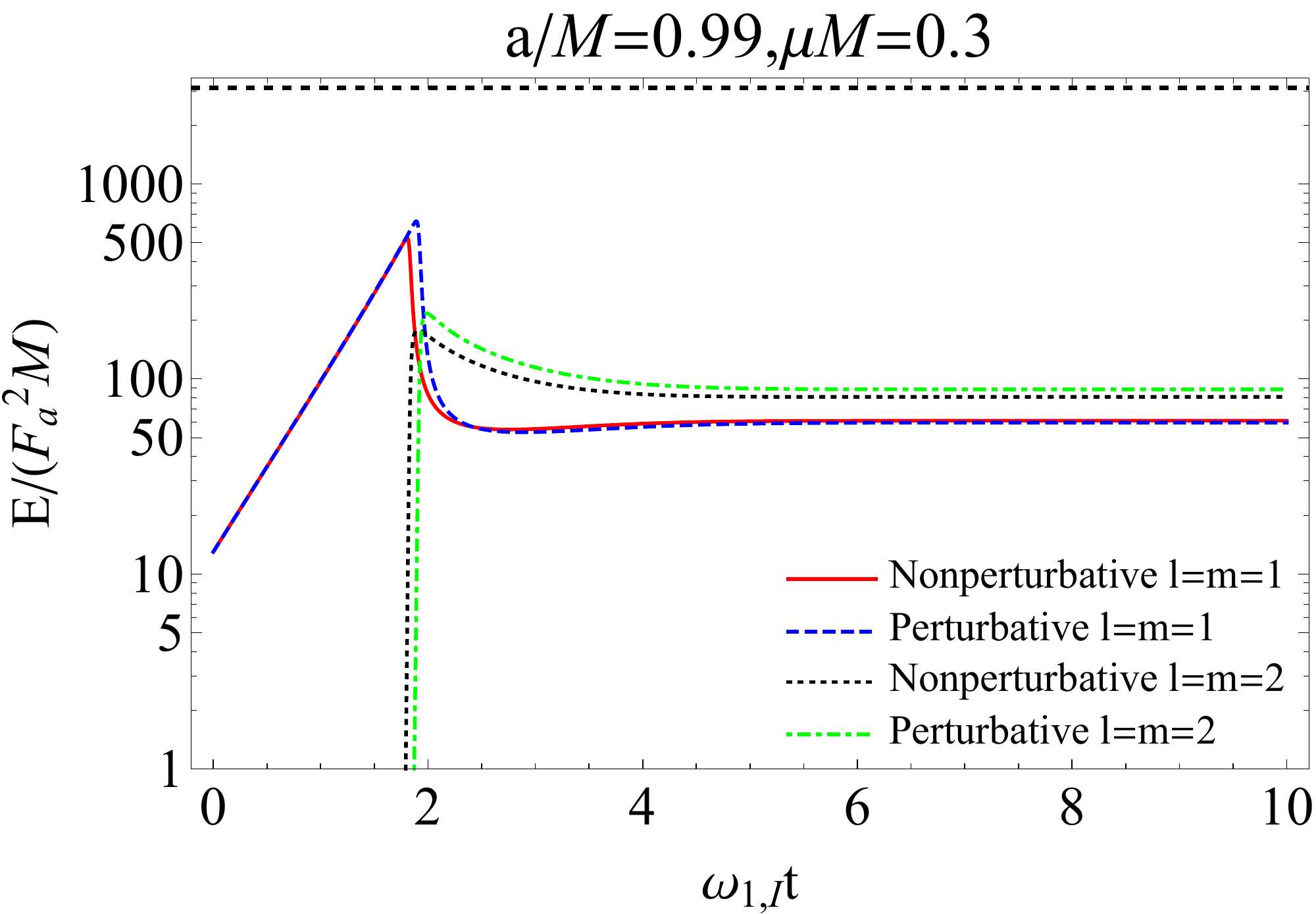}
 \caption{Comparison of time evolutions of the energy of the axion cloud in the non-perturbative and perturbative approaches. The left and right panels show the time evolution of the energy of the $l=m=1$ cloud with $\mu M = 0.42$ and that with  $\mu M = 0.3$, respectively. The red solid and black dotted lines correspond to the non-perturbative evolution of the $l=m=1$ and $l=m=2$ clouds, respectively. The blue dashed and green dot-dashed lines show the same quantities but calculated with the perturbative method. We take the initial energy of the cloud to be $E_1 \sim 0.27 F_a^2 M~(12 F_a^2 M) $ for $\mu M = 0.42~(0.3)$, and $E_2 = 10^{-8} F_a^2 M$. Time is normalized by $\omega_{1,I}$, the instability rate calculated from the linear analysis. Black dashed horizontal  line around $E =  3000 F_a^2M$ in the right panel indicates the energy where the $l=m=1$ cloud is predicted to become unstable by the non-perturbative single mode calculation.}
 \label{fig:nonpertevol}
\end{figure}

\subsection{Result}

In Fig.~\ref{fig:nonpertevol}, we show the time evolution of the amplitudes for $\mu M = 0.42$ and $0.3$ with $a/M = 0.99$, in a similar manner to Fig.~\ref{fig:emissionleading}. The solid curves represent the cases including the non-perturbative effects discussed in this section, while the dashed curves are the references obtained by the perturbative analysis discussed in the preceding section.  
In the early epoch, where the cloud energy is small, non-perturbative and perturbative calculations trace almost the same evolution track. 
As the cloud energy becomes larger, the self-interaction starts to be relevant. 
First, the acceleration of the instability occurs. As a result, the non-perturbative calculation predicts faster growth than the perturbative one. 
Then, the dissipation mechanism reviewed in the preceding section starts to 
reduce the energy, and eventually the whole condensate settles to a quasi-stationary configuration. 
To summarize, the overall qualitative picture obtained by the  perturbative calculation does not change in the present case. 
It would be worth mentioning that the whole field configuration settles to almost the same quasi-stationary one that is predicted by the perturbative calculation, because the cloud amplitudes are sufficiently small in the quasi-stationary state.

\begin{figure}[t]
 \centering
 \includegraphics[keepaspectratio, scale=0.5]{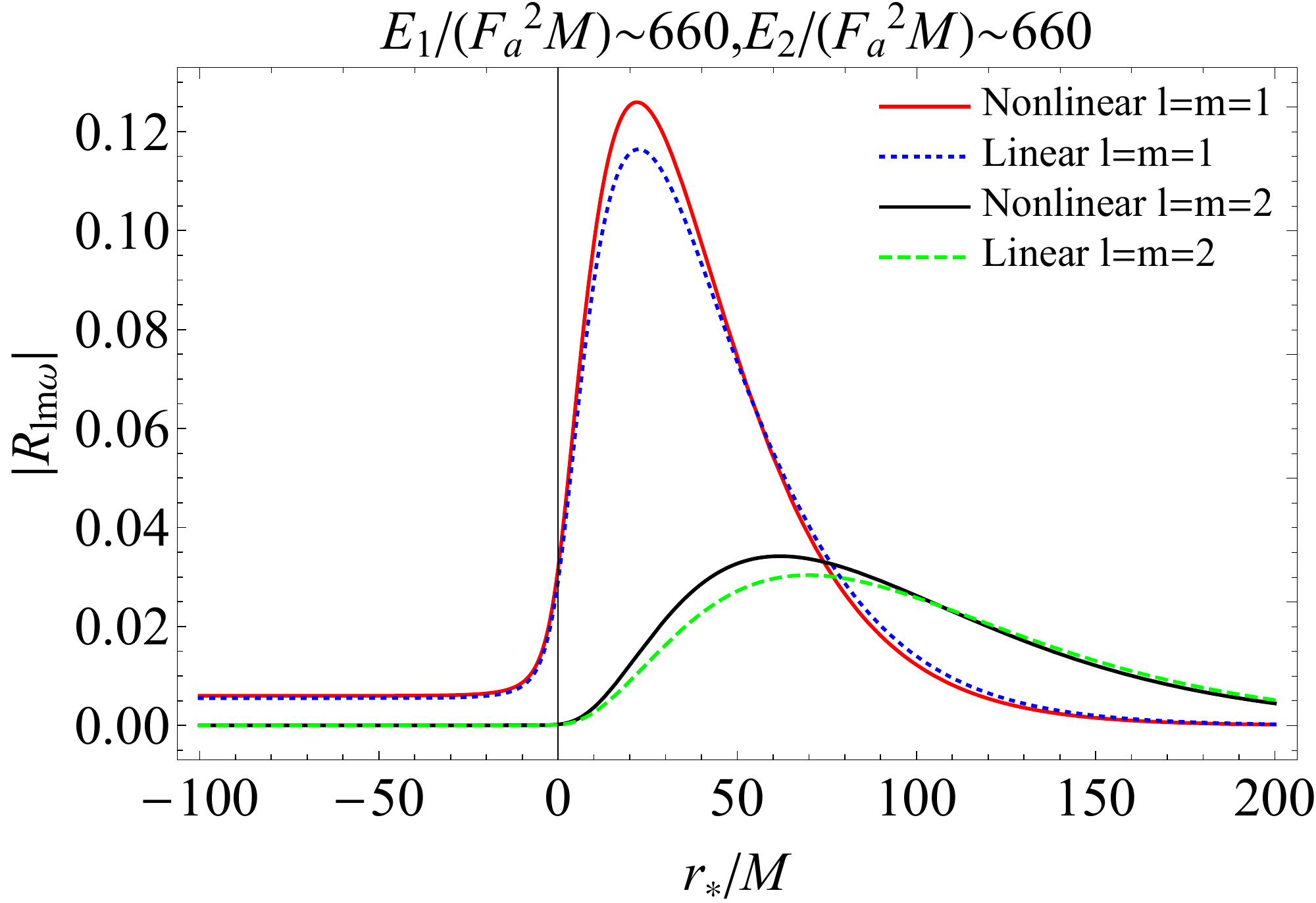}
 \caption{Snapshot of the absolute values of the radial mode functions in the perturbative and the non-perturbative evolutions. The red solid and black solid lines, respectively, show the absolute value of mode function for the $l=m=1$ mode and that for the $l=m=2$ mode, when the self-interaction is taken into account non-perturbatively. The blue dotted and green dashed lines represent the same quantities but when the self-interaction is neglected. We set the energy of the cloud to be $E_1 = E_2 \sim 660 F_a^2 M$, which is close to the peak value of $E_1$, in the right panel of Fig.~\ref{fig:nonpertevol}. The axion mass and the spin of the BH are set to $\mu M = 0.3$ and $a/M = 0.99$.}
 \label{fig:nonpertconfig}
\end{figure}

The quantitative difference arises in the peak energy, which depends on when the dissipation effect starts to overcome the energy gain by the superradiance. 
From Fig.~\ref{fig:nonpertevol}, we observe that the peak energy is reduced when we take into account all the effects of self-interaction. 
The enhancement mechanism of the dissipation can be understood as follows. Because of the self-interaction, two clouds attract each other (see Fig.~\ref{fig:nonpertconfig}). Then, the overlap between the two dominant modes gets larger. Because the dominant dissipative modes are 
sourced by the products of these two modes, the dissipation is enhanced by the increase of the overlap in the non-linear regime. 

\begin{figure}[t]
 \centering
 \includegraphics[keepaspectratio, scale=0.4]{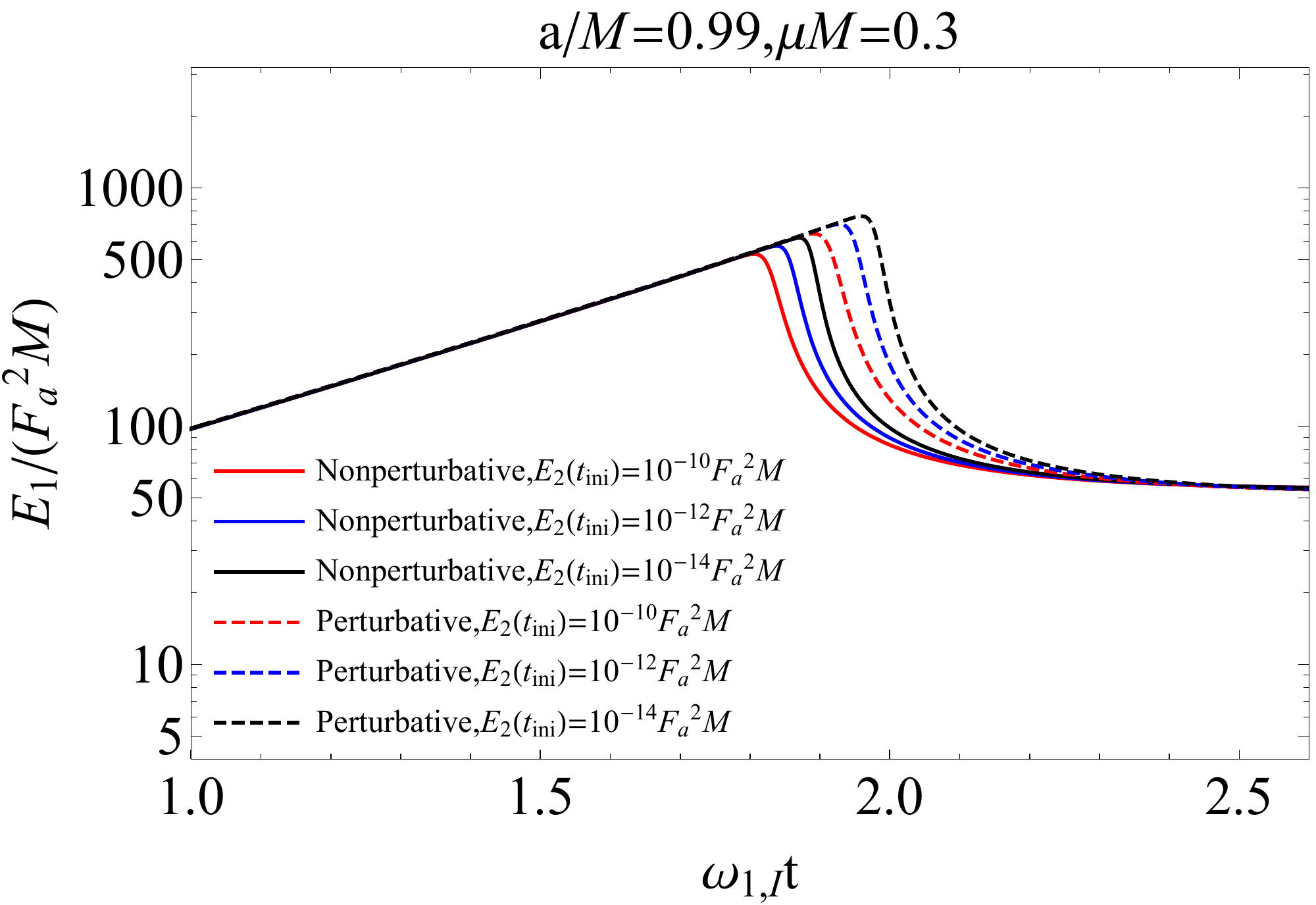}
 \caption{Dependence of the time evolution of the energy of the $l=m=1$ mode on the initial energy of the $l=m=2$ mode $E_2(\ti ) $. The red, blue, and black solid lines show the energies of the $l = m= 1$ cloud, starting with $E_2(\ti )/(F_a^2 M) = 10^{-8}, 10^{-12},$ and $10^{-14}$, respectively. 
 The dotted lines with the same color represent the case when the self-interaction is treated perturbatively. 
 The values of the axion mass and the BH spin are set to $\mu M = 0.42$ and $a/M = 0.99$.}
 \label{fig:nonpertinit}
\end{figure}

Next, we investigate how the evolution of the $l=m=1$ cloud depends on the energy of the $l=m=2$ cloud at the initial time, $E_2(\ti )$. In Fig.~\ref{fig:nonpertinit}, we show examples 
of the time evolution of the energy of the primary cloud for several values of $E_2(\ti )$. As $E_2(\ti )$ is reduced, the maximum energy of $E_1$ increases. Of course, in the limit of $E_2(\ti ) \to 0$, the evolution of the $l=m=1$ cloud is the same as the single cloud evolution, discussed in 
Ref.~\cite{Omiya:2022mwv}, and the bosenova can happen. 
Therefore, our main interest is in how the maximum energy of $E_1$ depends on the initial energy of the secondary cloud, $E_2(\ti )$. 
As it will turn out to be true that an extremely small initial amplitude of the secondary cloud is sufficient as the seed, it is not necessary to take into account the non-linear effect precisely. 
Hence, we approximately evaluate the maximum energy in the following manner. 
Before the $l=m=1$ cloud approaches the maximum, the amplitude of the $l=m=2$ cloud is kept small. Thus, we can approximate the evolution equations \eqref{eq:energyevonl} and \eqref{eq:angmomevonl} as
\begin{align}
    \frac{d E_1}{d t} + \frac{d E_2}{dt} &\sim - F_E^{\rm SR,1}(E_1) - F_E^{\rm diss}(E_1) E_2~,\label{eq:energyevonlapp}\\
    \frac{d J_1}{d t} + \frac{d J_2}{dt} &\sim - \frac{1}{\omega_{1,R}}F_E^{\rm SR,1}(E_1)~.\label{eq:angmomevonlapp}
\end{align}
Here, $F_E^{\rm SR,1}(E_1)$ is the energy flux at the horizon induced by the $l=m=1$ mode and $F_E^{\rm diss}(E_1)$ is the energy flux at the horizon by the $l=m=0$ mode, normalized by the energy of the $l=m=2$ cloud. Note that $F_E^{\rm SR,1}(E_1)$ is negative since the $l=m=1$ mode satisfies the  superradiance condition.
Since the configuration of each cloud is helically symmetric, we have the relation
\begin{align}
    \frac{\delta E_i}{\omega_{i,R}} = \frac{\delta J_i}{m_i}~.
\end{align}
Thus, Eqs.~\eqref{eq:energyevonlapp} and \eqref{eq:angmomevonlapp} are reduced to 
\begin{align}
    \frac{d E_1}{d t} &\sim -F_E^{\rm SR,1}(E_1) - 2 F_E^{\rm diss}(E_1) E_2~,\\
    \frac{d E_2}{dt} &\sim F_E^{\rm diss}(E_1) E_2~.
    \label{eq:dE2dt}
\end{align}
From the first equation, the maximum value of $E_1$, which is denoted by $E_1^{\rm max}$, is determined by
\begin{align}\label{eq:maxequation}
    -F_{E}^{\rm SR,1}(E_1^{\rm max}) &= 2 F_E^{\rm diss}(E_1^{\rm max}) E_2(t_{\rm max})~,
\end{align}
where $t_{\rm max}$ is the time when $E_1 = E_1^{\rm max}$ is achieved. 
Before reaching this maximum value, we would be able to use the approximation
\begin{align}
    \frac{d E_1}{d t} &\sim -F_E^{\rm SR,1}(E_1)~. 
\end{align}
Then, Eq.~\eqref{eq:dE2dt} can be formally solved as
\begin{align}
    E_2(t_{\rm max}) = E_2(\ti ) \exp\left(\int_{E_1(\ti )}^{E_1^{\rm max}} dE_1\ g(E_1)\right)~,
\end{align}
where 
\begin{align}
    g(E_1) &= -\frac{F_E^{\rm diss}(E_1)}{F_E^{\rm SR,1}(E_1)}~.
\end{align}
Using Eq.~\eqref{eq:maxequation}, we solve this relation for $E_2(\ti )$, to obtain
\begin{align}\label{eq:maxest}
    E_2(\ti ) = \frac{1}{2 g(E_1^{\rm max})} \exp\left(-\int_{E_1(\ti )}^{E_1^{\rm max}} dE_1\ g(E_1)\right)~,
\end{align}
which approximates the initial energy of the secondary cloud $E_2(\ti )$ as a function of the maximum energy of the $l=m=1$ cloud. 

\begin{figure}[t]
 \centering
  \includegraphics[keepaspectratio, scale=0.4]{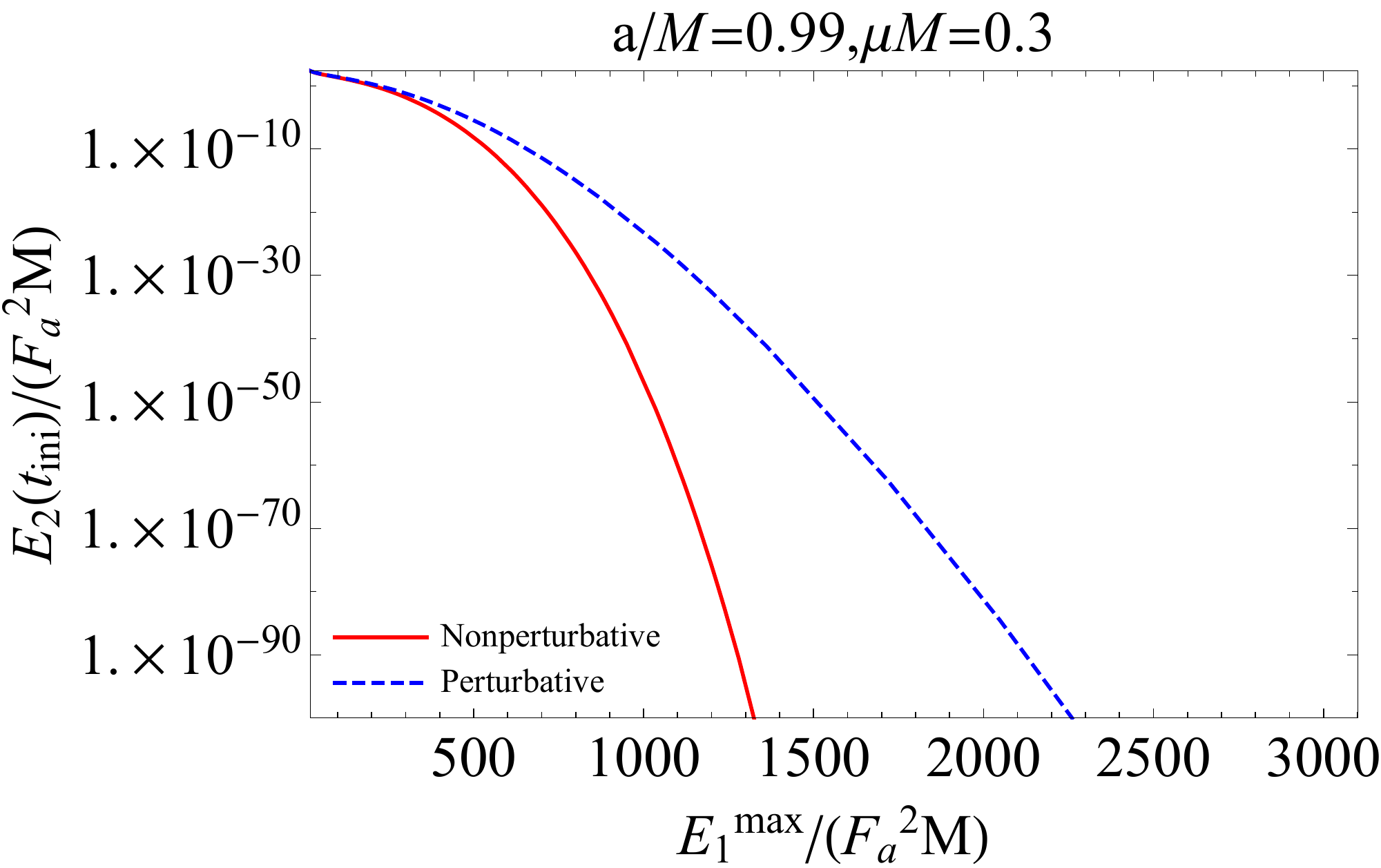}
 \caption{Relation between the maximum value of energy of the $l=m=1$ mode during the time evolution and the initial energy of the $l=m=2$ mode. The red solid and blue dashed curves show the relations between the initial energy of $l=m=2$ cloud $E_2(\ti )$ and the maximum energy of the $l=m=1$ cloud $E_1^{\rm max}$ obtained by solving Eq.~\eqref{eq:maxest} under the non-perturbative scheme and the perturbative approximation, respectively.}
 \label{fig:maxestimate}
\end{figure}

In Fig.~\ref{fig:maxestimate}, we show the relation between $E_2(\ti )$ and $E_1^{\rm max}$. In the range of the initial value of $E_2(\ti )$ shown in Fig.~\ref{fig:nonpertinit}, between $10^{-14}F_a^2 M$ and $10^{-10}F_a^2 M$, our analytic estimation reproduces the energy of the peak value within%, at most, 
\ 3\% error. 
As we can observe from Fig.~\ref{fig:maxestimate}, the peak value of the actual primary cloud energy is always smaller than the estimate by the perturbative calculation. 
This clearly indicates that between the two effects of the self-interaction; the enhancement of the energy dissipation and the acceleration of the growth, the former effect always dominates. 

To realize a bosenova, we need to prepare an extremely small value of $E_2(\ti )$. 
In fact, if we take $E_1^{\rm max} = E_1^{\rm BN} \sim 3100 F_a^2 M$, which is the threshold value for the onset of the dynamical instability~\cite{Omiya:2022mwv}, the integral in the exponent in Eq.~\eqref{eq:maxest} is evaluated as 
\begin{align}
    \int_{E_1(\ti )}^{E_1^{\rm BN}} dE_1\ g(E_1) \sim 3.9 \times 10^3 ~.
\end{align}
Therefore, $E_2(\ti )$ must be an extremely small quantity suppressed by the factor $\exp(- 3.9 \times 10^3)$. 
Even if the $l=m=2$ cloud starts with a single axion particle (the number of excited particle is unity), we have $E_2(\ti ) \sim 10^{-70} F_a^2 M$ (see Eq. \eqref{eq:initialcond}), which is much larger than the 
small initial energy required to allow bosenova. Since the BH Hawking temperature is comparable to the axion mass in the current setup, the thermal radiation of the axion field from the BH is not suppressed at all. 
Therefore, it is impossible to tune the state of the initial cloud to be completely vacant.

\begin{figure}[t]
 \centering
  \includegraphics[keepaspectratio, scale=0.5]{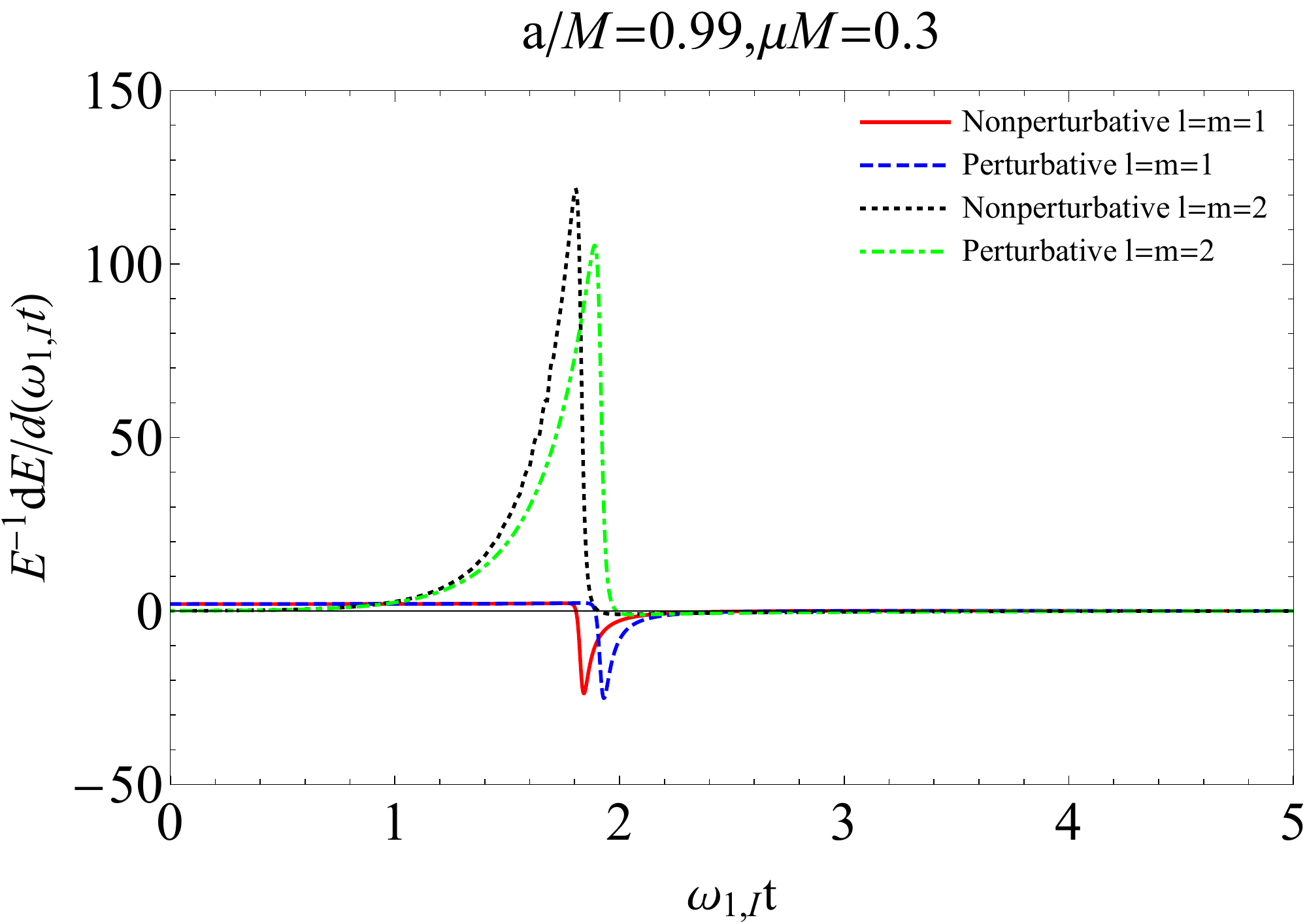}
 \caption{The time dependence of the energy change rates for the $l=m=1$ and the $l=m=2$ modes. The red solid and black dotted curves show 
 the change rates of the energy for the $l=m=1$ and $l=m=2$ modes, respectively, calculated by using the non-perturbative method. The blue dashed and green dot-dashed lines show the same quantities but calculated by using the perturbative method. We normalized the change rates of the energy by the growth rate of the $l=m=1$ modes under the linear approximation, $\omega_{1,I}$. The horizontal axis is the time also normalized by means of $\omega_{1,I}$. The initial conditions are the same as those in Fig.~\ref{fig:nonpertevol}.}
 \label{fig:timescale}
\end{figure}

Finally, we examine whether or not the adiabaticity assumed in our calculation is valid throughout the evolution. 
In Fig.~\ref{fig:timescale}, we show the change rate of the energy $E^{-1} dE/dt$, normalized by the linear growth rate $\omega_{1,I}$. As we observe from the figure, the growth rate is enhanced at most by a factor 100 compared to the linear growth rate. Since the linear growth rate is, at most, $10^{-7}$ of the inverse of the dynamical timescale, the adiabaticity is maintained by a large margin. This confirms the validity of the adiabatic approximation adopted in our numerical scheme.

\section{Late-time evolution caused by various perturbative processes}\label{sec:4}

In the preceding section, we confirmed that the whole axion condensate settles to a quasi-stationary one, even if we take into account non-linear effects in the context of the condensate composed of two clouds. 
In this section, we investigate whether other effects such as quantum processes, which cannot be captured by the classical analysis, and excitations of other modes (the overtone modes and the higher multipole modes), may or may not subsequently dissolve the quasi-stationary configuration that we have obtained. 

\begin{figure}[t]
 \centering
 \includegraphics[keepaspectratio, scale=0.3]{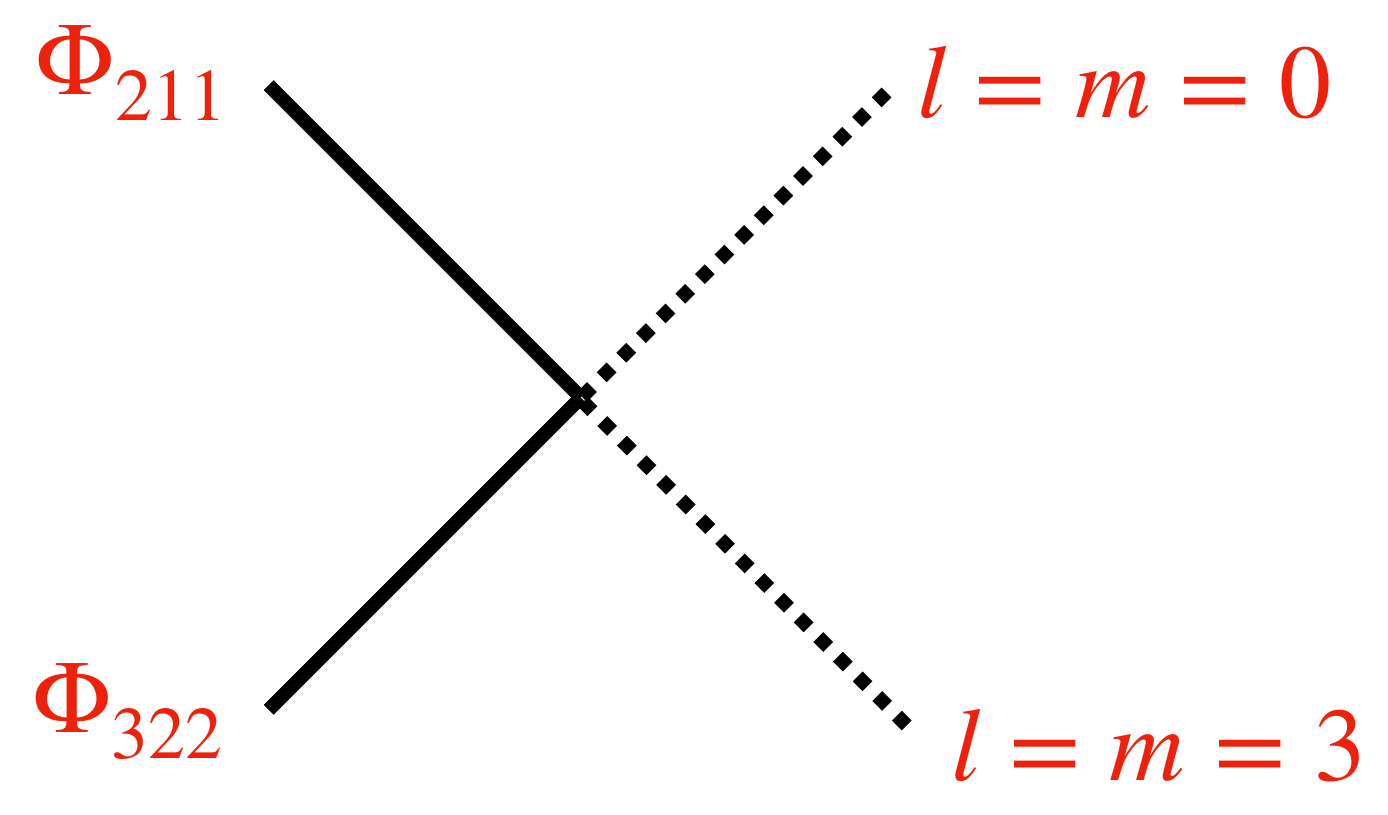}
 \caption{The simplest quantum process induced by the self-interaction.}
 \label{fig:quant}
\end{figure}

We first consider quantum processes. Here, ``quantum processes'' mean the ones that cannot be captured by the classical equations of motion \eqref{eq:eomfull2}. The simplest tree level process is shown in Fig.~\ref{fig:quant}. In order to see whether quantum processes can have an effect that can compete with the processes in Sec.~\ref{sec:2B}, we first identify the effective quantity corresponding to the reduced Planck constant $\hbar$. 
Recovering $G = M_{pl}^{-2}$ temporarily, we rewrite the action in terms of non-dimensional quantities as 
\begin{align}\label{eq:actionnd}
	S = \left(\frac{F_a}{M_{pl}}\right)^2\left(\frac{M}{M_{pl}}\right)^2 \int d^4\! \hat{x} \sqrt{-g} \left\{ - \frac{1}{2}g^{\mu\nu}\hat{\partial}_\mu \phi \hat{\partial}_\nu \phi - \hat{\mu}^2 \left(1 - \cos\phi\right)\right\}~,
\end{align}
where the quantities with ``$\ \hat{ }\ $'' are non-dimensional normalized by the length scale $G M$. For example, $\hat{\mu} = G \mu M$. 
  From this action, we find that the combination $F_a^2 M ^2/M_{pl}^4$ 
acts effectively as the inverse of the reduced Planck constant $\hbar^{-1}$. Therefore, quantum processes, such as the one shown in Fig.~\ref{fig:quant}, must be suppressed by the factor 
\begin{align}
    \left(\frac{M_{pl}}{F_a}\right)^2\left(\frac{M_{pl}}{M}\right)^2 \sim 10^{-96}\left(\frac{M_{pl}}{F_a}\right)^2\left(\frac{M_\odot}{M}\right)^2 ~.
\end{align}
This ratio can be also understood as the ratio between the mass of a single axion particle and the typical mass of a macroscopic cloud considered here, $\sim M F_a^2$. 
Thus, quantum processes are extremely suppressed, unless $F_a$ takes a significantly small value, which is unlikely for a string axion. Hereafter, we only focus on processes which can be described by the classical equations of motion.

Since we have confirmed that the quasi-stationary configuration can be well approximated by a linearized solution with its amplitude determined by the balance of competing perturbative processes, we can safely neglect the deformation of clouds by the self-interaction to discuss the subsequent evolution. In the following calculation, we use the linearized solution to describe the quasi-stationary configuration. For the same reason as before, we only consider the excitation of low frequency radiation and non-superradiant dissipative modes (see Sec.~\ref{sec:2B} for their definitions).

% \subsection{Excitation of other modes}\label{sec:4A}

Up to this point, we considered an axion condensate which starts with only two component clouds belonging to the $l=m=1$ and $l=m=2$ fundamental modes. However, quantum fluctuations should give a tiny seed to all modes. Therefore, the processes in which the source terms of the non-linear interaction include other modes, such as overtone and higher multipole modes can be relevant. These processes have been studied extensively in Ref.~\cite{Baryakhtar:2020gao} under the non-relativistic approximation, $\mu M \ll 1$. Here, we extend the analysis to the relativistic regime with $\mu M \sim 1$.

\begin{figure}[t]
 \centering
  \includegraphics[keepaspectratio, scale=0.3]{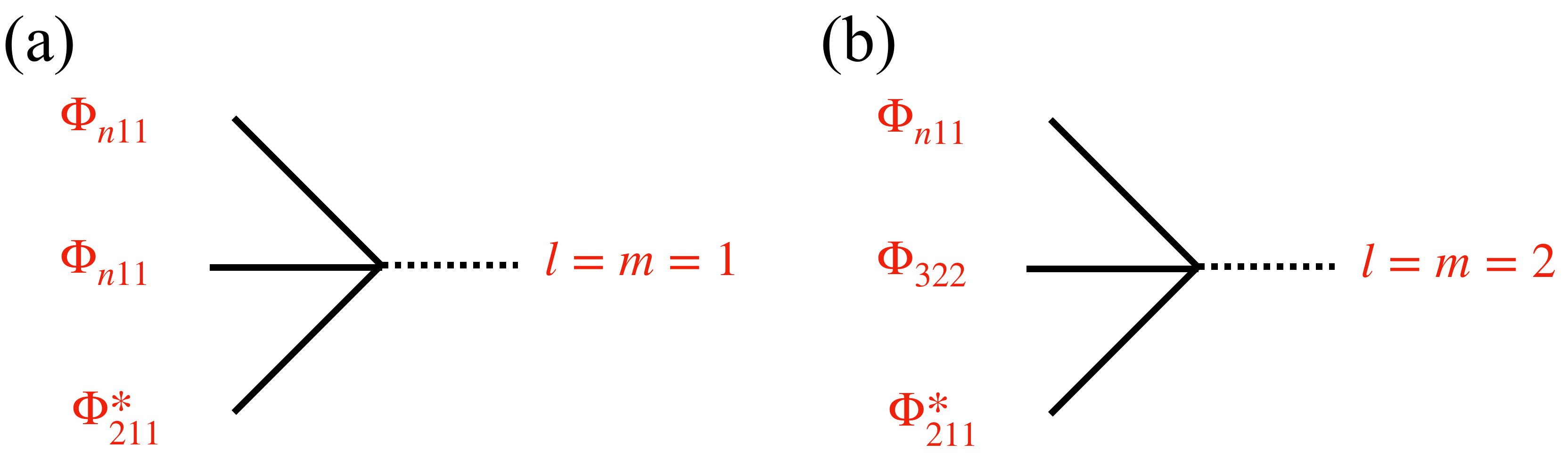}
 \includegraphics[keepaspectratio, scale=0.3]{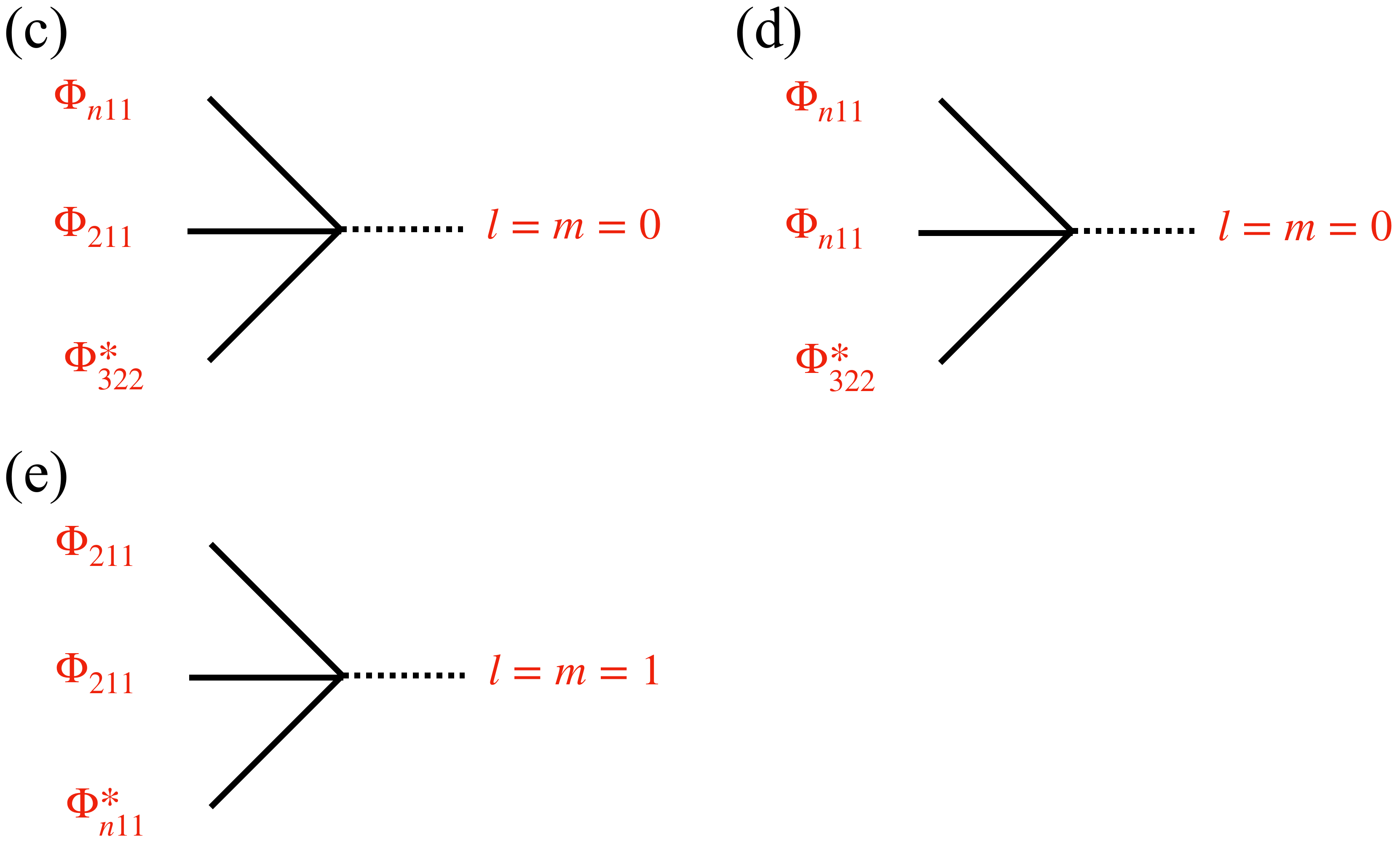}
 \caption{Diagrams for the dissipative processes when the $l=m=1$ overtone modes are taken into account. Top two diagrams ((a) and (b)) correspond to the processes that dissipate energy to infinity. Bottom three diagrams ((c), (d), and (e)) correspond to the processes that dissipate energy to the central BH. As in Fig.~\ref{fig:emissionleading}, we only consider the modes having the smallest possible $l$ in the right-hand side of the diagrams.}
 \label{fig:overtoneint}
\end{figure}

\subsection{Excitation of $l=m=1$ overtones}\label{sec:4A1}

First, we consider to include an overtone mode ($n > 2$) of the $l=m=1$ multipole with frequency $\omega_o$. 
Similarly to the interaction between the $(\omega,m) = (\omega_1,1)$ and $(\omega_2,2)$ modes discussed in Sec.~~\ref{sec:2B}, we can determine the possibly excited modes by writing down the source terms of the first-order perturbation equation with respect to the coupling. 
The source terms labeled by $(\omega,m)$ are relevant for dissolving the quasi-stationary configuration when (i) $\omega > \mu$,  
which includes excitations of the low frequency radiation $\omega \gtrsim \mu$, 
or (ii) $\omega < \mu$ with a sufficiently small $m$ to be non-superradiant. 
The processes in (i) contribute to the energy dissipation to infinity, while those in (ii) contribute to the energy dissipation to the BH. We focus on the processes with the excited mode having the lowest orbital angular momentum, since they give the leading contribution to the energy dissipation.
In Fig.~\ref{fig:overtoneint} we summarize the processes in a diagrammatic form.

The essential difference from the interaction between the $l=m=1$ mode and the $l=m=2$ mode is in that there are no processes which feed the clouds composed of $l=m=1$ overtone modes. This can be seen by the diagram (e) in Fig.~\ref{fig:overtoneint}. This process produces the $l=m=1$ fundamental mode from the overtone mode of $l=m=1$. Unlike the corresponding process in Sec.~\ref{sec:2B} (the left diagram in Fig.~\ref{fig:emissionleading}), which produces the $m=0$ mode, the right-hand side of the diagram (e) in Fig.~\ref{fig:overtoneint} is the $m=1$ mode, which is superradiant. 
Thus, the direction of the transition and hence that of the energy flow are opposite; the energy is not lost into but extracted from the BH. As a result, overtone modes dissipate through this process. This is the reason why we can conclude that $l=m=1$ overtones simply decay in the two mode approximation adopted in Sec.~\ref{sec:2B}.

Here, we should notice that a process that produces an overtone mode from the $l=m=2$ mode is absent, in contrast to the process to produce the $l=m=1$ fundamental mode (corresponding to the right diagram in Fig.~\ref{fig:emissionleading}). 
This is because the frequency of the mode to be excited, $2\omega_2- \omega_o < \mu$,  is not large enough to escape to infinity\footnote{Under the non-relativistic approximation, where the real part of the frequency can be approximately estimated by using the well-known result for the hydrogen atom,
\begin{align}
    2\omega_2 - \omega_o = 2\mu \left(1 - \frac{(\mu M)^2}{2\times 3^2}\right) -  \mu (1 - \frac{(\mu M)^2}{2n^2}) = \mu\left(1 - (\mu M)^2\left(\frac{1}{3^2} - \frac{1}{2 n ^2}\right)\right) < \mu~,
\end{align}
for $n \geq 3$.
}. 
In summary, the interaction between an $l=m=1$ overtone mode and the $l=m=2$ fundamental mode always dissipates the overtone mode. 
Therefore, overtone modes can grow by the superradiant instability, only 
in the early epoch before the dissipation due to the mode-mode interaction begins to balance.

\begin{figure}[t]
 \centering
 \includegraphics[keepaspectratio, scale=0.4]{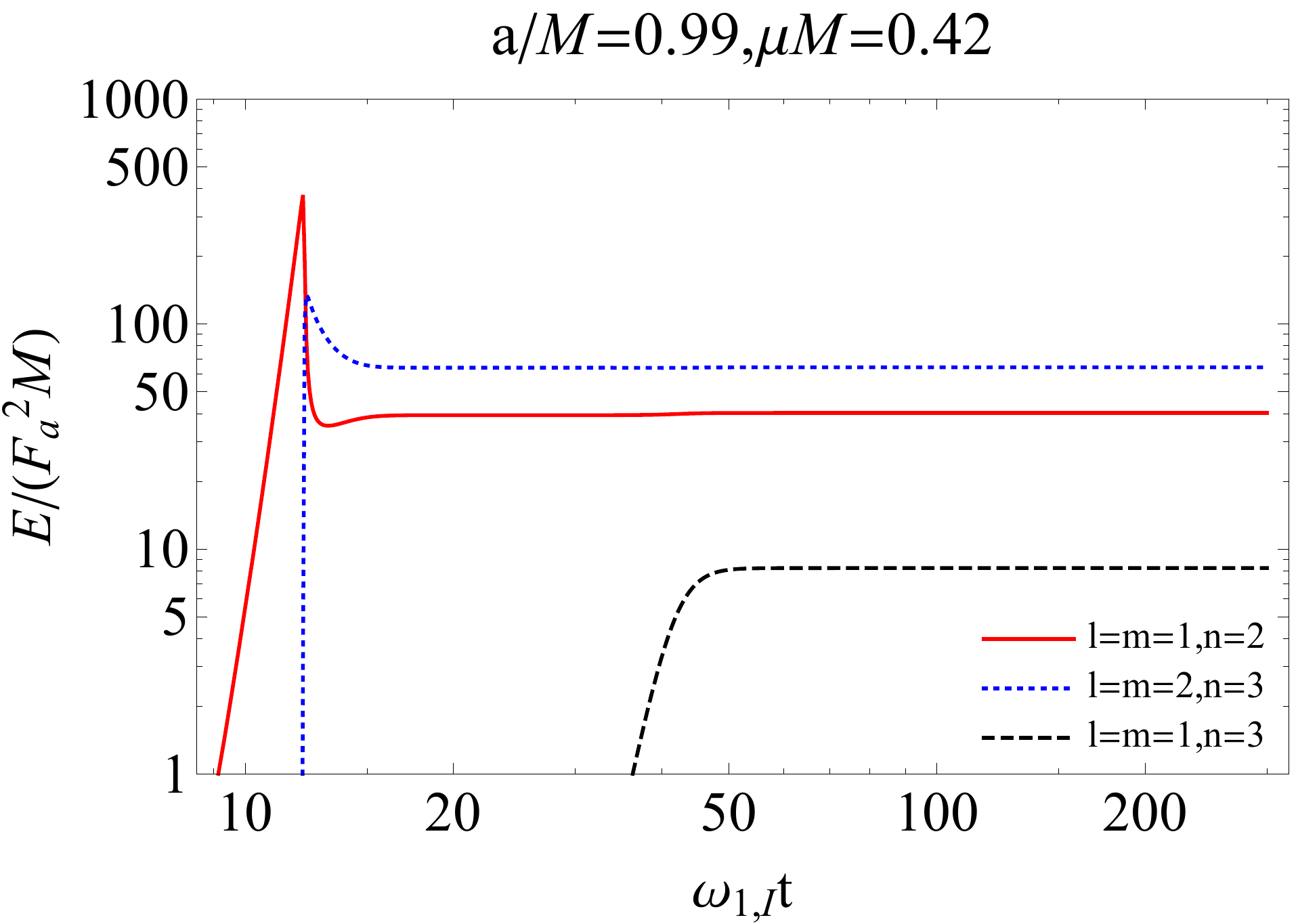}
  \includegraphics[keepaspectratio, scale=0.4]{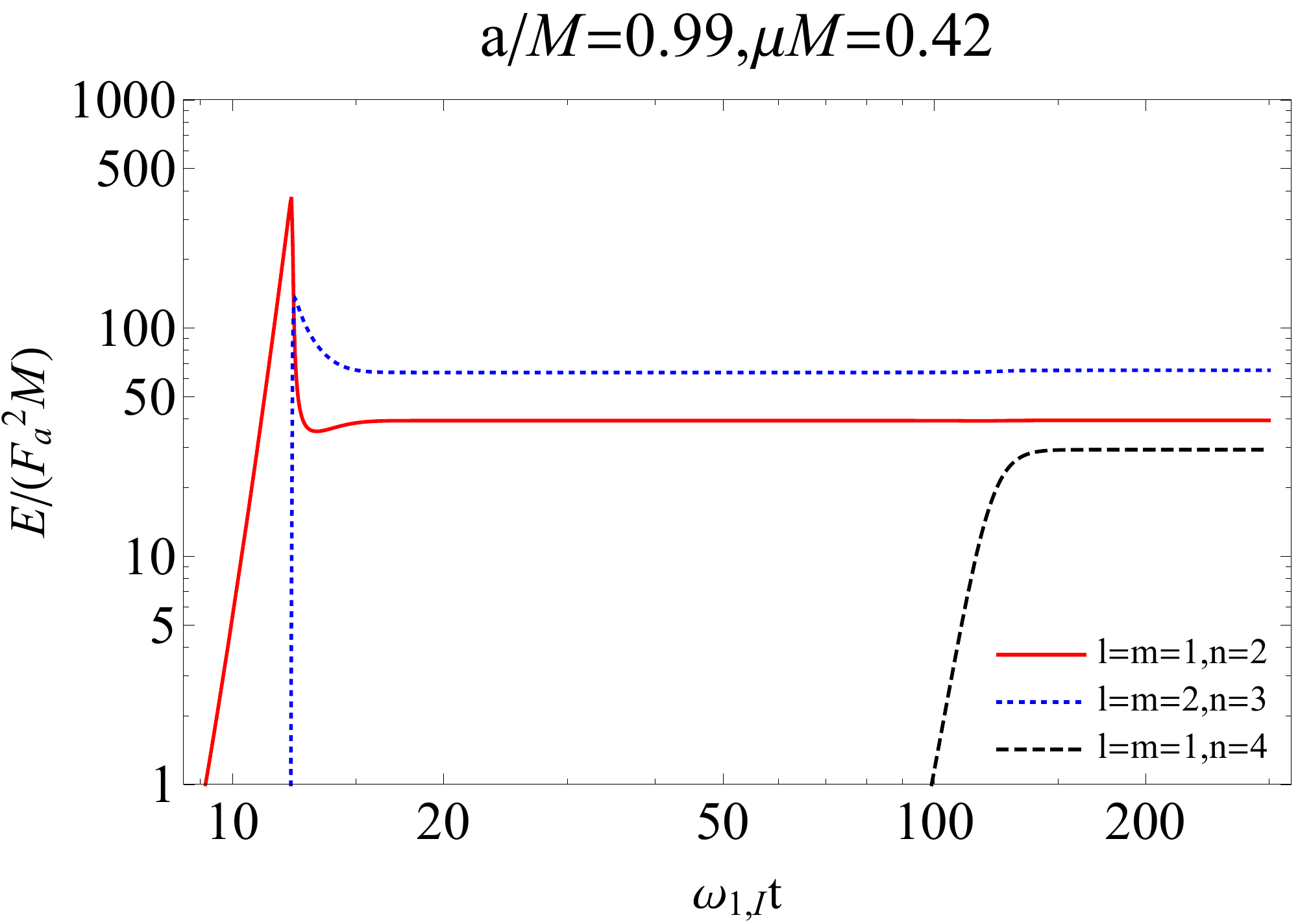}
    \includegraphics[keepaspectratio, scale=0.4]{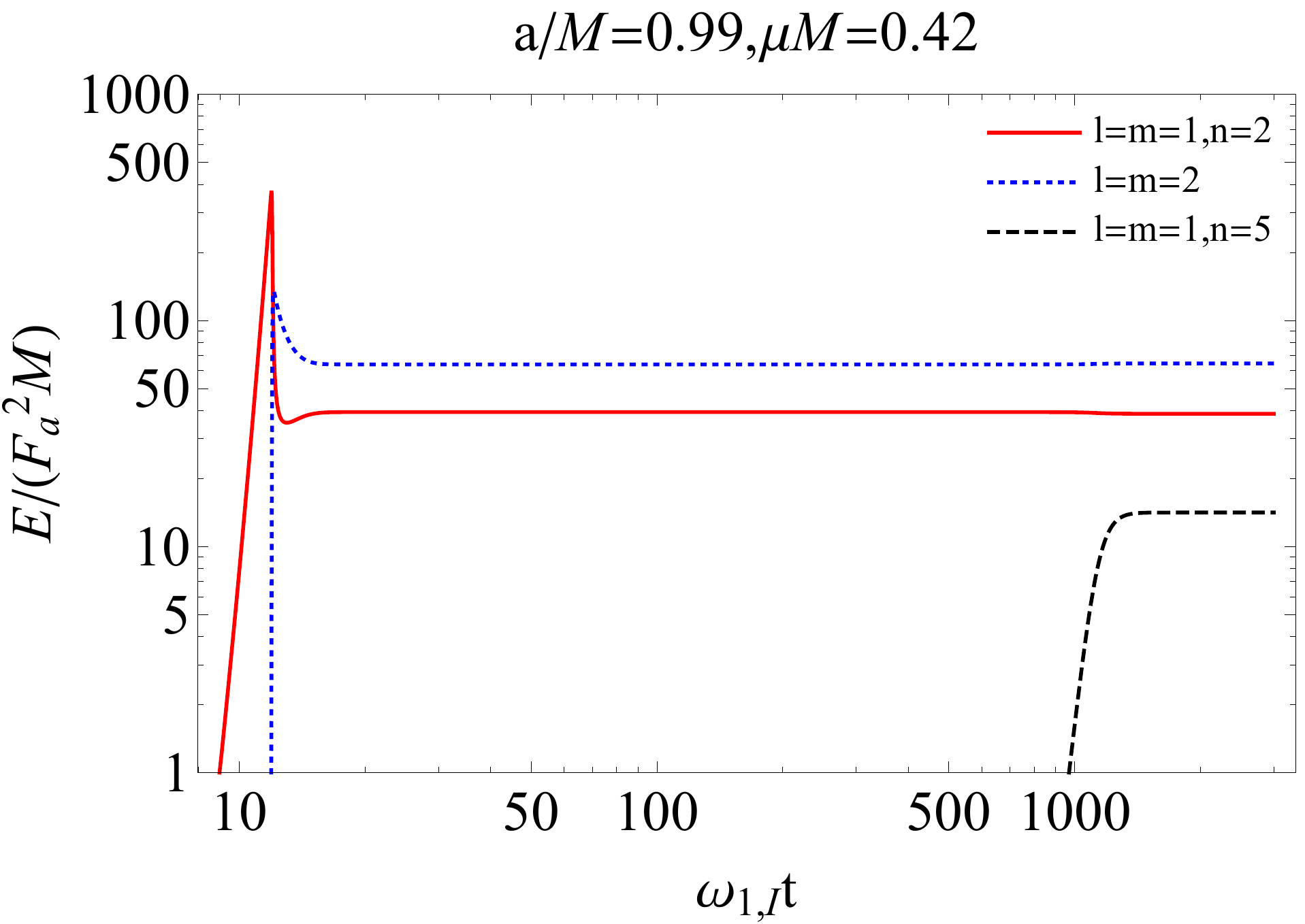}
 \caption{Examples of the time evolution including an $l=m=1$ overtone mode. The red solid and blue dotted lines correspond to the $l=m=1$ and $l=m=2$ fundamental modes. The black dashed line correspond to the overtone mode. From left to right, the label $n$ of the overtone modes is varied from $3$ to $5$. The horizontal axis is the time normalized by the superradiant instability rate of the $l=m=1$ fundamental mode. We set the axion mass and the BH spin to $\mu M = 0.42$ and $a/M = 0.99$.}
 \label{fig:overtone}
\end{figure}

The time evolution including an overtone can be derived from the local energy and angular momentum conservation laws, similarly to the method in Sec.~\ref{sec:2B}. The equations that govern the time evolution of three clouds are 
\begin{align}
    \frac{d E_1}{dt} =& 2 \omega_{1,I} E_1 - \frac{2 \omega_{1,R}}{2 \omega_{1,R} - \omega_{2,R}} F_0 E_1^2 E_2 + \frac{\omega_{1,R}}{2\omega_{2,R}-\omega_{1,R}} F_3 E_1 E_2^2\cr
    &+ \frac{\omega_{1,R}}{2 \omega_{o,R} - \omega_{1,R}}F_{1^*oo}E_1 E_o^2 + \frac{\omega_{1,R}}{\omega_{2,R} + \omega_{o,R} - \omega_{1,R}}F_{1^*2o}E_1 E_2 E_o\cr
    &- \frac{2\omega_{1,R}}{2 \omega_{1,R} - \omega_{o,R}}F_{11o^*}E_1^2 E_o - \frac{\omega_{1,R}}{\omega_{1,R} + \omega_{o,R} - \omega_{2,R}}F_{12^*o}E_1 E_2 E_o~,\label{eq:threemodelm1}\\
    \frac{d E_2}{dt} =& 2 \omega_{2,I} E_1 + \frac{\omega_{2,R}}{2 \omega_{1,R} - \omega_{2,R}} F_0 E_1^2 E_2 - \frac{2\omega_{2,R}}{2\omega_{2,R}-\omega_{1,R}} F_3 E_1 E_2^2\cr
    &- \frac{\omega_{2,R}}{\omega_{2,R} + \omega_{o,R} - \omega_{1,R}}F_{1^*2o}E_1 E_2 E_o \cr
    &+ \frac{\omega_{2,R}}{2 \omega_{o,R} - \omega_{2,R}}F_{2^*oo}E_2 E_o^2 + \frac{\omega_{2,R}}{\omega_{1,R} + \omega_{o,R} - \omega_{2,R}}F_{12^*o}E_1 E_2 E_o~,\label{eq:threemodelm2}\\
    \frac{d E_o}{dt} =& 2 \omega_{o,I} E_o - \frac{2\omega_{o,R}}{2 \omega_{o,R} - \omega_{1,R}}F_{1^*oo}E_1 E_o^2 - \frac{\omega_{o,R}}{\omega_{2,R} + \omega_{o,R} - \omega_{1,R}}F_{1^*2o}E_1 E_2 E_o\cr
    &+ \frac{\omega_{o,R}}{2 \omega_{1,R} - \omega_{o,R}}F_{11o^*}E_1^2 E_o - \frac{2\omega_{o,R}}{2 \omega_{o,R} - \omega_{2,R}}F_{2^*oo}E_2 E_o^2 - \frac{\omega_{o,R}}{\omega_{1,R} + \omega_{o,R} - \omega_{2,R}}F_{12^*o}E_1 E_2 E_o\label{eq:threemodeot}~.
\end{align}
Here, $E_o$ is the energy of the overtone mode. The coefficients $F_{abc}$ in Eqs.~\eqref{eq:threemodelm1} - \eqref{eq:threemodeot} are the values of the energy flux of the diagrams in Fig.~\ref{fig:overtoneint} for normalized amplitude of clouds in the source term. For example, $F_{1^*oo}$ corresponds to the first diagram in Fig.~\ref{fig:overtoneint}. The actual expression of $F_{abc}$ is similar to Eqs.~\eqref{eq:flux0} and \eqref{eq:flux3}.
We show the time evolution of the three mode system composed of $(n,l,m) = (2,1,1), (3,2,2)$ and 
$(n,1,1)$ with $n=3,4,5$ in Fig.~\ref{fig:overtone}. As expected from the above discussion, in the early epoch the overtone mode is built up by the superradiance, and eventually the growth saturates at the point where the dissipation balances the superradiance. 
The excitation of $n=5$ is delayed compared to the $n=4$ and $n=3$ overtones because the dissipation processes involving only one overtone mode in the source (the diagrams (b) and (c) in Fig, \ref{fig:overtoneint}) become comparable to the superradiant growth rate of the overtone for the quasi-stationary configuration. In fact, one can numerically confirm that the dissipation becomes faster than the growth due to the superradiance for $n \ge 6$. Therefore, the late time growth of $l=m=1$ overtones as shown in Fig.~\ref{fig:overtone} occurs only for $n \le 5$.
Interestingly, the time evolution of the energies of the $l=m=1$ and $l=m=2$ fundamental modes is almost unchanged.
This is because dissipation of the $l=m=1$ overtones are so strong that the overtones never dominate the whole condensate.

\subsection{Excitation of $l=m=2$ overtones}\label{sec:4B}

\begin{figure}[t]
 \centering
  \includegraphics[keepaspectratio, scale=0.3]{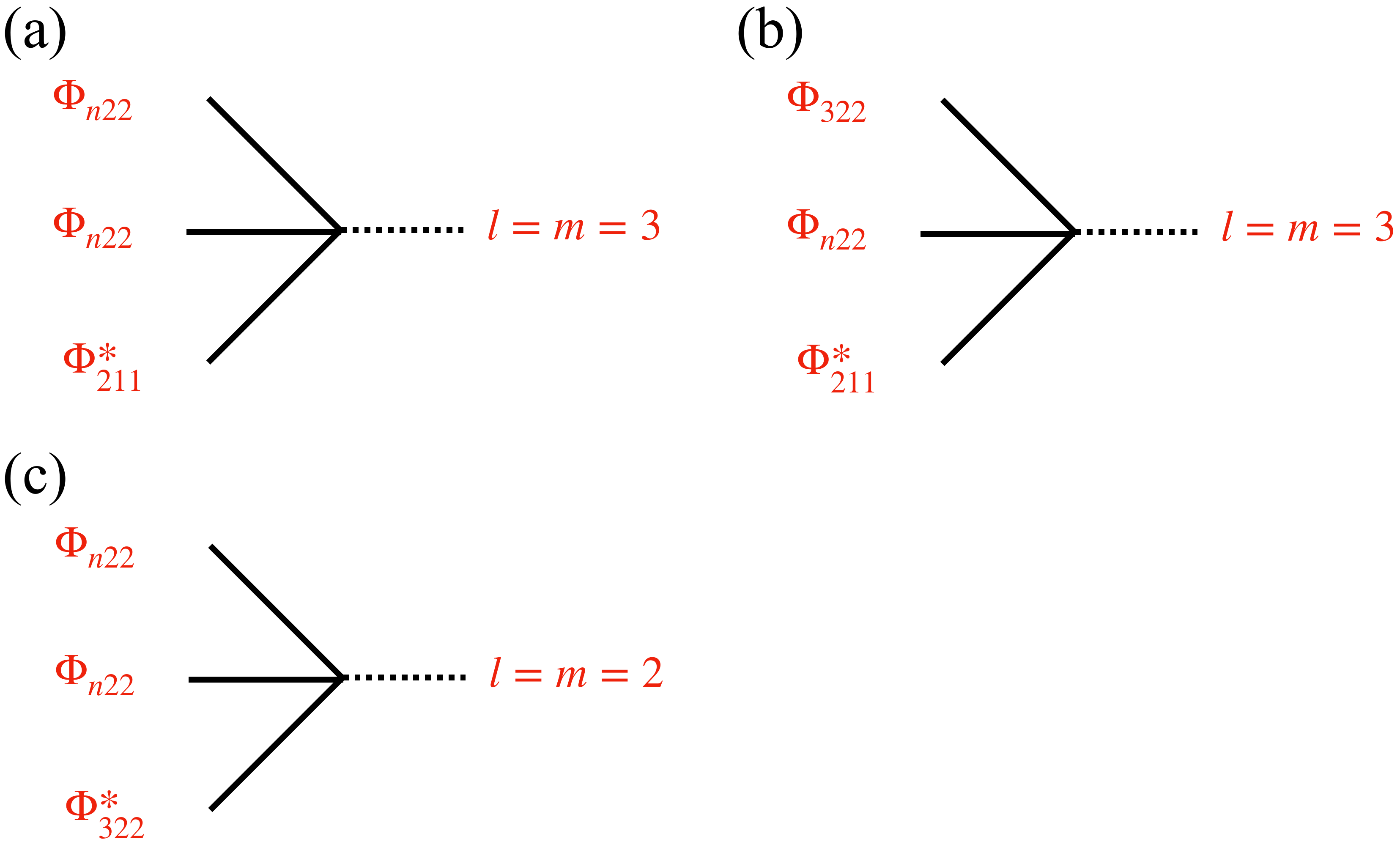}
 \includegraphics[keepaspectratio, scale=0.3]{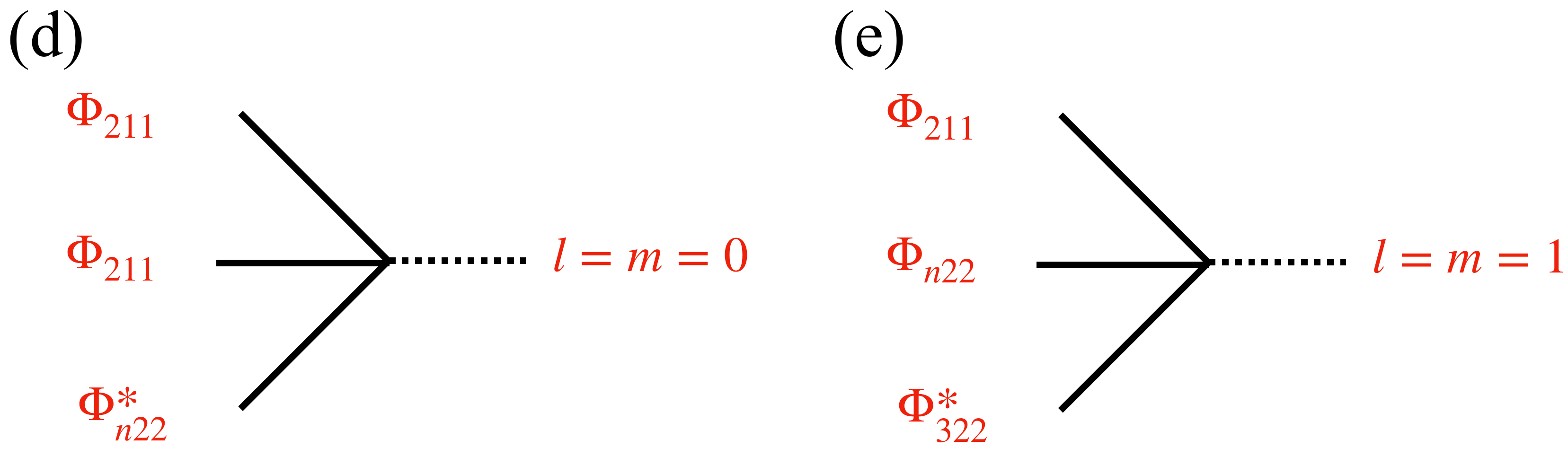}
 \caption{The diagrams involved when overtone modes of $l=m=2$ are included. Top three diagrams ((a), (b), and (c)) correspond to the processes which dissipate energy to infinity. Bottom two diagrams ((d) and (e)) correspond to the processes that dissipate energy to the BH. Since $2\omega_{n22} - \omega_{322}>\mu$ holds only for $n \ge 5$, the process specified by the diagram (c) can dissipate energy only for $n \ge 5$.}
 \label{fig:overtonelm2int}
\end{figure}

Next, we consider the excitation of $l=m=2$ overtones. The relevant processes in this case are summarized in the diagrammatic form in Fig.~\ref{fig:overtonelm2int}. The difference from $l=m=1$ overtones is in that there exists a process which excites $l=m=2$ overtones by the energy dissipation to the BH (the diagram (d) in Fig.~\ref{fig:overtonelm2int}). However, the flux induced by the diagram (d) in Fig.~\ref{fig:overtonelm2int} is small compared to the process which produces the $l=m=2$ fundamental mode (the left diagram of Fig.~\ref{fig:emissionleading} in Sec.~\ref{sec:2B}). 
Therefore, the growth of $l=m=2$ overtones are suppressed compared to that of the $l=m=2$ fundamental mode. For this reason, the condensate first reaches the quasi-stationary configuration composed of the $l=m=1$ and $l=m=2$ fundamental modes, and at this point, the energy of $l=m=2$ overtones are much smaller than that of the fundamental modes.

\begin{figure}[t]
 \centering
 \includegraphics[keepaspectratio, scale=0.4]{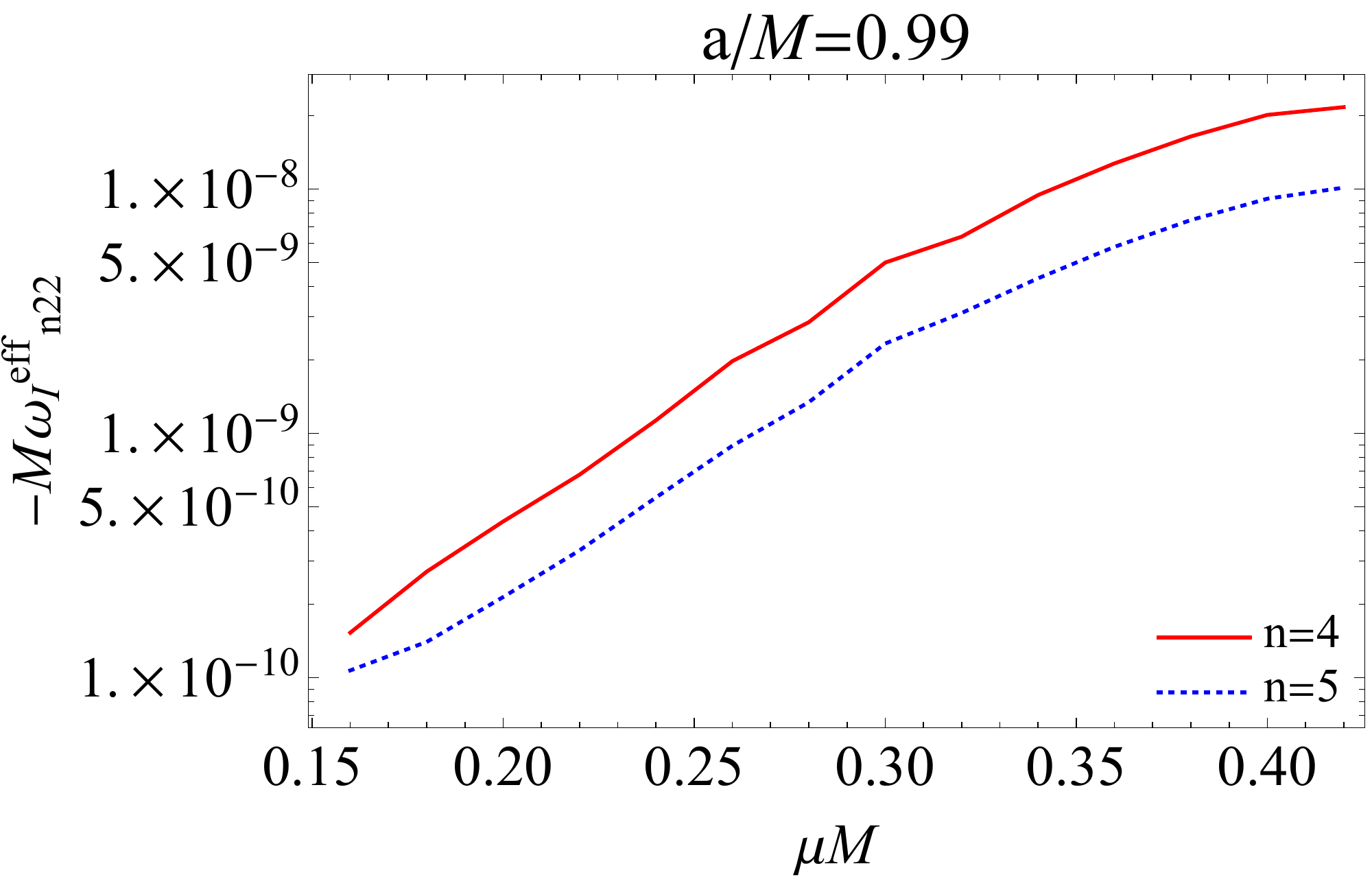}
 \caption{The red solid and blue dotted lines correspond to the $\omega^{\rm eff}_{n22,I}$ for $n=4$ and $n=5$, respectively. The spin of the central BH is set to $a/M = 0.99$.}
 \label{fig:wIeff}
\end{figure}

\begin{figure}[t]
 \centering
 \includegraphics[keepaspectratio, scale=0.4]{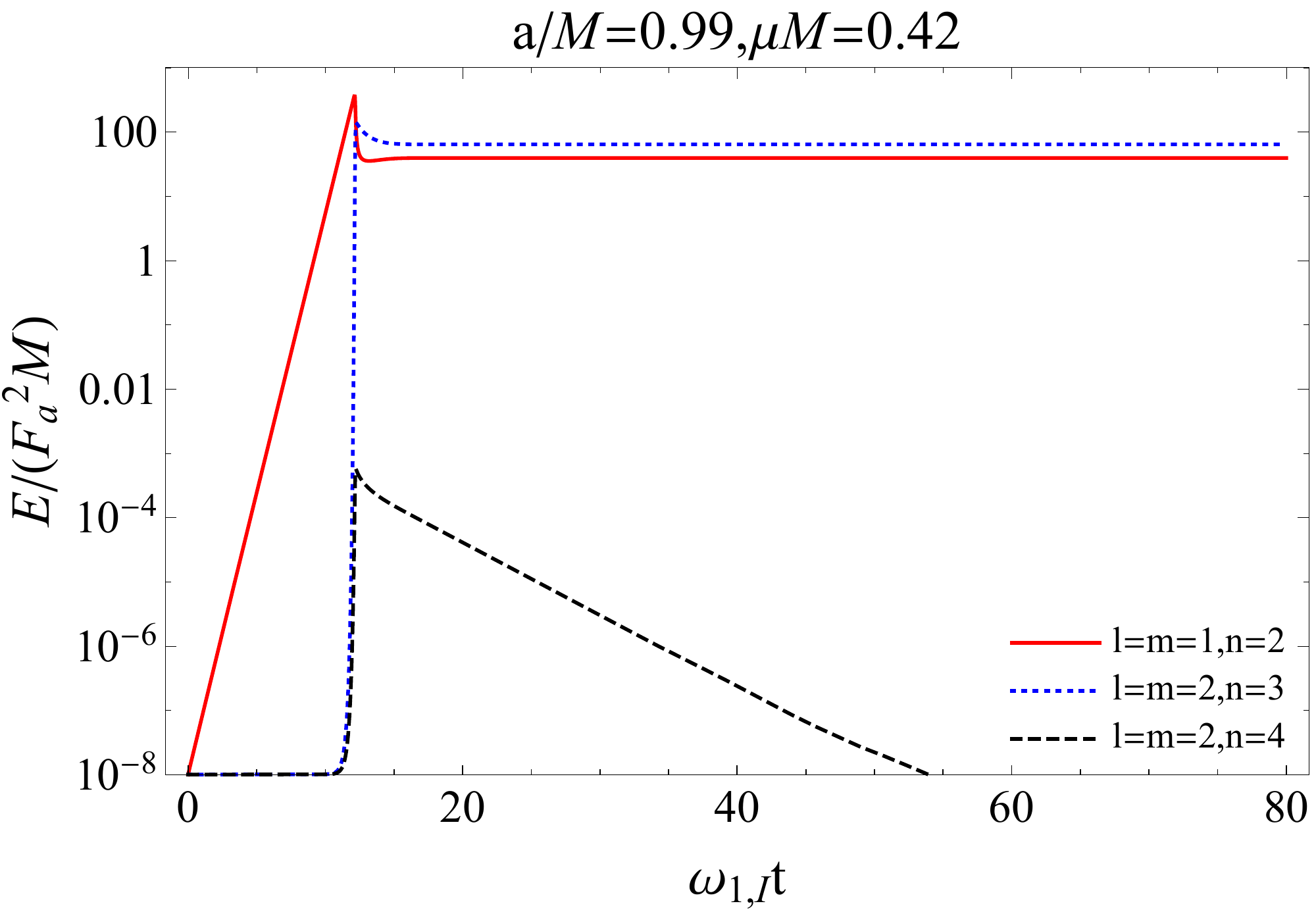}
  \includegraphics[keepaspectratio, scale=0.4]{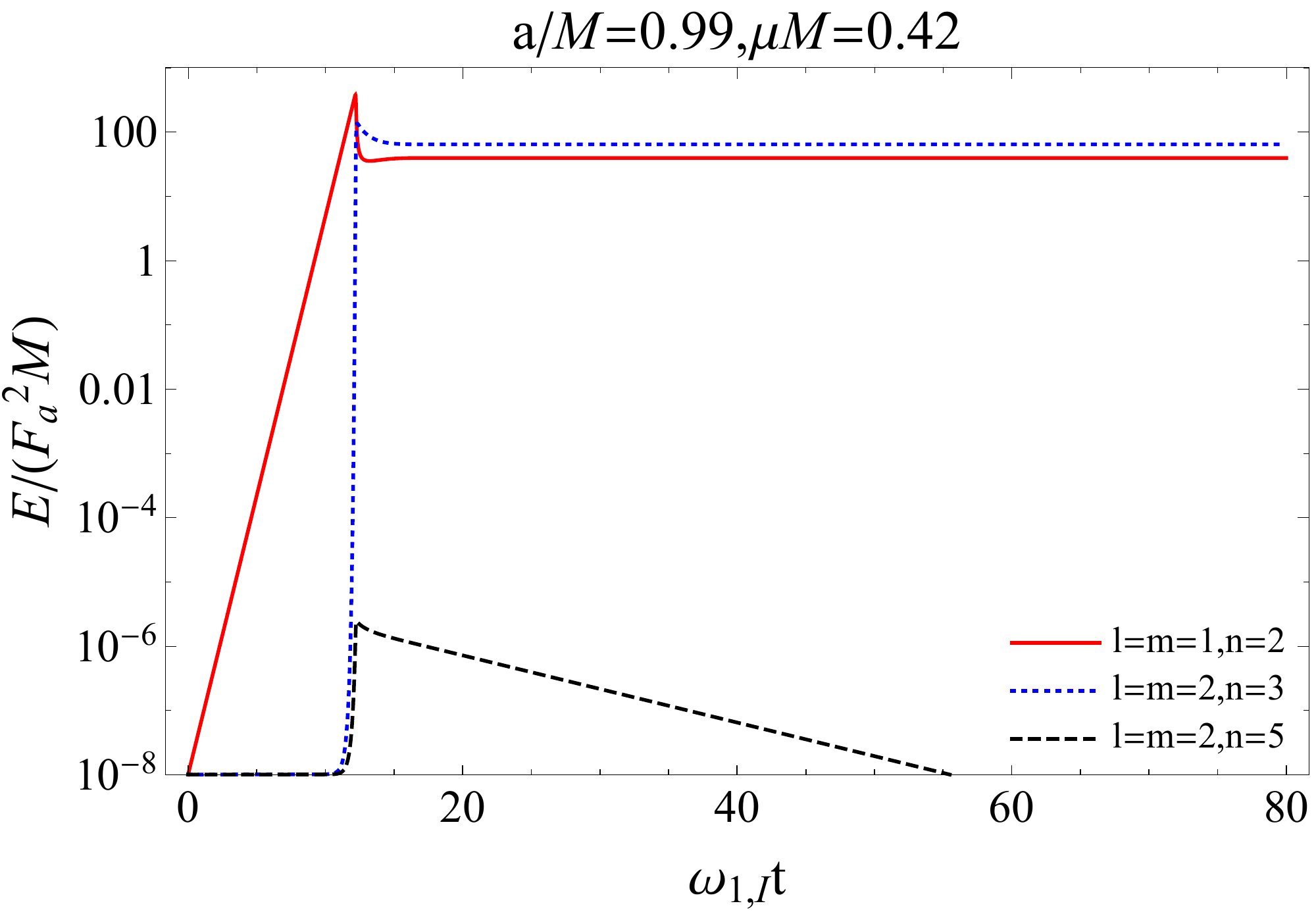}
 \caption{The same figures as Fig.~\ref{fig:overtone}, but with $l=m=2$ overtones instead of $l=m=1$ overtones. The black dashed lines in the left and right panels correspond to the $l=m=2$ overtone with $n=4$  and $n=5$, respectively.}
 \label{fig:overtonelm2}
\end{figure}

In order for $l=m=2$ overtones to continue to grow after the fundamental modes reach the quasi-stationary configuration, the growth rate of $l=m=2$ overtones must be maintained to be larger than the dissipation rate. Since the energy of $l=m=2$ overtones is still small after the saturation of the growth of the fundamental modes, considering the terms linear in $E_o$ in Eq.~\eqref{eq:threemodeot} is enough. Keeping only the terms linear in $E_o$, we obtain
\begin{align}
    \frac{d E_o}{dt} \sim& \left[2 \omega_{o,I}  - \Biggl(\frac{\omega_{o,R}}{\omega_{2,R} + \omega_{o,R} - \omega_{1,R}}F_{1^*2o}+\frac{\omega_{o,R}}{\omega_{1,R} + \omega_{o,R} - \omega_{2,R}}F_{12^*o}\right)E^{\rm sat}_1 E^{\rm sat}_2 \cr
    &\qquad + \frac{\omega_{o,R}}{2 \omega_{1,R} - \omega_{o,R}}F_{11o^*}\left(E_1^{{\rm sat}}\right)^2 \Biggr] E_o \cr
    \equiv& 2 \omega_{n22,I}^{\rm eff} E_o~.\label{eq:weff}
\end{align}
Here, $E^{\rm sat}_{1}$ and $E^{\rm sat}_{2}$ are the energies of the $l=m=1$ and $l=m=2$ clouds in the quasi-stationary state, determined by the perturbative analysis in Sec.~\ref{sec:2B}. If $\omega_{o,I}^{\rm eff} < 0$, the $l=m=2$ overtone cannot grow. In Fig.~\ref{fig:wIeff}, we show the behavior of $\omega_{o,I}^{\rm eff}$ for $n=4$ and $5$. As we observe from the figure, $\omega_{n22,I}^{\rm eff}$ is always negative. Therefore, we do not expect $l=m=2$ overtones to grow. In Fig.~\ref{fig:overtonelm2}, we show examples of the time evolution for the three mode system composed of $(n,l,m) = (2,1,1),(3,2,2),$ and $(n,2,2)$ with $n=4,5$. We observe the qualitative behavior as explained above. Due to the dissipation, the $l=m=2$ overtone clouds decrease after the cloud of the $l=m=1$ fundamental mode becomes sufficiently large.

\begin{figure}[t]
 \centering
  \includegraphics[keepaspectratio, scale=0.3]{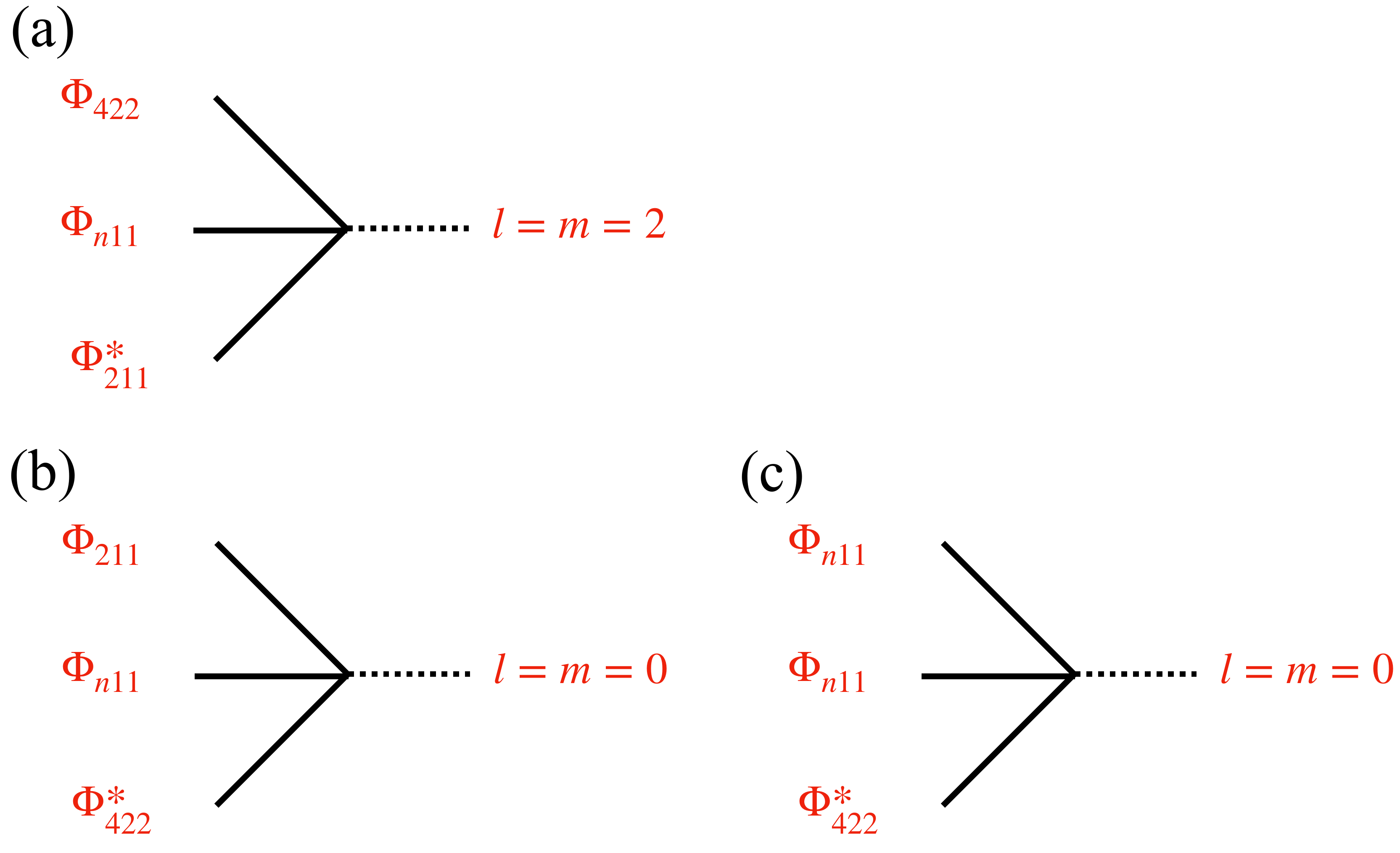}
 \caption{Diagrams for the interaction between the $l=m=1$ overtone modes and the $l=m=2$ overtone modes. The top diagram (a) describes a process which dissipates energy to infinity. Bottom two diagrams ((b) and (c)) describe processes which dissipate energy to the BH. }
 \label{fig:overtonelm2int2}
\end{figure}

\begin{figure}[t]
 \centering
 \includegraphics[keepaspectratio, scale=0.4]{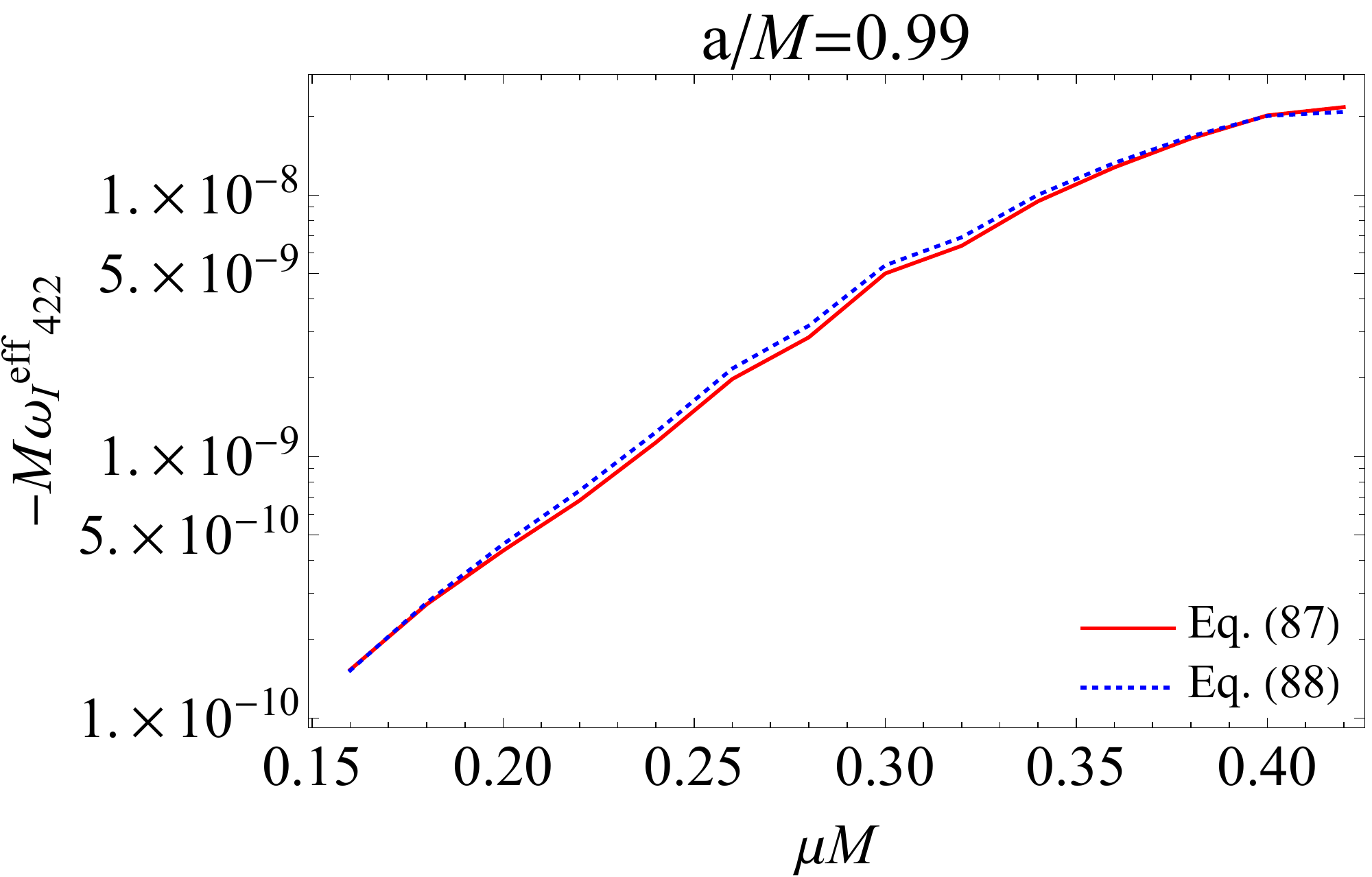}
 \caption{The red solid and blue dotted lines show $\omega^{\rm eff}_{422,I}$, calculated by Eqs.~\eqref{eq:weff} and \eqref{eq:weffmod}, respectively. When calculating $\omega^{\rm eff}_{422,I}$ with Eq.~\eqref{eq:weffmod}, we take into account the cloud composed of the $l=m=1, n=4$ mode that has the largest energy in the quasi-stationary configuration at a late time among clouds composed of $l=m=1$ overtones. The spin of the central BH is set to $a/M = 0.99$.}
 \label{fig:wIeffhigh}
\end{figure}

 However, it is too early to conclude that $l=m=2$ overtones do not always grow. As we showed in Sec.~\ref{sec:4A1}, $l=m=1$ overtones can grow after the saturation of the growth of the fundamental modes (see Fig.~\ref{fig:overtone}). Then, there exists an interaction between the $l=m=1$ and $l=m=2$ overtones. We show the relevant processes for the $n=4,l=m=2$ mode in Fig.~\ref{fig:overtonelm2int2}. With these processes, $\omega_{n22,I}^{\rm eff}$ in Eq.~\eqref{eq:weff} is modified to 
 \begin{align}
     2\omega_{n22,I}^{\rm eff} =& 2 \omega_{o_2,I}  - \left(\frac{\omega_{o_2,R}\,F_{1^*2o_2}}{\omega_{2,R} + \omega_{o_2,R} - \omega_{1,R}}+\frac{\omega_{o_2,R}\,F_{12^*o_2}}{\omega_{1,R} + \omega_{o_2,R} - \omega_{2,R}}\right)E^{\rm sat}_1 E^{\rm sat}_2 + \frac{\omega_{o_2,R}\,F_{11o_2^*}}{2 \omega_{1,R} - \omega_{o_2,R}}\left(E_1^{{\rm sat}}\right)^2 \cr
     & \qquad + \Biggl(\frac{\omega_{o_2,R}\,F_{1o_1o_2^*}}{\omega_{1,R} + \omega_{o_1,R} - \omega_{o_2,R}} - \frac{\omega_{o_2,R}\,F_{1^*o_1o_2}}{\omega_{o_1,R} + \omega_{o_2,R} - \omega_{1,R}}\Biggr) E^{\rm sat}_1 E^{\rm sat}_{o_1} + \frac{\omega_{o_2,R}\,F_{o_1o_1o_2^*}}{2 \omega_{o_1,R} - \omega_{o_2,R}}\left(E_{o_1}^{{\rm sat}}\right)^2 ~.\label{eq:weffmod}
 \end{align}
 Here, the subscripts $o_1$ and $o_2$ correspond to the  $l=m=1$ and $l=m=2$ overtones, respectively. In addition, $E_{1}^{\rm sat}, E_{2}^{\rm sat},$ and $E_{o_1}^{\rm sat}$ stand for the energies of the $(n,l,m) = (2,1,1), (3,2,2), (n,1,1)$ modes at the quasi-stationary configuration of the system composed of these three modes. If the second line in Eq.~\eqref{eq:weffmod} is larger in magnitude than the negative contribution from the first line, $l=m=2$ overtones can grow overcoming the dissipation. In Fig.~\ref{fig:wIeffhigh}, we show $\omega_{422,I}^{\rm eff}$ which is also negative, and hence this mode cannot grow. This result is explained by the smallness of the energy of $l=m=1$ overtones $E_{o_1}^{\rm sat}$ compared to the fundamental modes $E_{1,2}^{\rm sat}$, which leads to the smallness of the fluxes whose source involves $l=m=1$ overtones, compared to the fluxes sourced solely by the fundamental modes.

\begin{figure}[t]
 \centering
 \includegraphics[keepaspectratio, scale=0.3]{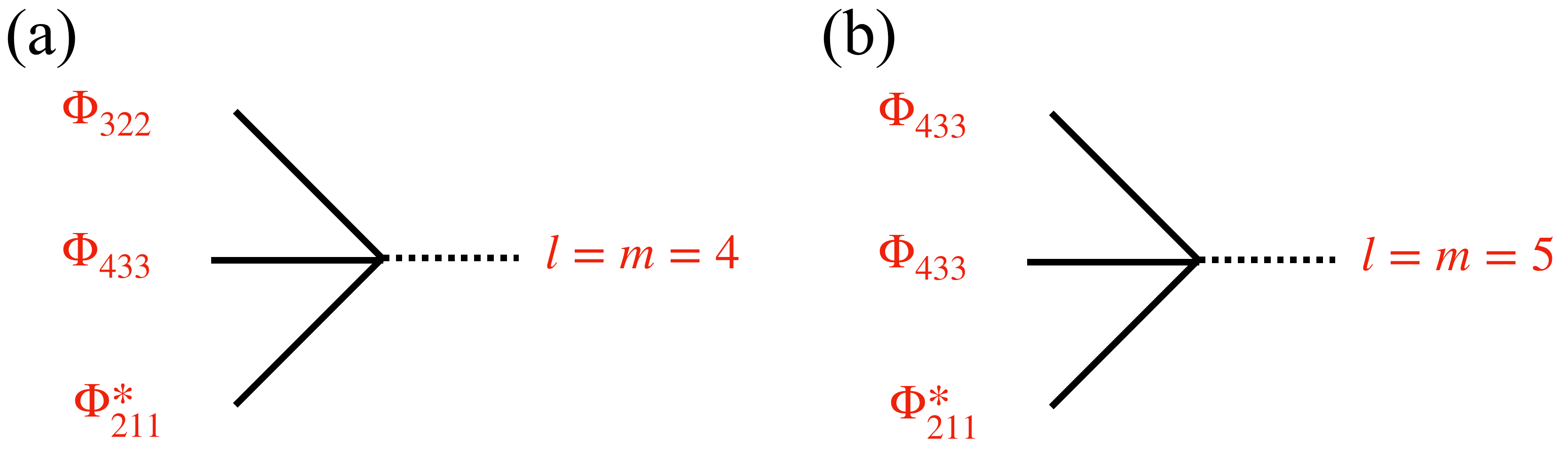}
 \includegraphics[keepaspectratio, scale=0.3]{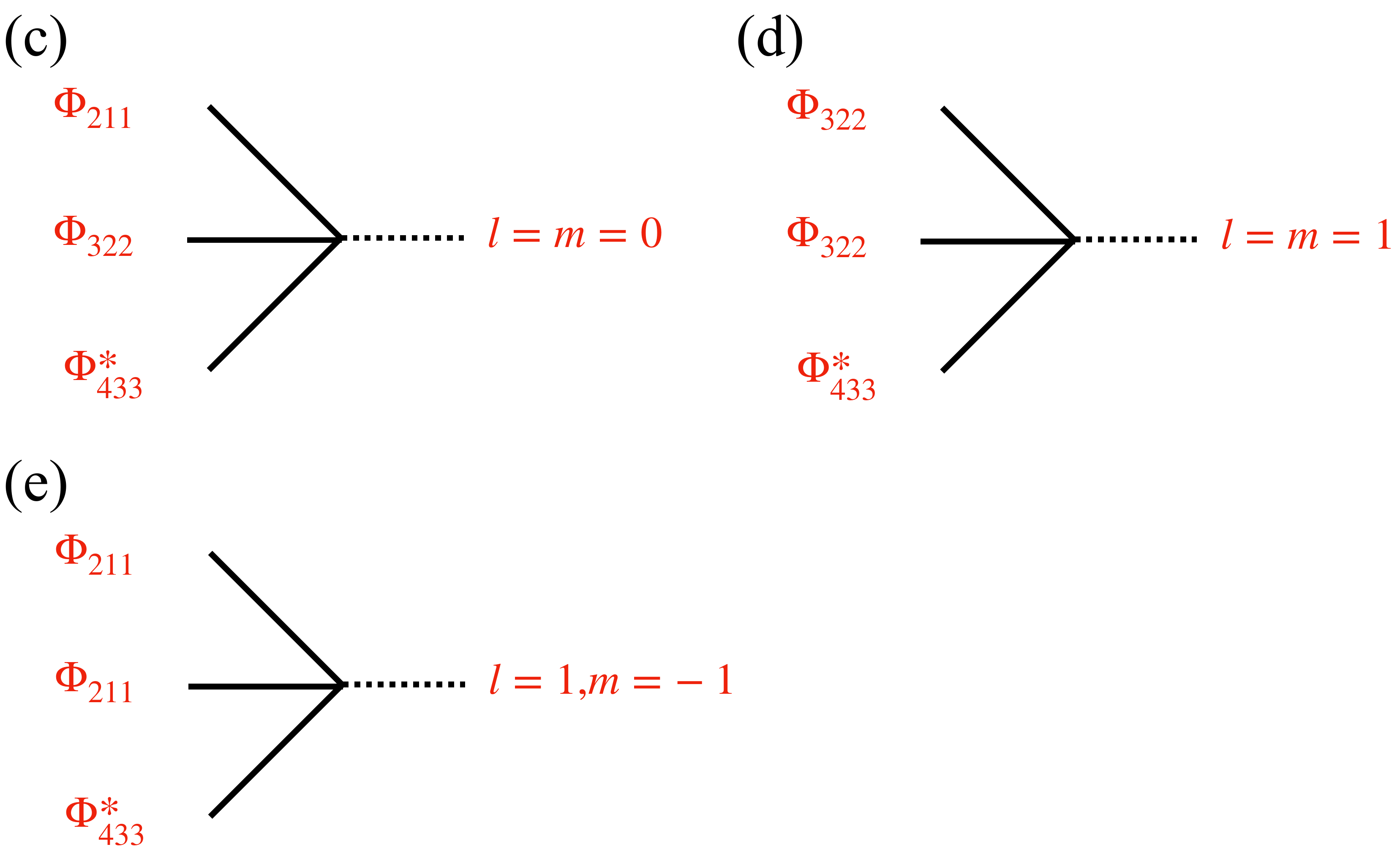}
 \caption{Relevant processes involving $l=m=1$, $l=m=2$ and $l=m=3$ fundamental modes. Top two diagrams ((a) and (b)) describe the processes in which the energy of the excited unbounded modes are dissipated to infinity, and the bottom three ((c), (d), and (e)) describe the ones in which 
 the excited modes are absorbed by the horizon.}
 \label{fig:lm3int}
\end{figure}

\subsection{Excitation of higher multipole modes}\label{sec:4A3}

\begin{figure}[t]
 \centering
 \includegraphics[keepaspectratio, scale=0.6]{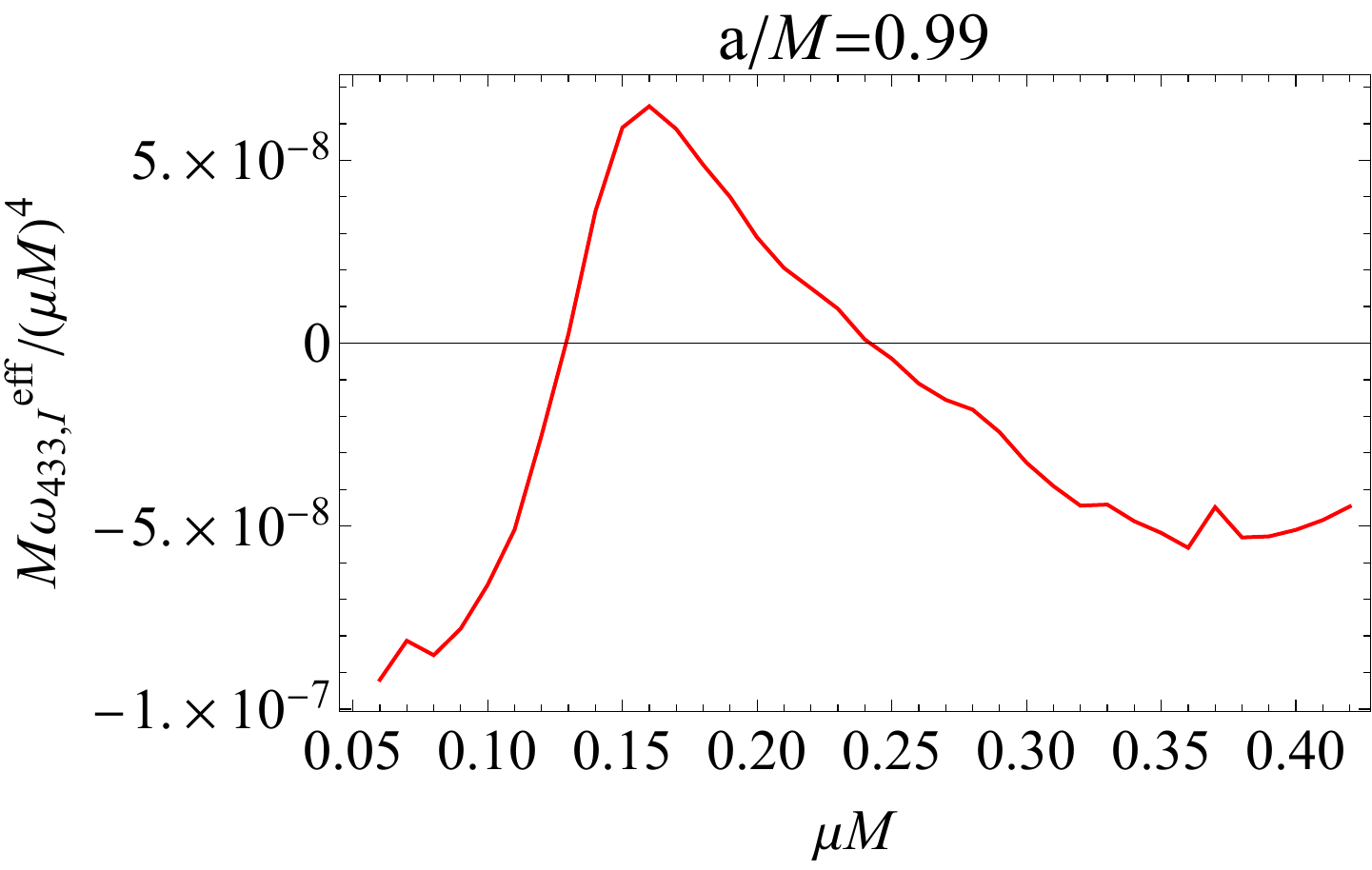}
 \caption{Mass dependence of $\omega^{\rm eff}_{433,I}/(\mu M)^4$ as shown in Fig.~\ref{fig:wIeff} for $\omega^{\rm eff}_{422,I}$. The spin of the central BH is set to $a/M = 0.99$.}
 \label{fig:wIefflm3}
\end{figure}

\begin{figure}[t]
 \centering
 \includegraphics[keepaspectratio, scale=0.4]{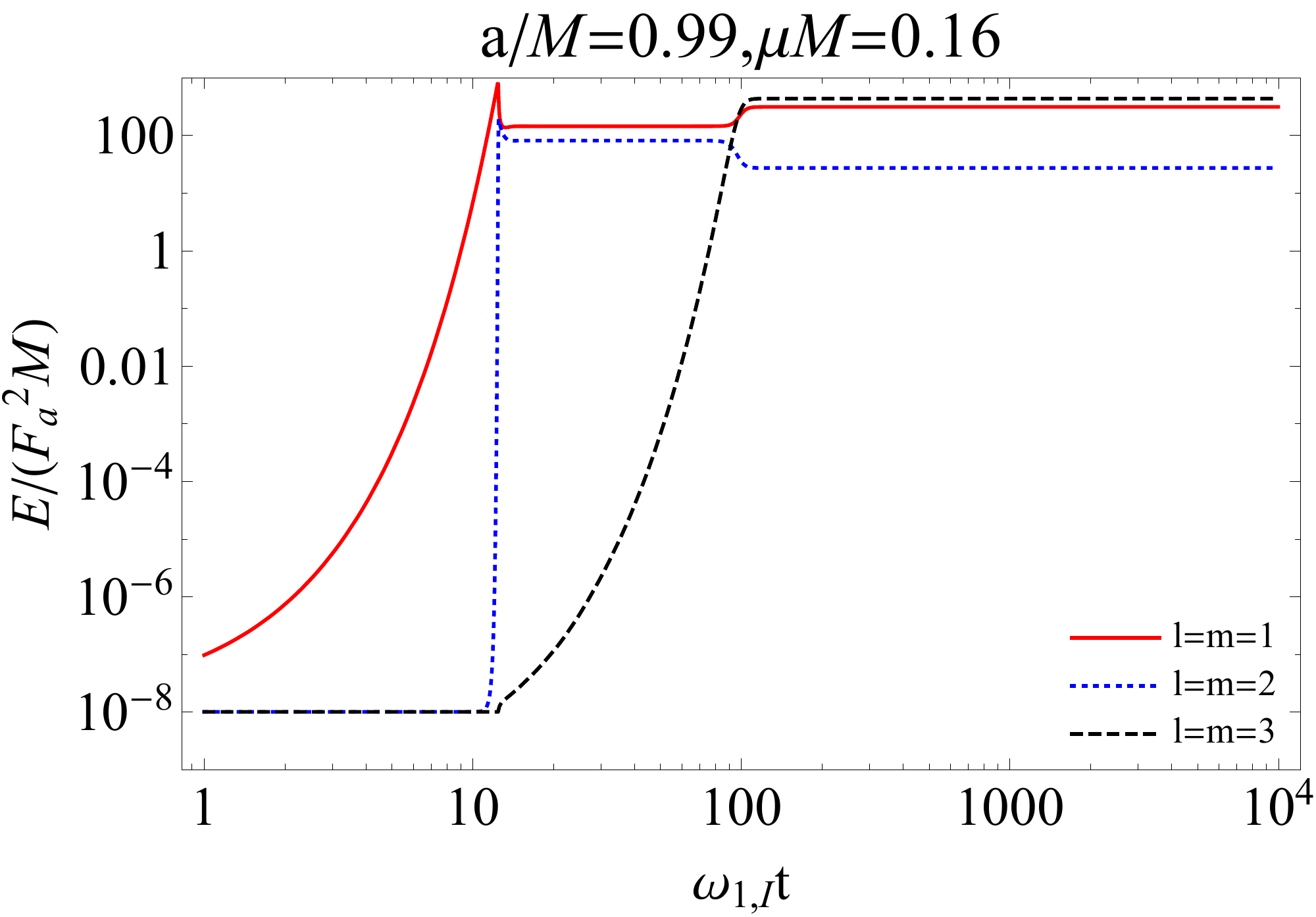}
  \includegraphics[keepaspectratio, scale=0.4]{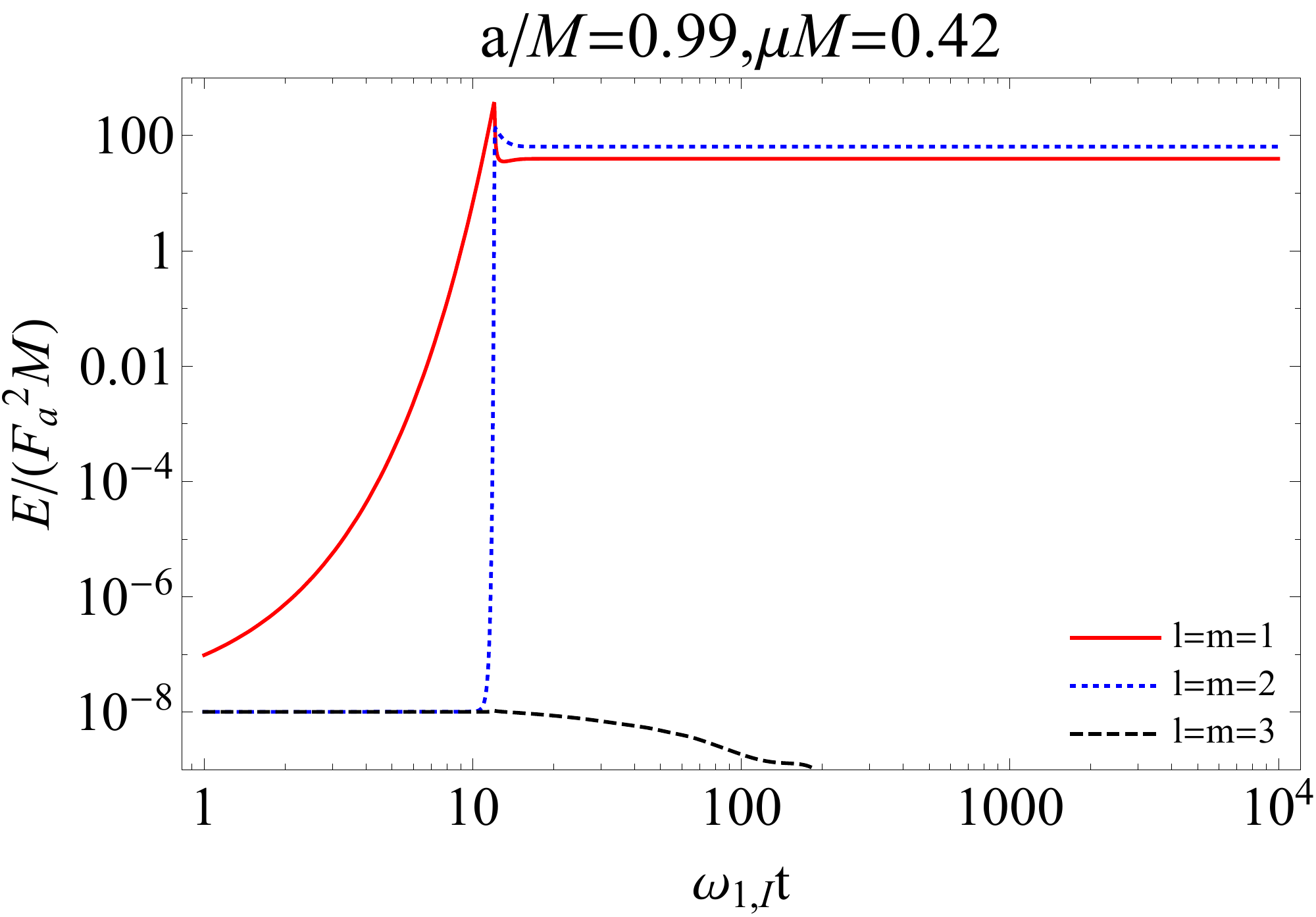}
 \caption{The same figures as Fig.~\ref{fig:overtone}, but with the fundamental $l=m=3$ mode as the third mode, instead of $l=m=1$ overtones. The left and right panels correspond to the case with $\mu M = 0.16$ and $0.42$, respectively.}
 \label{fig:fndlm3}
\end{figure}

Next, we consider the excitation of higher multipole modes. 
The analysis is parallel to the excitation of overtone modes. For concreteness, let us focus on the higher multipole mode with $l=m=3$ and $n= 4$. In Fig.~\ref{fig:lm3int}, we show the relevant diagrams that we include in the numerical evolution. 
Similarly to the $l=m=2$ overtone modes, the $l=m=3$ mode can be produced from the interaction with the $l=m=1$ and $l=m=2$ fundamental modes (diagrams (c) and (e) in Fig.~\ref{fig:lm3int}). At the same time, the $l=m=3$ mode dissipates through the process described by the diagram (a). The $l=m=3$ mode can grow only in the parameter region where the production rate exceeds the dissipation rate. In a similar manner to $\omega_{n22,I}^{\rm eff}$ in Eq. \eqref{eq:weff}, we define $\omega_{433,I}^{\rm eff}$ as 
\begin{align}\label{eq:wefflm3}
    2 \omega_{433,I}^{\rm eff} &\equiv 2 \omega_{433,I}  + \left(-\frac{\omega_{433,R}}{\omega_{2,R} + \omega_{433,R} - \omega_{1,R}} F_{1^*2(433)}+\frac{\omega_{433,R}}{\omega_{1,R} + \omega_{2,R} - \omega_{433,R}}F_{12(433)^*}\right)E^{\rm sat}_1 E^{\rm sat}_2 \cr
    &\qquad + \frac{\omega_{433,R}}{2 \omega_{1,R} - \omega_{433,R}}F_{11(433)^*}E_1^{{\rm sat}2} ~.
\end{align}
In Fig.~\ref{fig:wIefflm3}, we show the value of $\omega_{433,I}^{\rm eff}$ as a function of $\mu M$. The sign of $\omega_{433,I}^{\rm eff}$ is determined as a result of competition between the leading two comparable contributions represented by the diagrams (a) and (c). 
For most of the axion mass $\mu M$, $\omega_{433,I}^{\rm eff}$ is negative, {\it i.e.}, the dissipative process (a) dominates. In such cases, we do not have the growth of the $l=m=3$ cloud (see, for example, the right panel of Fig.~\ref{fig:fndlm3}). 
On the other hand, for $0.12 \lesssim \mu M \lesssim 0.24$, 
we have $\omega_{433,I}^{\rm eff} > 0$, and hence the growth of the $l=m=3$ cloud is expected.
In the left panel of Fig.~\ref{fig:fndlm3}, we show the evolution of cloud energies for $\mu M = 0.16$, 
where the $l=m=3$ cloud eventually dominates the whole condensate.

\begin{figure}[t]
 \centering
 \includegraphics[keepaspectratio, scale=0.4]{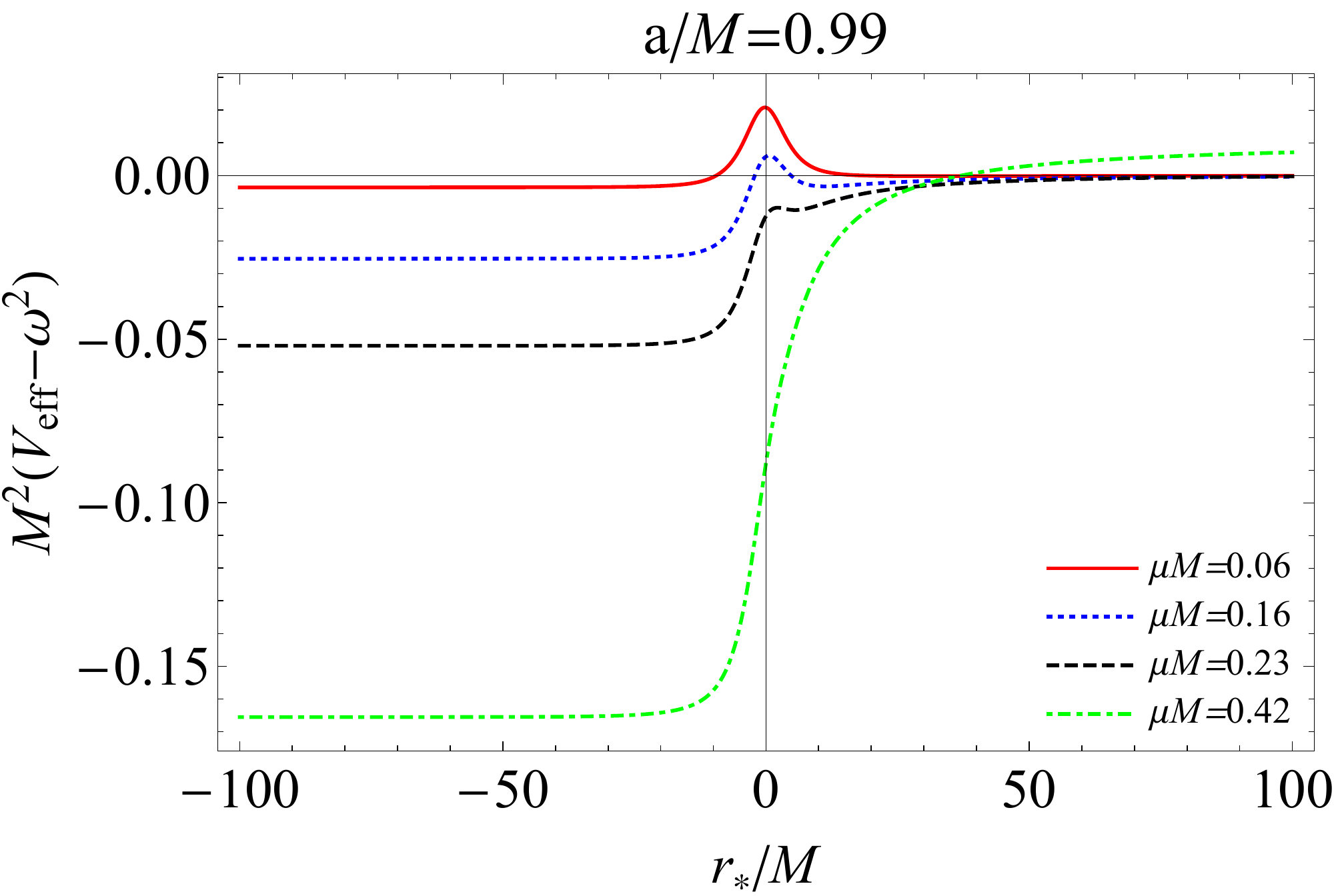}
 \caption{The effective potential for the $l=m=0$ mode function. The red solid, blue dotted, black dashed, and green dotted-dashed lines corresponds to the effective potential of the $l=m=0$ mode function with $\mu M = 0.06,0.16,0.23,0.42$.  The spin of the central BH is set to $a/M = 0.99$.}
 \label{fig:potential}
\end{figure}

The behavior of $\omega_{433,I}^{\rm eff}$ can be understood by the behavior of the effective potential of the $m=0$ mode function. The effective potential is defined by writing the equation of motion for the radial mode function \eqref{eq:EOMrad} in the Schr\"odinger form (see \cite{Zouros:1979iw} for the actual expression of the effective potential)
\begin{align}
    \frac{d^2 u}{dr_*^2} + (\omega^2 - V_{\rm eff})u = 0~,
\end{align}
with
\begin{align}
    u \equiv \sqrt{r^2 + a^2}R_{lm\omega}~.
\end{align}
In Fig.~\ref{fig:potential}, We show the effective potential of the $l=m=0$ mode excited by the process (c) with different $\mu M$. When $\mu M \lesssim 0.1$, the frequency of the $m=0$ mode is much smaller than the potential barrier. Thus, the energy flux to the horizon is suppressed. On the other hand, when $\mu M \gtrsim 0.1$, the mass of the axion starts to lower the potential barrier of the $m=0$ modes. Then the amplitude of the $m=0$ mode functions around the peak of the clouds gets significantly enhanced. As a result, the overlap integral of the $m=0$ mode and the clouds becomes large and the dissipative process involving the $m=0$ mode becomes relatively faster than that involving the $m=3$ mode. 
However, when $\mu M  \gtrsim 0.2$, the potential barrier becomes smaller than the frequency of the $m=0$. Therefore, enhancement of the amplitude of the $m=0$ mode function becomes moderate and flux to the horizon is no larger than that to the infinity.

\section{Primary cloud composed of a higher multipole mode}\label{sec:5}

So far, we considered axion condensate starting with the dominance of the $l=m=1$ fundamental mode. 
However, this is the case only when the axion mass satisfies $\mu M\lesssim 0.45$ (see Fig.~\ref{fig:omegaI}). In this section, we consider the case with a larger axion mass (or a larger BH mass). In such a situation, the fastest growing mode is the fundamental mode of $l=m=2$, {\it i.e.}, $(l_1,m_1) = (2,2)$.

\begin{figure}[t]
 \centering
 \includegraphics[keepaspectratio, scale=0.3]{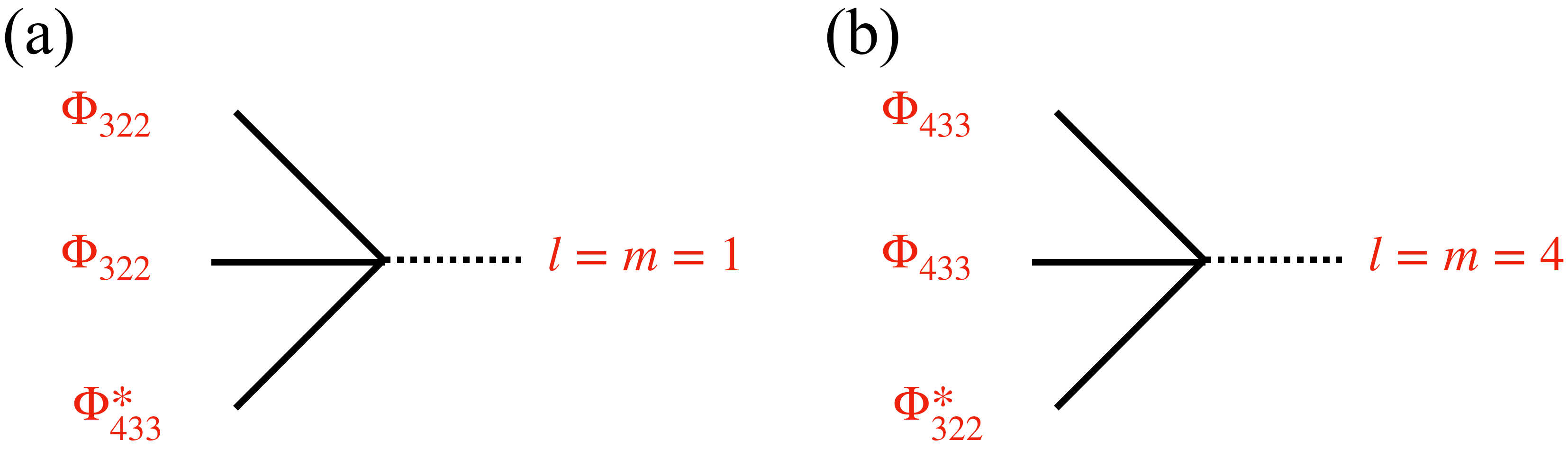}
 \caption{The interaction between the $l=m=2$ and the $l=m=3$ clouds when the primary cloud is the $l=m=2$ mode.}
 \label{fig:hlintleading}
\end{figure}

\begin{figure}[t]
 \centering
 \includegraphics[keepaspectratio, scale=0.5]{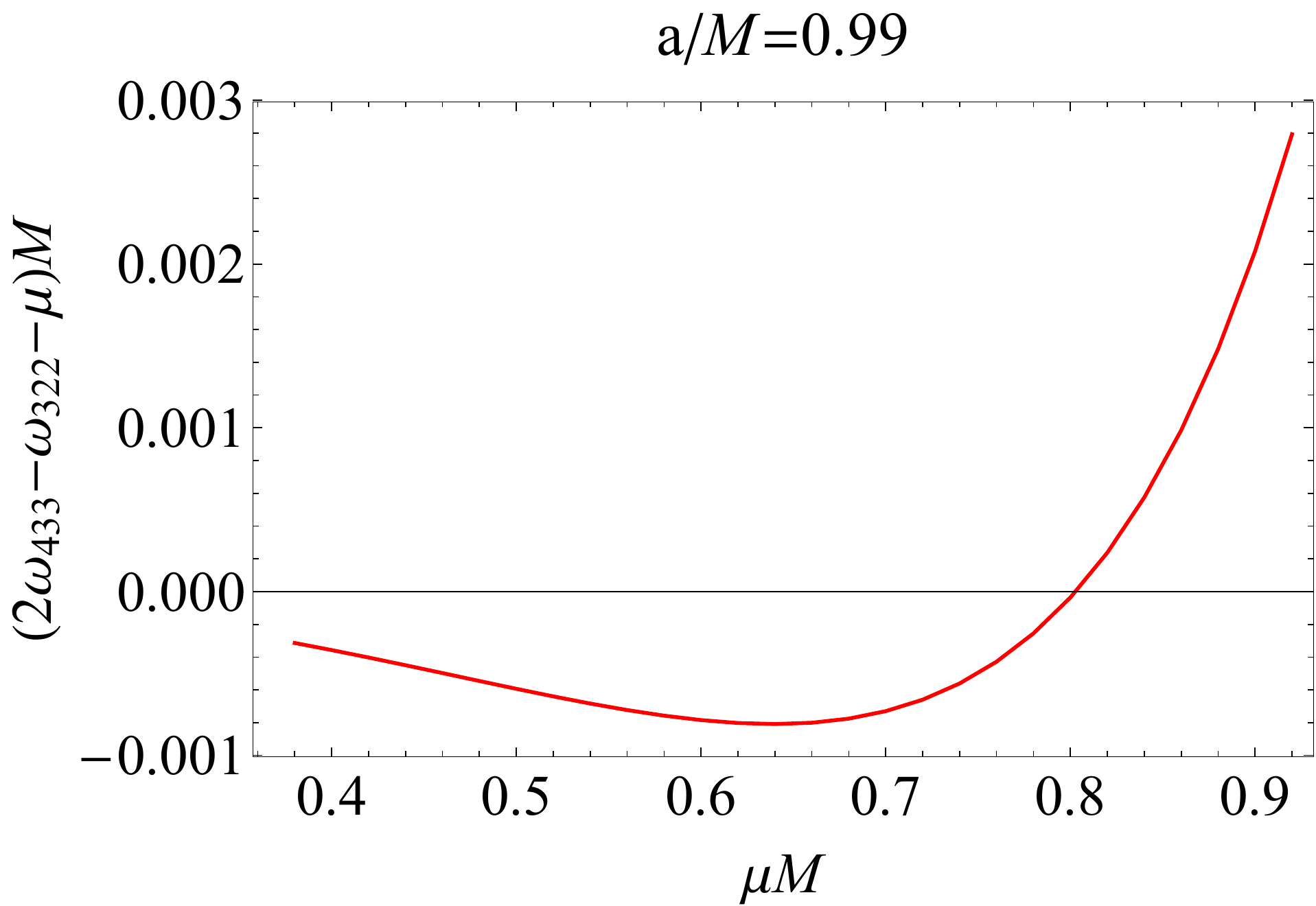}
 \caption{The frequency of the excited mode presented in the right diagram (b) of Fig.~\ref{fig:hlintleading}.}
 \label{fig:w3lm2}
\end{figure}

\begin{figure}[t]
 \centering
 \includegraphics[keepaspectratio, scale=0.4]{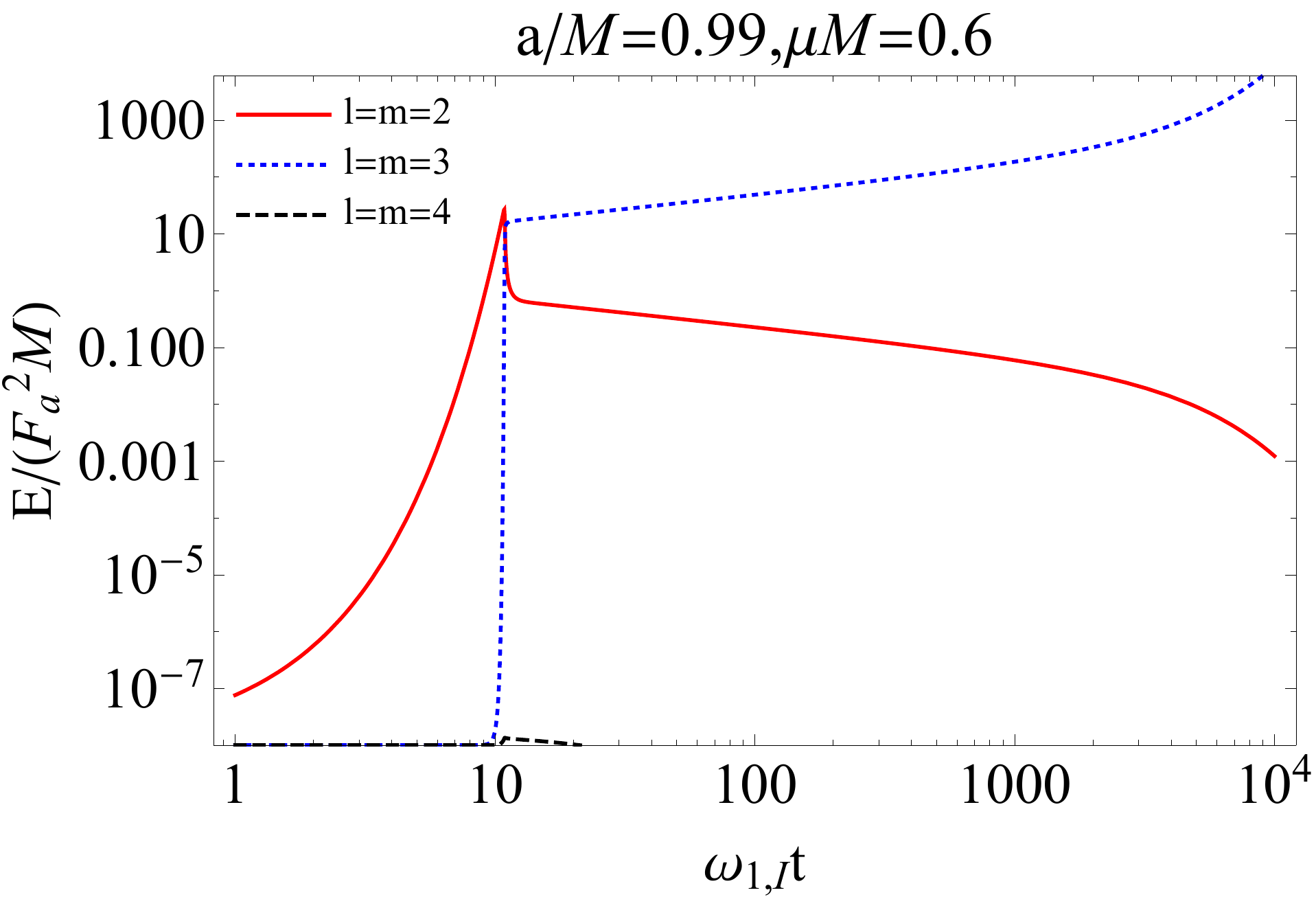}
 \includegraphics[keepaspectratio, scale=0.4]{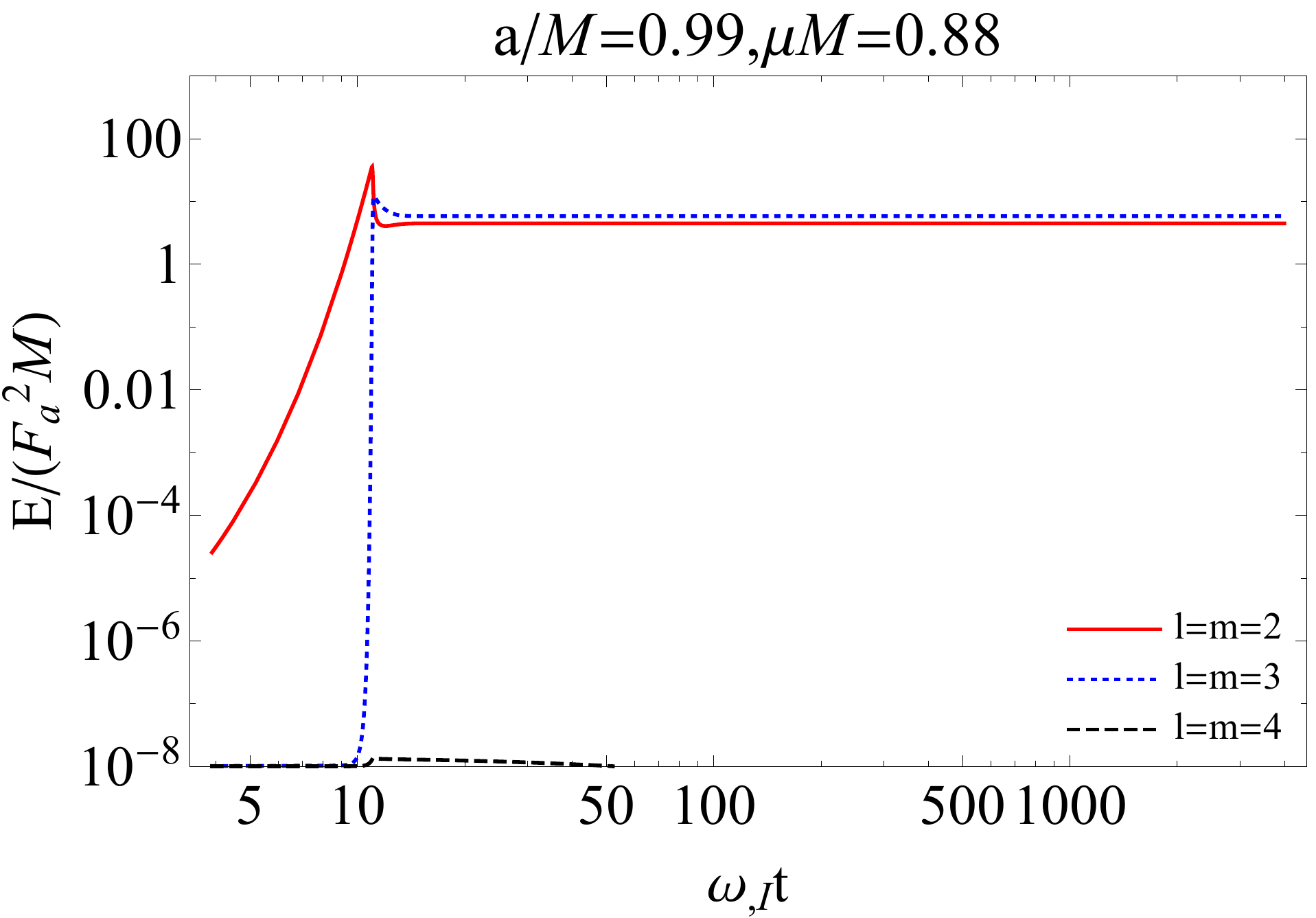}
 \caption{Time evolution of energy of the modes $l=m=2$, $l=m=3$, and $l=m=4$ when the primary cloud is composed of the $l=m=2$ mode. The red solid, blue dotted, and black dashed lines correspond to the fundamental mode of $l=m=2$, $l=m=3$, and $l=m=4$, respectively. The left and the right panels show the cases with $\mu M = 0.6$ and $0.88$, respectively.  We set the BH spin to $a/M = 0.99$.}
 \label{fig:hlevolution}
\end{figure}

In parallel with Sec.~\ref{sec:2B}, we choose the secondary cloud to be the adjacent higher multipole mode, {\it i.e.}, $(l_2,m_2) = (3,3)$. The relevant process is shown in Fig.~\ref{fig:hlintleading}. 
Whether the diagram (b) in Fig.~\ref{fig:hlintleading} contributes to the energy dissipation depends on the axion mass $\mu$.
As a reference, let us consider the case in which the non-relativistic approximation is valid ($\mu M \ll 1$). Then, the real part of the axion cloud frequency can be well approximated by the energy of the hydrogen atom,
\begin{align}
	\omega_1|_{l_1 = m_1 = 2} \sim \mu \left(1 - \frac{(\mu M )^2}{2\times 3^2} \right)~,\\
	\omega_2|_{l_1 = m_1 = 3} \sim \mu \left(1 - \frac{(\mu M )^2}{2\times 4^2} \right)~, 
\end{align}
and the frequency of the excited mode presented in the right diagram (b) of Fig.~\ref{fig:hlintleading} is given by
\begin{align}
	2 \omega(l=m=3) - \omega(l=m=2) \sim \mu - \mu  \frac{(\mu M )^2}{144} < \mu~.
\end{align}
Therefore, this mode is bounded by the gravitational potential and cannot carry the energy to infinity. In addition, the mode is superradiant and thus contribute to the dissipation of $l=m=2$ mode. We can numerically confirm that this inequality is true for $\mu M \lesssim 0.8$ (see Fig.~\ref{fig:w3lm2}). Therefore, saturation of superradiance cannot be expected for an axion mass in this range, and the  $l=m=3$ cloud keeps growing. An interesting point is that the growth of the $l=m=3$ cloud is accelerated by the presence of the $l=m=2$ cloud. On the contrary, for $\mu M \gtrsim 0.8$, everything is parallel to the case of the $l_1 = m_1 = 1$ mode dominance; the condensate settles into a quasi-stationary configuration composed of the $l_1 = m_1 = 2$ and $l_2 = m_2 = 3$ clouds. In Fig.~\ref{fig:hlevolution}, we show examples of time evolution for $\mu M =0.6$ and $\mu M = 0.88$ with $a/M = 0.99$.

The qualitative picture presented above does not change even if we include the $l=m=4$ mode (see Fig.~\ref{fig:hlevolution}). In fact, the $l=m=4$ mode cannot be excited, by the same mechanism that prevents the excitation of the $l=m=2$ overtones for condensate starting with the $l=m=1$ fundamental mode (see Sec.~\ref{sec:4B}). In summary, clouds starting from the dominance of a higher multipole mode with $l \ge 2$ would terminate the growth due to the superradiant instability only when the axion mass marginally satisfies the superradiance condition. In the case of initial $l=m=2$ dominance, only when the axion mass falls in the range, $0.8 \lesssim \mu M \lesssim 0.92$, the superradiant instability is regulated by dissipation.

\section{Summary and discussion}\label{sec:6}

In this paper, we numerically investigated the non-linear cloud evolution with self-interaction by using the adiabatic approximation. First, the case in which the $l=m=1$ fundamental mode is the mode with the maximum growth rate ($\mu M \lesssim 0.45$) was examined. In this case, the dissipation due to the interaction between the $l=m=1$ and $l=m=2$ fundamental modes is strong enough that the condensate ceases to grow and settles into a quasi-stationary state. In particular, since the self-interaction for the typical cosine-type potential is attractive, different modes attract each other and the energy dissipation due to the mode-mode interaction is enhanced. Therefore, even if the secondary $l=m=2$ cloud starts to evolve from a small magnitude expected by quantum fluctuations, dissipation inevitably becomes strong enough to terminate the growth of the primary cloud. 

Furthermore, we investigate whether the quasi-stationary state is stable or not under the influence of the other self-interaction effects. For example, there could be interactions with other modes and dissipation due to quantum mechanical processes. However, these processes are much slower than the classical processes between the $l=m=1$ and $l=m=2$ fundamental modes, and do not significantly alter the quasi-stationary state composed of these two modes. Our results show that the final configuration of the cloud is composed mainly of the $l=m=1$ and $l=m=2$ fundamental modes, augmented by the $l=m=1$ overtones with a smaller amplitude. The $l=m=2$ overtones cannot grow because of the dissipation induced by the interaction. The higher multipole modes with $l= 3$ can be excited only for the axion mass satisfying $0.12 \lesssim \mu M \lesssim 0.24$. Since the $l=m=3$ modes can be the dominant mode for this case, further interaction involving the $l=m=3$ mode can significantly alter the configuration. For this case, calculation beyond the three state approximation adopted in this paper is necessary.

Finally, the case in which the $l=m=2$ fundamental mode is the fastest growing mode ($0.45 \lesssim \mu M \lesssim 0.92$) is also investigated. In this case, the evolution of the condensate qualitatively depends on the axion mass; When the axion mass is smaller than 0.8, a quasi-stationary state is not achieved, unlike the case starting with the $l=m=1$ fundamental mode. This is because the order of the magnitudes of energy level splittings are different and therefore the dissipative processes that are at work for the $l=m=1$ case can be prohibited. In this case, the $l=m=3$ mode grows to have a very large amplitude, and the non-linear calculations are required to follow the actual evolution. Unfortunately, because of the high numerical cost, we could not present the results for this case, leaving it for future work. When the mass is larger than 0.8, the shift of the energy spectrum changes due to the relativistic effects becoming sufficiently large to open up channels for dissipation. In this case, the final state becomes a superposition of the clouds composed of the $l=m=2$ and $l=m=3$ fundamental modes. We summarize our results in Tab. \ref{tab:result}.

\begin{table}[h]
 \caption{Summary of our results, classified by the size of the gravitational coupling. The same figure is referenced for $\mu M \lesssim 0.12$ and $0.24 \lesssim \mu M \lesssim 0.45$, since both cases follow qualitatively the same evolution track.}
 \label{tab:result}
 \centering
  \begin{tabular}{|c|c|c|}
   \hline
    $~$The range of gravitational  $~$& $~$Dominant modes $(n,l,m)$ in the$~$& $~$Example of $~$ \\
    $~$coupling $\mu M$ $~$& $~$quasi-stationary configuration$~$&$~$time evolution$~$\\
   \hline \vspace{-10pt} &&\\ \hline  \vspace{-6pt}
    & Dominant:$(2,1,1) + (3,2,2)$ & \\ \vspace{-6pt}
    {$\mu M \lesssim 0.12$} & & {Fig. \ref{fig:overtone}}\\
    &Subdominant:$(n,1,1),n \le 5$ & \\ \hline
   $0.12\lesssim \mu M\lesssim 0.24$ & $(2,1,1) + (3,2,2) + (4,3,3)$ & Fig. \ref{fig:fndlm3}(left panel)\\ \hline  \vspace{-6pt}
    & Dominant:$(2,1,1) + (3,2,2)$ &\\  \vspace{-6pt}
    {$0.24 \lesssim \mu M \lesssim 0.45$} && 
    {Fig. \ref{fig:overtone}}\\ 
  &Subdominant:$(n,1,1),n \le 5$ & \\ \hline
    $0.45\lesssim \mu M\lesssim 0.8$ & $(4,3,3)$ & Fig. \ref{fig:hlevolution}(left panel)\\ \hline
    $0.8\lesssim \mu M\lesssim 0.92$ & $(3,2,2) + (4,3,3)$ & Fig. \ref{fig:hlevolution}(right panel)\\
   \hline
  \end{tabular}
\end{table}

We should mention that the spin of the central BH is fixed to $a/M = 0.99$ in all of our calculations. This is because our interest is mainly in the relativistic regime, where the gravitational coupling is large. One might worry whether the spin of the central BH changes the result. However, we expect that our conclusion is robust under the change of the BH spin since the previous calculation in \cite{Omiya:2022mwv} shows that BH spin only has a small effect on the evolution of the self-interacting cloud, except for the presence of the superradiant instability and its rate. Nevertheless, a detailed analysis with different BH spin is necessary, as we will see below.

In this paper, the background is fixed to the Kerr spacetime, and the spin-down of the BH and the gravitational wave emission by the condensate are ignored. We make a few comments on these points since they are important for actual observations. First, we discuss the gravitational wave emission. After the condensate reaches a quasi-stationary state, several modes are simultaneously excited. In such a situation, there are two types of gravitational wave emission processes: the pair annihilation and the level transition. Since the level transition signal has a much lower frequency than the pair annihilation~\cite{Arvanitaki:2010sy}, we expect 
simultaneous continuous gravitational wave emissions in different frequency bands (see \cite{Siemonsen:2019ebd} in the case of spin-1 superradiance). 
Numerical calculations are required to estimate the amplitude of emitted gravitational waves, and we also leave them for future work.

Next, we discuss the BH spin-down. The evolution of the BH spin would be described by 
\begin{align}
    \frac{d J_{BH}}{dt} = - \frac{2m_1\omega_{1,I}}{\omega_{1,R}} E_{1} -  \frac{2m_2 \omega_{2,I}}{\omega_{2,R}} E_{2} \sim -\frac{2m_1\omega_{1,I}}{\omega_{1,R}} E_{1}~.
\end{align}
Therefore, the spin-down timescale $\tau_{\rm spin}$ is given by
\begin{align}
 	\tau_{\rm spin}^{-1} = \frac{1}{J_{BH}}\frac{d J_{BH}}{dt} \sim 2m_1 \omega_{1,I} \frac{E_{\rm 1}}{aM} \sim 10^2 m_1 \omega_{1,I}{F_a}^2~.
\end{align}
Here, we adopt $E_1 \sim 10^2 F_a^2 M $ taken from our numerical calculation.
In the case of the string axion ($F_a \sim 10^{-3}$),  we have $\tau_{\rm spin} \sim 10^4\omega^{-1}_{1,I}$, which is sufficiently long compared to the instability timescale. Therefore, the spin-down of the BH can be neglected in our calculation. This indicates that the condensate settles to the quasi-stationary state presented in this paper.

However, the effect of the spin-down should be taken into account when we consider the gravitational wave emission. Since the quasi-stationary state is supported by the superradiant instability, the overall amplitude of the condensate decreases as the BH spins down. As a result, the amplitudes of gravitational waves also decrease. In particular, the condensate gradually disappears when the superradiance condition ceases to be satisfied. Therefore, to discuss the actual observability of gravitational waves, we need to carefully examine the effect of the spin-down. We also leave this issue for future work.

\begin{acknowledgments}
TT is supported by JSPS KAKENHI Grant Number JP17H06358 (and also JP17H06357), \textit{A01: Testing gravity theories using gravitational waves}, as a part of the innovative research area, ``Gravitational wave physics and astronomy: Genesis'', and also by JP20K03928. 
%HO is supported by the Japan Society for the Promotion of Science (JSPS).
HO is supported by Grant-in-Aid for JSPS Fellows JP22J14159.
T. Takahashi was supported by JST, the establishment of university fellowships towards the creation of science technology innovation, Grant Number JPMJFS2123.
H. Y is in part supported by JSPS KAKENHI Grant Numbers JP22H01220 and JP21H05189, and is partly supported by Osaka Central Advanced Mathematical Institute (MEXT Joint Usage/Research Center on Mathematics and Theoretical Physics JPMXP0619217849).
\end{acknowledgments}

\appendix

\section{Calculation of the energy and angular momentum  flux}\label{sec:AppA}

In this appendix, we calculate the energy flux $F^{E}_{\rm tot}$ (Eq.~\eqref{eq:enecons})and angular momentum flux $F^J_{\rm tot}$ (Eq.~\eqref{eq:angcons2}). Let us first calculate the energy flux. The energy momentum tensor of the scalar field is given by
\begin{align}
    T_{\mu\nu}(\phi) = \partial_\mu\phi \partial_\nu\phi 
    + g_{\mu\nu} \left(-\frac{1}{2}(\partial\phi)^2 -V(\phi)\right)~.
\end{align}
Substituting to Eqs.~\eqref{eq:enefluxH} and \eqref{eq:enefluxI}, we obtain
\begin{align}
    F_{\mathcal{H}^+} &= M r_+ \int d\cos\theta\, d\varphi\ \partial_t \phi (\partial_t \phi + \Omega_H \partial_\varphi \phi)|_{r\to r_+}~,\\
    F_{\mathcal{I}^+} &=\int d\cos\theta\, d\varphi \ r^2 \partial_t\phi \partial_r\phi|_{r\to \infty}~.
\end{align}
Up to the first order in the perturbative expansion, $\phi$ is given by
\begin{align}
    \label{eq:pertsol1}
    \phi = \sqrt{E_1}\phi_1 + \sqrt{E_2}\phi_2 + E_1 \sqrt{E_2} \phi^{(1)}_0 + \sqrt{E_1}E_2 \phi^{(1)}_3~,
\end{align}
where
\begin{align}
    \phi^{(1)}_0 &= -\frac{\mu^2}{2!}e^{-i\omega_0 t}\sum_l S_{l0\omega_0}(\theta)\int d r'd\cos\theta' ({r'}^2 + a^2 \cos^2\!\theta') S_{l0\omega_0}(\theta')G_{l0\omega_0}(r,r')R_1(r')^2 R_2^*(r') S_1(\theta')^2 S^*_2(\theta') + {\rm c.c.}~,\\
    \phi^{(1)}_3 &= -\frac{\mu^2}{2!}e^{-i\omega_3 t + 3\varphi}\sum_l S_{l3\omega_3}(\theta) \int d r'd\cos\theta' ({r'}^2 + a^2 \cos^2\!\theta') S_{l3\omega_3}(\theta') G_{l3\omega_3}(r,r')R^*_1(r') R_2(r')^2 S^*_1(\theta') S_2(\theta')^2 + {\rm c.c.}~.
\end{align}

The first, second, and third terms in Eq.~\eqref{eq:pertsol1} contribute to the flux at the horizon $F_{\mathcal{H}^+}$ and the fourth term contributes to the flux to infinity $F_{\mathcal{I}^+}$. In particular, the first and the second terms are responsible for the superradiant instability of the $l=m=1$ and the $l=m=2$ modes~\cite{Zouros:1979iw,Dolan:2007mj}, and they contribute to the flux at the horizon, $F_{\mathcal{H}^+}$, as
\begin{align}
    -2 \omega_{1,I}E_1 - 2\omega_{2,I} E_2~.
\end{align}
To evaluate the contribution from the $\phi^{(1)}_0$, we need the asymptotic behavior of the $\phi^{(1)}$. Using the expression \eqref{eq:modegreen} and the asymptotic behavior of the mode functions \eqref{eq:BC}, the field is evaluated as  
\begin{align}
    \phi_0^{(1)} \to -\frac{\mu^2}{2}e^{-i\omega_0 t - i \omega_0 r_*}\sum_l Z_{l}^{\mathcal{H}^+} S_{l0\omega_0}(\theta) + {\rm c.c.}~, 
\end{align}
near $r = r_+$ with
\begin{align}
    Z_{l}^{\mathcal{H}^+} =\frac{1}{ W_{l0}(\omega_0)}\int d r'd\cos\theta' ({r'}^2 + a^2 \cos^2\!\theta') S_{l0\omega_0}(\theta')R^{\rm up}_{l0\omega_0}(r')R_1(r')^2 R_2^*(r') S_1(\theta')^2 S^*_2(\theta')~.
\end{align}
Using the orthonormality relation,
\begin{align}
    \int d\cos\theta\, S_{lm\omega} S_{l'm\omega}^* = \delta_{ll'}~,
\end{align}
we obtain
\begin{align}\label{eq:flux02}
        M r_+ \int d\cos\theta d\varphi\ \partial_t \phi_0^{(1)} (\partial_t \phi_0^{(1)} + \Omega_H \partial_\varphi \phi_0^{(1)})|_{r\to r_+} =& \left(2 \pi M \mu^4 r_+  \omega_0^2 \sum_{l}|Z_{l}^{\mathcal{H}^+}|^2\right)E_1^2 E_2\cr
        \equiv& F_0 E_1^2 E_2~.
\end{align}

The flux to infinity is calculated similarly. The asymptotic behavior of $\phi^{(1)}_3$ in the limit $r\to \infty$ is given by 
\begin{align}
    \phi_3^{(1)} \to& -\frac{\mu^2}{2}\frac{e^{-i\omega_3 t + 3 i \varphi + i \sqrt{\omega_3^2 - \mu^2} r_*}}{r} \sum_l Z_{l}^{\mathcal{I}^+} S_{l3\omega_3}(\theta) + {\rm c.c.}~,
\end{align}
with 
\begin{align}
    Z_{l}^{\mathcal{I}^+} = \frac{1}{W_{l3}(\omega_3)}\int d r'd\cos\theta' ({r'}^2 + a^2 \cos^2\!\theta') R^*_1(r') R_2(r')^2 S^*_1(\theta') S_2(\theta')^2 ~.
\end{align}
Similarly to the derivation of Eq.~\eqref{eq:flux02}, we obtain
\begin{align}
        F_{\mathcal{I}^+} =& \left(\pi \mu^4\omega_3\sqrt{\omega_3^2 - \mu^2} \sum_{l}|Z_{l}^{\mathcal{I}^+}|^2\right)E_1 E_2^2\cr
        \equiv& F_3 E_1 E_2^2~.
\end{align}
The net energy flux is obtained by adding all the fluxes. The result is
\begin{align}\label{eq:totenergyfluxpert}
    F^E_{\rm tot} =& F_{\mathcal{H}^+} + F_{\mathcal{I}^+}\cr
    =& -2 \omega_{1,I}E_1 - 2\omega_{2,I} E_2 + F_0 E_1^2 E_2 + F_3 E_1 E_2^2~.
\end{align}

The angular momentum flux of the wave with given $(\omega,m)$ is given by  multiplying the energy flux by $m/\omega$~\cite{Brito:2015oca}. 
Since the respective terms in Eq.~\eqref{eq:totenergyfluxpert} have $(\omega,m) = (\omega_1,1),(\omega_2,2),(\omega_0,0),$ and $(\omega_3,3)$, the total angular momentum flux is given by
\begin{align}
    F^J_{\rm tot} =& -2 \omega_{1,I}\frac{1}{\omega_{1,R}}E_1 - 2\omega_{2,I} \frac{2}{\omega_{2,R}}E_2 +  \frac{3}{\omega_3}F_3 E_1 E_2^2~.
\end{align}

% Create the reference section using BibTeX:
\bibliography{axionref}

\end{document}